\documentclass[fleqn,usenatbib]{mnras}

\usepackage{newtxtext,newtxmath}
\usepackage{fontawesome}

\usepackage[T1]{fontenc}

\DeclareRobustCommand{\VAN}[3]{#2}
\let\VANthebibliography\thebibliography
\def\thebibliography{\DeclareRobustCommand{\VAN}[3]{##3}\VANthebibliography}

\usepackage{graphicx}	
\usepackage{amsmath}	
\usepackage{tabularx}
\usepackage[table,dvipsnames]{xcolor}
\usepackage[normal]{threeparttable}
\usepackage[nolist]{acronym}
\usepackage{multirow}
\usepackage{subfig}
\usepackage{cleveref}
\usepackage{hyperref}
\usepackage{xspace}
\usepackage{caption}

\crefformat{section}{\S#2#1#3}

\newcommand{\simba}[0]{\mbox{\textsc{Simba}}\xspace}
\newcommand{\simbaNG}[0]{\mbox{\textsc{Simba}}}
\newcommand{\hyenas}[0]{\mbox{\textsc{Hyenas}}\xspace}
\newcommand{\moxha}[0]{\mbox{\textsc{MOXHA}}\xspace}

\newcommand{\GIZMO}[0]{{ \textsc{Gizmo}}\xspace}

\newcommand{\SOXS}[0]{{ \textsc{SOXS}}\xspace}
\newcommand{\PyXSIM}[0]{{ \textsc{PyXSIM}}\xspace}

\newcommand{\yT}[0]{{ \textsc{yT\xspace}}\xspace}

\newcommand{\hMsun}{{$M_\odot /h$}}

\newacro{icm}[ICM]{intracluster medium}
\newacro{igrm}[IGrM]{intragroup medium}
\newacro{ism}[ISM]{interstellar medium}
\newacro{igm}[IGM]{intergalactic medium}
\newacro{agn}[AGN]{active galactic nucleus}
\newacro{smbh}[SMBH]{supermassive black hole}
\newacro{bcg}[BCG]{brightest cluster galaxy}
\newacro{bgg}[BGG]{brightest group galaxy}
\newacro{cgm}[CGM]{circumgalactic medium}
\newacro{gmc}[GMC]{giant molecular cloud}
\newacro{lem}[LEM]{Light Elements Mapper}
\newacro{cmb}[CMB]{Cosmic Microwave Background}
\newacro{bhl}[BHL]{Bondi-Hoyle-Lyttleton}
\newacro{grmhd}[GRMHD]{general relativistic}
\newacro{cxb}[CXB]{Cosmic X-ray Background}
\newacro{cie}[CIE]{collisional ionization equilibrium}
\usepackage[dvipsnames]{xcolor}

\newcommand{\commentblock}[1]{}

\title[\hyenas: X-ray Bubbles and Cavities]{\hyenas: X-ray Bubbles and Cavities in the Intra-Group Medium}

\author[F. J. Jennings et al.]{
Fred J. Jennings,$^{1}$\thanks{E-mail: \href{mailto:Fred.Jennings@ed.ac.uk}{Fred.Jennings@ed.ac.uk}}
Arif Babul,$^{2,3,4}$
Romeel Davé,$^{1,5,6}$
Weiguang Cui, $^{1,7,8}$
 and Douglas Rennehan $^9$
\\
$^{1}$Institute for Astronomy, University of Edinburgh, Royal Observatory,  Blackford Hill, Edinburgh, EH9 3HJ,  United Kingdom \\
$^{2}$Leverhulme Visiting Prof., Institute for Astronomy, University of Edinburgh, Royal Observatory,  Blackford Hill, Edinburgh, EH9 3HJ,  United Kingdom \\
$^{3}$Department of Physics \& Astronomy, University of Victoria, BC V8X 4M6, Canada \\
$^{4}$Infosys Visiting Chair Professor, Department of Physics, Indian Institute of Science, Bangalore 560012, India \\
$^{5}$University of the Western Cape, Department of Physics and Astronomy, Bellville, Cape Town 7535, South Africa \\
$^{6}$South African Astronomical Observatories, Observatory, Cape Town 7925, South Africa \\
$^{7}$Departamento de F\'{i}sica Te\'{o}rica, Universidad Aut\'{o}noma de Madrid, M\'{o}dulo 15, E-28049 Madrid, Spain  \\
$^{8}$Centro de Investigaci\'{o}n Avanzada en F\'isica Fundamental (CIAFF), Facultad de Ciencias, Universidad Aut\'{o}noma de Madrid, 28049 Madrid, Spain \\
$^{9}$Center for Computational Astrophysics,
Flatiron Institute, 162 Fifth Ave, New York, NY 10010, USA
}

\date{Accepted For Publication in MNRAS 2024 November 11. Received 2024 November 11; in original form 2024 July 19}

\pubyear{2015}

\begin{document}
\label{firstpage}
\pagerange{\pageref{firstpage}--\pageref{lastpage}}
\maketitle

\begin{abstract}
We investigate the role of the \simba feedback model on the structure of the Intra-Group Medium (IGrM) in the new \hyenas suite of cutting-edge cosmological zoom-in simulations. Using 34 high-resolution zooms of halos spanning from $10^{13}-10^{14}$ $M_\odot$ at $z=0.286$, we follow halos for 700 Myr, over several major active galactic nuclei (AGN) jet feedback events. We use the \moxha package to generate mock \textit{Chandra} X-ray observations, as well as predictive mocks for the upcoming LEM mission, identifying many feedback-generated features such as cavities, shock-fronts, and hot-spots, closely mimicking real observations.  
Our sample comprises $105$ snapshots with identified cavities, $50$ with single bubbles and $55$ with two, and spans three orders of magnitude in observed cavity enthalpies, from $10^{41}-10^{44}$ erg/s. Comparing semi-major axis length, midpoint radius, and eccentricity to a matched sample of observations, we find good agreement in cavity dimensions with real catalogues. We estimate cavity power from our mock maps following observational procedures, showing that this is typically more than enough to offset halo cooling, particularly in low-mass halos, where we match the observed excess in energy relative to cooling. Bubble enthalpy as measured with the usual midpoint pressure typically exceeds the energy released by the most recent jet event, hinting that the mechanical work is done predominantly at a lower pressure against the IGrM. We demonstrate for the first time that X-ray cavities are observable in a modern large-scale simulation suite and discuss the use of realistic cavity mock observations as new halo-scale constraints on feedback models in cosmological simulations.
\end{abstract}




\section{Introduction}

The physical process(es) driving the quenching of massive galaxies remains one of the key mysteries in understanding the formation of today's galaxy population.  Currently, models that are successful in creating a quenched galaxy population all invoke a process referred to as ``radio mode" or ``maintenance mode" feedback~\citep{SomervilleDave2015} from active galactic nuclei (AGN), which involves relativistic jets emanating from massive central black holes that heat the surrounding circum-galactic medium and counteracts the expected cooling of hot gas~\citep{BabulBalogh2002,BruggenRuszkowski2005,RoychowdhuryRuszkowski2005, SijackiSpringel2006,NusserSilk2006}. However, the exact mechanism by which jets transfer energy to the surrounding gas to prevent cooling remains uncertain~\citep{NusserSilk2006,Fabian2012, BabulSharma2013,CieloBabul2018}. Jets are seen to be highly collimated, occasionally terminating in a lobe, yet somehow this must counteract the cooling from a spherical halo of hot gas.  The leading idea for how this happens is that the jet lobes inflate cavities or ``bubbles" of overpressurised gas, generating pressure waves in the surrounding hot atmosphere that then sphericalises the jet energy input over time.  X-ray observations indicate the presence of such bubbles in clusters and groups~\citep{BirzanRafferty2004,PanagouliaFabian2014,McDonaldGaspari2018} lending credence to this scenario, yet the link between jets, bubbles, halo gas heating, and quenching remains tenuous.

Galaxy demographics indicate that widespread quenching must occur in the regime of galaxy group \citep{RasmussenMulchaey2012,CoendaMartinez2018,SalernoMartinez2019}.  Isolated galaxies and those in poor groups such as the Local Group tend to be star-forming, but moving to slightly larger halos finds a growing fraction of quenched galaxies \citep{WetzelTinker2012}, and there exists a similar trend also in the properties of their central galaxies \citep{WeinmannvandenBosch2006,GozaliaslFinoguenov2016,EinastoEinasto2024}. Satellites and centrals also appear to be tightly correlated in terms of their star-forming properties \citep{WeinmannvandenBosch2006}.  Hence if black hole jets are the mechanism that drives or maintains quenching, then it should be operating on this halo mass scale.  Indeed many galaxy groups are observed to contain a hot atmosphere seen in the X-rays, i.e. an \ac{igrm} \citep{Tully2015}, which has a complex, multiphase structure with cold, precipitated clouds and filaments co-existing with a mass-dominant hot phase \citep{SharmaMcCourt2012,PrasadSharma2015, PrasadSharma2017, McCabeBorthakur2021, McCabeBorthakur2023, SaeedzadehJung2023}. 

The inferred cooling timescales in the centres of groups result in estimated cooling rates of a few tens of $\mathrm{\  M}_\odot/$yr, in contrast to the observed star formation rates and molecular gas masses in these systems which puts the cooling rate at $\la 1-10 \mathrm{\  M}_\odot/$yr \citep{McDonaldGaspari2018}. Hence despite the multi-phase nature of the \ac{igrm}, cooling appears to be strongly suppressed throughout most of the hot gas.

To better understand how the jet energy suppresses cooling, many groups have undertaken numerical simulations of this process.  Early work focused on constructing idealised halos of hot gas, and understanding how jets can inflate bubbles  \citep{VernaleoReynolds2006}, excite gravitational modes as cavities transform into overdensities \citep{OmmaBinney2004, ReynoldsBrenneman2005}, and drive sound waves which heat the medium \citep{RuszkowskiBruggen2004, FabianReynolds2005, BambicReynolds2019}. Subsequent work aimed at simulating isolated halos but containing cosmologically motivated substructure.

In these models, the plasma in the black hole accretion disk is heated and uplifted via mechanical jets emanating from a central \ac{smbh} \citep{PopeBabul2010,DuboisDevriendt2010,GaspariRuszkowski2012,GaspariBrighenti2013, PrasadSharma2015, PrasadSharma2017, PrasadSharma2018}. These jets then inflate large cavities, or bubbles, of hot plasma in the \ac{igrm}, which shock the denser medium, heating the gas and preventing cooling.  Hence in these idealised situations, it appears radio mode feedback is able to sphericalise jet energy input as long as the jets are able to change direction on timescales that are short compared to the dynamical and cooling timescales of the halo \citep{BabulSharma2013,CieloBabul2018}.

Separately, cosmological simulations of galaxy formation have attempted to incorporate some version of jet feedback, with the primary motivation of quenching massive galaxies.  \citet{SijackiSpringel2007} included the impact of jet-driven bubbles via inflating bubbles by hand in the surrounding medium, which showed some promise but was not able to fully quench galaxies.  \citet{DuboisPeirani2016, KavirajLaigle2017} directly modelled the jets accounting for black hole spin within a cosmological setting in the HorizonAGN simulation~\citep{DuboisPichon2014}, but likewise had trouble coupling the energy sufficiently to produce a realistic quenched galaxy population.  More recently, the effects of AGN feedback have been included into cosmological simulations such as IllustrisTNG~\citep{WeinbergerSpringel2017} and \textsc{Obsidian} \citep{RennehanBabul2023}, both of which have been able to produce a realistic quenched population. Both models randomise the jet direction at every timestep in order to help distribute energy in an isotropic sense. Locally isotropic \ac{agn} feedback has been shown in the \textsc{Romulus} \citep{TremmelKarcher2017} simulations to be able to still produce directional jets with varying direction via interaction with an anisotropic gas profile in the cluster core \citep{TremmelQuinn2019}.

The \simba simulation successfully employs stably bipolar jets to produce a quenched galaxy population in good agreement with observations~\citep{DaveAngles-Alcazar2019, SzpilaDave2024}. \simba stands apart in terms of its model as being one that can achieve this whilst freeing the black hole jet axis to reorient based on the angular momentum of gas within the accretion kernel, without requiring enforced isotropy as several other simulations do.  By comparing to models run without jets, it is clear that galaxy quenching is primary driven by the action of the jets, although the effect of an additional feedback mode associated with X-ray photon pressure is not negligible~\citep{Cui2021}.  Hence, \simba is at least one example of a full cosmological galaxy formation simulation that is able to directly associate the quenching of massive galaxies with the action of bipolar AGN jet feedback on the \ac{icm} and \ac{icm}/\ac{igrm}, e.g. via heating and driving outflows.  This begs the question however on the exact nature of the energy transfer: Did \simba achieve this by inflating bubbles in the surrounding gas and transferring energy into the hot \ac{igrm} in a manner consistent with observations?  This is the question we seek to answer in this work.

Bubbles and cavities have now been observed in X-ray observations of many groups and clusters \cite{}, and allow one to make estimates of the jet power, the time since the feedback event occurred, and the suppression of cooling within the central regions of groups. There are however many aspects of \ac{agn} feedback which we do not understand well. The energy output of individual feedback events and their frequency, how exactly the jetted material propagates through, and couples to, the hot \ac{igrm}, and to what extent cooling flows are disrupted, are still not fully understood. One of the best ways to investigate the nature of \ac{agn} feedback and its effects is through halo-scale, or, ideally, cosmological simulations. Feedback is typically implemented in two modes; a hot radiative, or "quasar", mode, operating at high Eddington fractions, and a high-velocity "jet" mode (mechanical feedback) operating at low Eddington ratios \citep{BestHeckman2012, HeckmanBest2014} (though see \citealt{RennehanBabul2023} for an example of an alternative implementation).

Simulations of \ac{agn} feedback in (generally idealised) clusters show that the action of the jet on the \ac{igrm} or \ac{icm} gas is complex and non-linear. Feedback has been shown to both shut-off star formation due to gas removal when active \citep{SijackiSpringel2006,SijackiSpringel2007,McCarthySchaye2010,LeBrunMcCarthy2014,BeckmannDevriendt2017,RaoufSilk2019,Cui2021,PiotrowskaBluck2022}, reducing stellar mass fractions in-line with observation. However, in some circumstances, it can promote star formation locally after the fact due to gas compression near shocks and through the drawing out of low-entropy gas into filaments \citep{NayakshinZubovas2012,Silk2013, SilkDiCintio2014,ZubovasBourne2017,HuskoLacey2022}. Star formation is shown to be suppressed in and around radio lobes \citep{ShabalaKaviraj2011}, and the promotion of star formation has been corroborated by observations of sites of active star formation around X-ray cavities and jet-induced shocks in multi-wavelength studies \citep{CrockettShabala2012, VantyghemMcNamara2018}.

Generally groups and clusters are simulated in idealised halos, either as single systems or at best as a limited number of them. The computational cost of running such simulations limits the sample size, and the idealised initial conditions are in attempt to be able to draw general conclusions about feedback, without findings being muddied by a complex halo thermodynamic and velocity structure upon initialisation. In reality, groups are significantly affected by their environment via gas inflow and accretion from the cosmic web, tidal forces from neighbouring halos or subhalos, and mergers. Furthermore, a representative sample of halos in terms of mass, assembly time, and \ac{bcg} morphology should be simulated for a true statistical sample of feedback events. Finally, the feedback parameters in these studies should be calibrated such that they produce the correct, observed galaxy populations in cosmological simulations. The local properties of jets can then be studied self-consistently and be predictive, as the calibration is done on a global scale.  This is what we aim to do.

An ideal platform for such a study is to use high-resolution zoom simulations of a cosmological region encompassing a halo and its direct environment, with a simulation model which is proven to reproduce realistic galaxy populations. This gives the greatest balance between including large scale processes such as halo accretion and subhalo mergers, as well as providing sufficiently high resolution to be able to accurately simulate halo centers and resolve sub-kiloparsec-scale gas physics. To this end we employ \hyenas, a cutting-edge suite of zoom-in simulation boxes with halos selected based on virial mass and formation time, simulated with the \simba model, which has a proven track-record in reproducing observables in cosmological-scale boxes.

Our paper is organised as follows; In \hyperref[sec:Simulation]{\S2}, we detail the \hyenas simulation set-up as well as the group population we work with. In \hyperref[sec:Cavity Analysis]{\S3} we outline the selection of cavities, and our analysis of them. We also present measurements of bolometric cooling luminosities  and accretion rates, linking them both to cavity properties. \hyperref[sec:Case Studies]{\S4} contains a few brief case-studies of groups selected from our sample, and provides comparison between \textit{Chandra} and \textit{LEM} observations. We conclude this work with a discussion of our findings in \hyperref[sec:Discussion]{\S5}

\section{The Simulations} \label{sec:Simulation}
\subsection{Hyenas}

The simulations in this paper are drawn from the \hyenas \texttt{Level-1} suite of simulation boxes \citep{CuiJennings2024}, with the dark matter particle masses of $8.1 \times 10^6$ \hMsun~ and gas particle masses of $1.5 \times 10^6$ \hMsun (almost one order of magnitude resolution better than \simba). The initial conditions are selected from a large dark matter box (L = $200\,\mathrm{cMpc}\,h^{-1}$) spanning a range in mass and halo formation time, with the aim of choosing a representative sample of halos. The galaxy formation model, including the feedback prescription, is essentially identical to \simba \citep{DaveAngles-Alcazar2019}.  We now outline the most important features of the \hyenas (\simba) model in the context of this project, with a main focus on the active galactic nuclei (\ac{agn}) feedback implementation. For complete details of all aspects of the model, we direct the reader to the original code paper \citep{DaveAngles-Alcazar2019}.

The \simba model is unique within the set of current cosmological simulations in its treatment of accretion onto black holes. Accretion is split by the local gas properties into a \ac{bhl} mode and a torque-limited accretion mode \citep{HopkinsQuataert2010,HopkinsQuataert2011} depending on the gas temperature. 

The hot gas ($T > 10^5$K) is accreted onto black holes using a \ac{bhl} accretion model under the assumption of spherical symmetry. The mass accretion rate is computed \cite{DaveAngles-Alcazar2019};
\begin{equation}
    \dot{M}_\text{Bondi} = \varepsilon_m \frac{4 \pi G^2 M^2_\text{BH} \rho}{(v^2 + c_s^2)^{3/2}},
\end{equation}
where $\rho$ is the density of the hot gas within the accretion kernel, $c_s$ is an averaged sound speed of this gas, and $v$ is an average relative velocity compared to the black hole itself. $\varepsilon_m$ is a tunable efficiency factor, set to $0.1$ in \simba . 

For cold gas ($T < 10^5$K), the torque-limited accretion model~\citep{Angles-AlcazarDave2017}, yields an accretion rate

\begin{equation*} 
    \dot{M}_\text{Torque} \approx \varepsilon_T f_d^{5/2} \times \left(\frac{M_\text{BH}}{10^8M_\odot}\right)^{1/6} \left(\frac{M_\text{enc}(R_0)}{10^9M_\odot}\right)  \left(\frac{R_0}{100 \text{pc}}\right)^{-3/2} \times 
\end{equation*}
\begin{equation}\label{eqn:torque-limited-accretion}
    \left(1+\frac{f_0}{f_\text{gas}}\right)^{-1} M_\odot \text{ yr}^{-1},
\end{equation}

\noindent where $f_d$ is the disk mass fraction (including stellar and gas components), $M_\text{enc}(R_0)$ is the total sum of these two components, $f_\text{gas}$ is the gas mass fraction in the disk, and $R_0$ is the distance from the black hole which encloses the nearest $256$ distinct gas elements. $f_0$ is a factor that describes the amplitude of epicyclic oscillations of a test particle around a perturbed circular orbit in the central accretion disk - see \citep{HopkinsQuataert2011} for a full definition. A kinematic decomposition is used during the simulation to separate the disk and spheroidal components, each comprised of gas and star particles.  The \ac{ism} in the \simba galaxy model is artificially pressurized in order to resolve the Jeans mass, so for accretion we treat all \ac{ism} gas ($n_\mathrm{H} > 0.13\,\mathrm{cm}^{-3}$) as "cold".  We assume $\varepsilon_T=0.1$ which is tuned to match the amplitude of the black hole mass -- stellar mass (or velocity dispersion) relation at $z\approx 0$.

The accretion rate is modulated by a radiative efficiency factor $\eta = 0.1$, and is given in its final form as $\dot{M}_\mathrm{BH} = (1-\eta)\times (\dot{M}_\text{Torque} + \dot{M}_\text{Bondi})$. We impose upper limits on the rates of both components of the total accretion rate based on the Eddington accretion rate $\dot{M}_\text{Edd}(M_\text{BH})$ for each black hole.

The feedback itself is also implemented in two modes; \textit{kinetic} and \textit{X-ray}. The kinetic mode is implemented through a sub-grid model which is split into two sub-modes, determined by the Eddington fraction $f_\text{Edd} \equiv \dot{M}_\text{BH}/\dot{M}_\text{Edd}$ of the accreting black hole.  For high Eddington ratios ($\gtrapprox0.2$) the \ac{agn} drives winds with velocities on the order of $10^3$ $\mathrm{km}\,\mathrm{s}^{-1}$, which consist of molecular and warmly ionized gas. We call this the \textit{radiative} mode.  At lower Eddington fractions a \textit{jet} mode is instead active, with hot gas being driven at velocities of order $10^4$ kms$^{-1}$. Jet-mode feedback is effective at evacuating gas from the central regions through entrainment \citep{BorrowAngles-Alcazar2020}. Both modes of kinetic feedback are implemented in a bipolar manner, with the jet axis calculated according to the angular momentum of the inner gas disk. 

We parameterise the radiative wind velocity as a function of the black hole mass

 \begin{equation} \label{eqn:quasar velocity}
     v_\text{w,EL} = 500\left(1+ \frac{\log_{10} M_\text{BH} - 6}{3}\right) \text{   km s}^{-1},
 \end{equation}
 
\noindent with the gas being ejected at the \ac{ism} temperature. If the Eddington fraction $f_\mathrm{Edd} < 0.2$, the feedback instead occurs via the jet mode and the velocity is boosted to

\begin{equation}
    v_\text{w,jet} = v_\text{w,EL} + 7000 \log_{10} (0.2/f_\text{Edd}) \text{  km s}^{-1}.
\end{equation}

\noindent We impose an upper limit of $7000$ kms$^{-1}$ at $f_\text{Edd}<0.02$. Additionally, jets will not activate unless the black hole mass reaches $\gtrsim 10^{7.5} M_\odot$, in order to stop small, but slowly-accreting black holes from generating jets. Gas ejected in jets is automatically raised to the virial temperature of the halo in \simba.

The mass outflow rate in the winds $\dot{M}_w$ is determined in \simba by requiring that the momentum outflow is given by $\dot{P}_w = 20 L_\text{BH}/c$ where $L_\text{BH} = \eta \dot{M}_{BH} c^2$ is the bolometric luminosity of the black hole. The outflow rate is then set by $\dot{M}_w = \dot{P}_w / v_w$ such that the mass-loading scales inversely with jet velocity. The energy in jets is given by $\dot{E} = \frac{1}{2}\dot{P}_w v_w = 10 L_\text{BH} \frac{v_w}{c} = 10 \eta \dot{M}_{BH} v_w c$. The effective feedback efficiency therefore is given by $\epsilon_f \equiv 10 \frac{v_w}{c}$. For maximum jet velocities the feedback efficiency is $\sim 0.27$.

The final \ac{agn} feedback mode present in \simba\ is in the form of \textit{X-ray} feedback, which models the radiation pressure from X-rays off the accretion disk. This feedback mode is only active when full-velocity jet feedback is occurring, and so is again determined by the mass accretion rate. Additionally, X-ray heating is only active if $f_{gas} \equiv M_{gas}/M_* < 0.2$. \simba\ heats gas within the kernel of the BH, with the energy input scaled by the inverse square of the distance from the black hole. X-rays heat the non-ISM gas directly, whereas for ISM gas, which would be expected to just cool quickly, only half the energy is added as heat, with the other half being used to give the gas a kick radially outwards. The gas heating rates due to the X-rays is given in \cite{ChoiOstriker2012} as

\begin{equation}
    \dot{E} = n^2(S_1 + S_2),
\end{equation}

\noindent where $n$ is the local proton number density, $S_1$ is the power contribution from Compton heating, and $S_2$ is the contribution coming from the sum of photoionization heating. The radiation-pressure change in momentum of the kicked gas is straightforwardly calculated by $\dot{p} = \dot{E}'/c$ where $E'$ is half the total energy available.

The free parameters in the feedback model are broadly tuned to match observations of the galaxy stellar mass function evolution and the stellar mass--black hole mass relation. The X-ray feedback mode is included primarily to obtain enough fully-quenched galaxies at $z=0$.  \simba\ has been tested against a large number of other galaxy, halo, and intergalactic medium properties, with good alignment with available observations~\citep[e.g.][]{DaveAngles-Alcazar2019} at a level comparable or better than other similar cosmological simulations. For example, the X-ray scaling relations and baryonic content in groups and clusters \citep{JenningsDave2023}, the gas content of the \ac{cgm} including observed $\mathrm{H I}$ equivalent width and $\mathrm{O VI}$ absorbers (including the observed dichotomy around quenched versus star-forming galaxies) \citep{ApplebyDave2021,BradleyDave2022}, the fraction of baryons locked into the \ac{igm} at low redshift \citep{SoriniDave2022}, black hole scaling relations, the satellite galaxy stellar mass function and color-magnitude relationship, and stellar properties of the \ac{bcg} \citep{Cui2021}, richness of environments of radio galaxies \citep{ThomasDave2021}, general agreement in the fraction of rapidly quenched quiescent galaxies at $z=1$ \citep{ZhengDave2022}, and good agreement with the number density of quenched galaxies between $z=3-4$ \citep{SzpilaDave2024}.

\subsection{Resimulation technique} \label{sec:resimulation}

In order to capture the short timescales of the jet and bubble-inflation processes, we select a \hyenas snapshot produced at $z=0.286$ and restart the simulation from that point. This allows us to increase the snapshot cadence to every $10$ Myr. We run the simulation for an additional $700$ Myr in total. We check that the gas, stellar, and black hole properties are consistent after restarting, yet still allow a grace period of a few tens of Myr before the earliest analysis is done. This is to mitigate any potential transient effects which may be present due to restarting from snapshot initial conditions within \GIZMO.

\begin{figure}
	\includegraphics[width=\columnwidth]{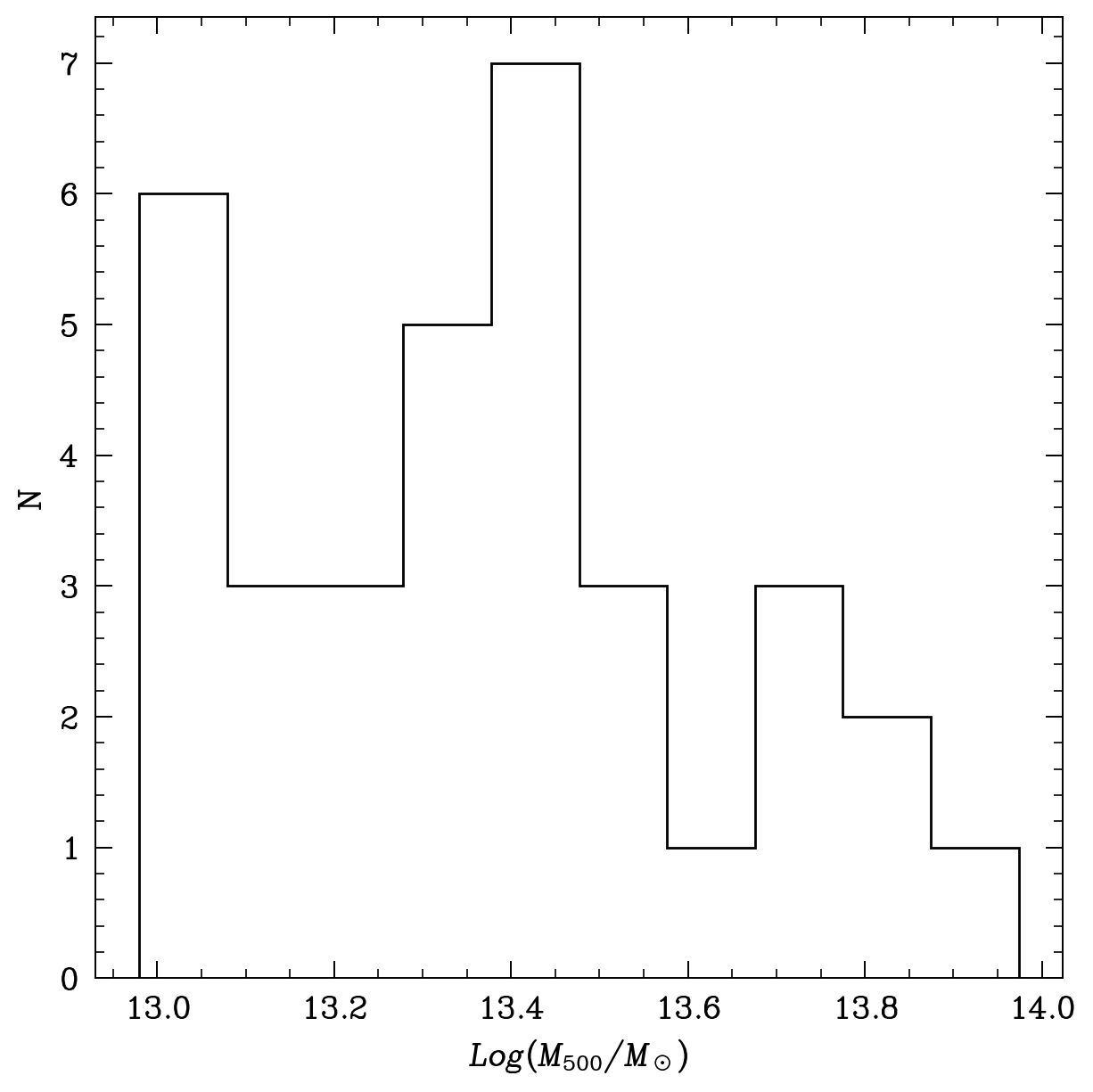}
    \caption{The initial mass distribution of halos for which we find bubbles as measured at the first snapshot at $z=0.286$.}
    \label{fig:initial m500 hist}
\end{figure}

In Fig. \ref{fig:initial m500 hist} we plot the distribution of initial halo masses at the first snapshot at redshift $z=0.286$. The range of halo masses that we simulate (and find bubbles) is $10^{13} M_\odot$ to $10^{14} M_\odot$.

\subsection{X-ray Maps} \label{sec:xray_maps}

To make comparisons to existing catalogues of X-ray cavities, we produce mock \textit{Chandra} observations using the \moxha\ package~\citep{JenningsDave2023}. We use the \textit{Chandra ACIS-I cycle-0} specification\footnote{Response files as known in \SOXS: \textit{acisi\_aimpt\_cy0.rmf, acisi\_aimpt\_cy0.arf}}, which does not suffer from the degredation of effective area at low energies due to contamination build-up over operating time compared to later cycles \citep{PlucinskyBogdan2018}. Our motivation for using \textit{Chandra} is that there is a wealth of observed bubbles in the literature, and the instrument has excellent angular resolution ($0.492''$ per pixel) which is needed to spatially resolve cavities. We artificially remove the \textit{Chandra} chip gaps before observation, so as to emulate a mosaic-ed observation. This means that our full field of view is $16.6' \times 16.6'$, with approximately $4$ times the single-chip area. 

In order to offer insights into X-ray cavity detection using the next generation of X-ray observatories, we produce mock observations using the \textit{Line Emission Mapper (LEM)} probe, with the 2eV spectral resolution specification\footnote{Response files as known in \SOXS:: \textit{lem\_2ev\_110422.rmf}, \textit{lem\_300522.arf}}. We present some comparison X-ray maps in \hyperref[sec:Bondi_vs_Cavity_Power]{\S5} and also perform spectral fits for all halos for both \textit{Chandra} and \textit{LEM}, comparing the accuracy of their recovered source luminosities when fitting to a jet-disturbed halo.

We create the X-ray maps using \PyXSIM to generate photons from an APEC model \citep{SmithBrickhouse2001} and project them along the line of sight.  We include the effects of neutral hydrogen absorption from the Milky Way, and use \SOXS to observe these photons via convolution with the \textit{Chandra} and \textit{LEM} instrumental responses. Furthermore, we select an observation redshift for each halo that results in the field of view on the \textit{Chandra} mosaic being $2 \times 0.6 R_{500}$ from edge-to-edge, where $R_{500}$ is measured at the initial redshift $z=0.286$. This is to standardise the observations over the whole order-of-magnitude range of halo masses, and prevent biasing against observations of small X-ray bubbles in low-mass groups due to reduced resolution which would occur if we instead held redshift constant. Our field of view choice results in a range of observational redshifts: $0.0377-0.0172$. We use an exposure time of $75$ks, which is consistent with real observations of clusters with identifiable X-ray cavities.

\subsection{Thermodynamic Profiles}\label{sub:thermodynamic profiles}

We require radial thermodynamic profiles so that we can sample the temperature, pressure, and (electron) density at given radius within a halo. To this end, we generate emissivity-weighted electron density and temperature profiles from the ISM-filtered gas particle data and fit the parameterised models below.

For the temperature we follow \cite{VikhlininKravtsov2006}:

\begin{equation}
    T_{3D}(r) = T_0 t_\text{cool}(r)t(r)
\end{equation}

\noindent where outside the central cooling region the profile is dominated by

\begin{equation}
    t(r) = \frac{(r/r_\mathrm{t})^{-a}}{\Big[1 + (r/r_\mathrm{t})^b \Big]^{c/b}},
    \label{eqn:kT_model}
\end{equation}

\noindent and the central regions are described by

\begin{equation}
    t_\text{cool}(r) = \frac{(x+ T_\text{min} /T_0)}{x+1} \ \ \ \ \ \ \  \ , \ \ \ \ \ \ \\ x = \Bigg( \frac{r}{r_\text{cool}} \Bigg)^{a_\text{cool}}.
\end{equation}

\noindent Note that in our groups sample, the central regions are generally only at-best partially captured on orders of $\sim$ a few kpc scales and below. 

For the electron density $n_e$, we use the modified $\beta$ model from \cite{VikhlininKravtsov2006}:

\begin{equation}
    n_\mathrm{e}^2 = n_\mathrm{e,0}^2 \frac{(r/r_\mathrm{c})^{-1}}{(1+r^2/r_\mathrm{c}^2)^{3\beta }} \frac{1}{ \left( 1 + (r/r_\mathrm{s})^\gamma \right) ^\frac{\varepsilon}{\gamma}}.
    \label{eqn:ne_model}
\end{equation}

We fit both the temperature and density profiles individually to the corresponding emissivity-weighted radial profiles out to $100$ kpc. For this work, we use the emissivity between $0.2-3.0$ keV (source-frame) for the profiles. We reject fitted profiles if the Coefficient of Determination value (for model values $f_i$ and data $y_i$ this is $R = 1 - \frac{\sum (y_i - f_i)^2}{\sum (y_i - \bar{y})^2}$, which we evaluated in logarithmic space) is $<0.5$ for either or both fits. In this case, the halo is discounted from further quantitative analysis. We furthermore obtain a radial pressure profile by multiplying our best-fit temperature and electron density profiles. In doing this we assume a mean molecular weight per electron of $\mu_e = 1.155$ in order to mimic the usual assumptions made when these profiles are derived from observations.

\section{Cavity Analysis} \label{sec:Cavity Analysis}
\subsection{Selecting Cavities}

The first step in our study of X-ray cavities is finding them. They are mainly characterised by depressions in surface brightness on scales of a kiloparsec to a few tens-of-kpc close to the cluster or group centre. Several observational techniques have become standard in the field \citep[see for example][]{Hlavacek-LarrondoMcDonald2015}, and we use them in this work to match the observations as closely as possible.

We search for cavities in lightly-smoothed exposure-corrected X-ray flux maps. We focus on the region within $0.3R_{500}$ (measured from the minimum potential position, which coincides with the central black hole position), where the intensity contrast is likely to be greatest. Furthermore, we generate a residual X-ray map, extracting the mean pixel flux of a lightly-smoothed X-ray flux map in radial bins about the cluster center. We fit a King model \citep{King1966} to this mean radial flux profile

\begin{equation}
    I(r) = I_0 \left( 1 + (1/r_0)^2 \right)^{- \beta} + C_0,
\end{equation}

\noindent where $I_0, r_0, C_0$ are the normalization, core radius, and a constant variable, respectively --- all are free parameters in the fit. We then subtract off the King model counts from the raw X-ray counts map, and divide by the same King model, before lightly smoothing.

Next, we generate unsharped maps. These are designed to highlight features on a particular scale, by subtracting a strongly-smoothed image from a lightly-smoothed counterpart. We use Gaussian smoothing on several different pairs of scales. The main scales we use to search for cavities are $(60,2)$kpc and $(80,5)$kpc.

Finally, we run the machine-learning cavity detection pipeline \textsc{cadet} \footnote{\href{https://github.com/tomasplsek/CADET}{https://github.com/tomasplsek/CADET}} \citep{plšek2023cavity} on the exposure-corrected flux maps. \textsc{cadet} comprises a convolutional neural network that has been trained on \textit{Chandra} images to make pixel-wise predictions for the identification of cavities. We apply \textsc{cadet} to the flux image subjected to various different scales of cropping in order to test the robustness of a given detection and to highlight cavities on different scales. We find that \textsc{cadet} does well in identifying cavities in the vast majority of cases, and in several cases can recover a very good match to the true region of low density from comparison with the true projected maps even when the corresponding surface brightness depression is not clear in the smoothed and/or filtered X-ray maps.

In all we therefore have four different types of spatial map used for cavity detection, that \textit{would} be available to an observer; X-ray flux ($0.5-7$ keV), King model-subtracted residual, unsharped, and \textsc{cadet} pixel-wise predictions.

We selected cavities based predominantly on mock X-ray quantities, but occasionally we use the true thermodynamic maps (and their time-evolution) to inform the direction and rough size and shape of bubbles, in order to reduce false positive detections and poor shape fits to the surface brightness depressions. Therefore, our search is not a truly blind test. However, we point out that due to computational and time limitations, we are not afforded access to other signatures of bubbles such as (a) tesselated spectrally-fitted spatial maps of temperature and density over the group, or (b) a multi-wavelength analysis looking for radio emission in the directions of cavities.

For each cavity where we observe double-cavities, we select only bubbles launched at approximately the same time. Older, relic bubbles are ignored when they appear in snapshots where younger bubbles/s are observed for the purposes of calculating the \textit{recent} AGN power; these relic bubbles may still be analysed earlier in their lifetimes corresponding to a preceding snapshot and period of jet activity, if they are clear at that earlier time. 

\begin{figure*}
\captionsetup[subfigure]{labelformat=empty}
 \subfloat[]{\includegraphics[width=\textwidth]{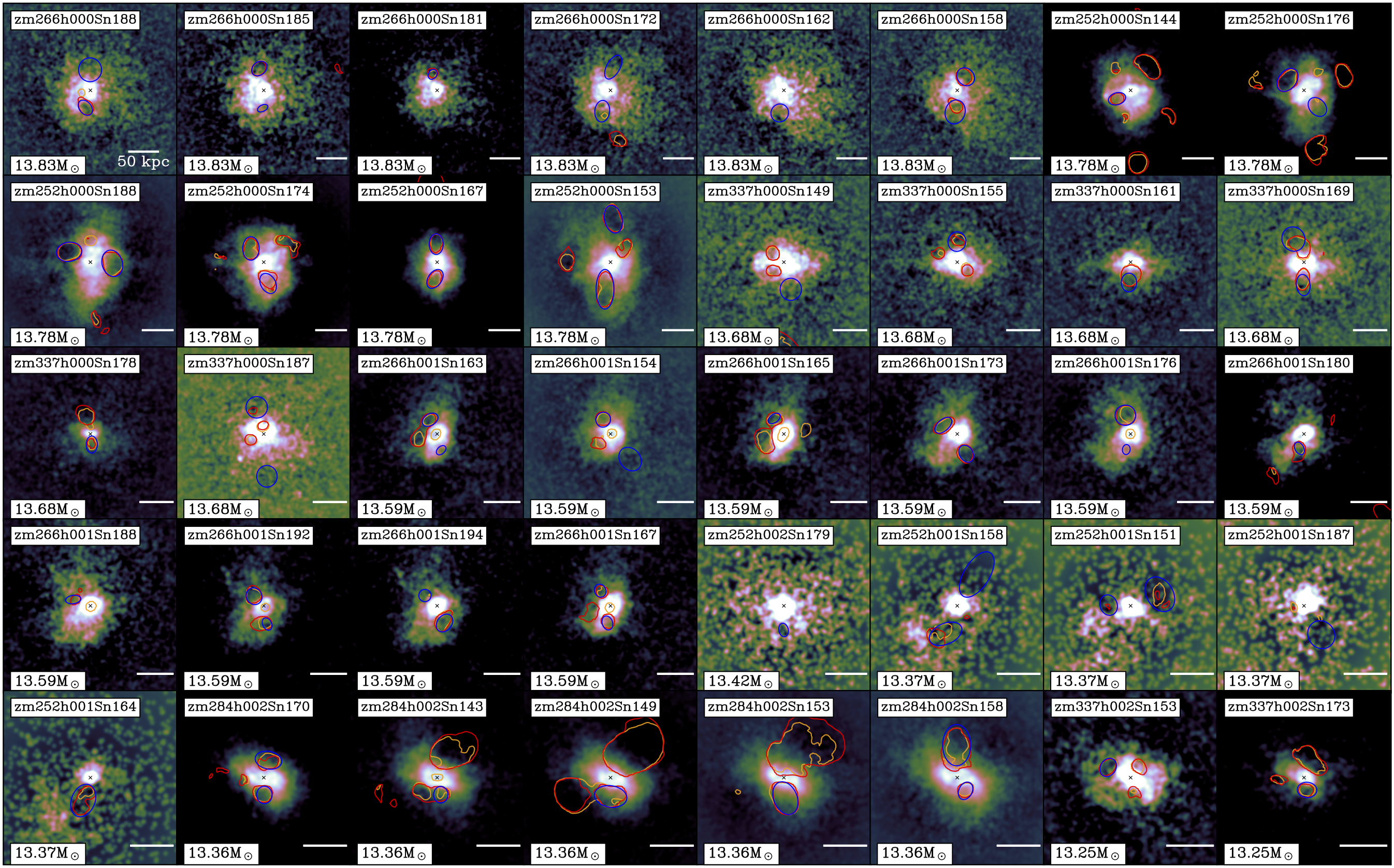}}\hfill
\caption{Unsharp-masked maps are show for a selection of our galaxy groups. They are generated by subtracting a strongly-smoothed X-ray map from a lightly-smoothed one; we use $\sigma =60,2$ here respectively. The orange and red contours show the \textsc{cadet} predictions at a significance level of 90\% when run on two different cutout scales ($1024$ and $1280$ \textit{Chandra} pixels respectively), and we show the fitted ellipses that we identify in blue. The scale bars denote $50$ kpc. Note that the colormap normalisation varies between each plot and is designed to maximise contrast in the radius range occupied by the bubble/s. Some apparent cavities are not designated so as they are ruled out via consultation of maps and movies of the jets in the true thermodynamic maps, or are relic cavities not associated with the most recent bouts of jet activity, but possibly included in the analysis of a previous snapshot. } \label{fig:unsharp mosaic}
 \end{figure*}

\begin{figure*}
\captionsetup[subfigure]{labelformat=empty}
 \subfloat[]{\includegraphics[width=\textwidth]{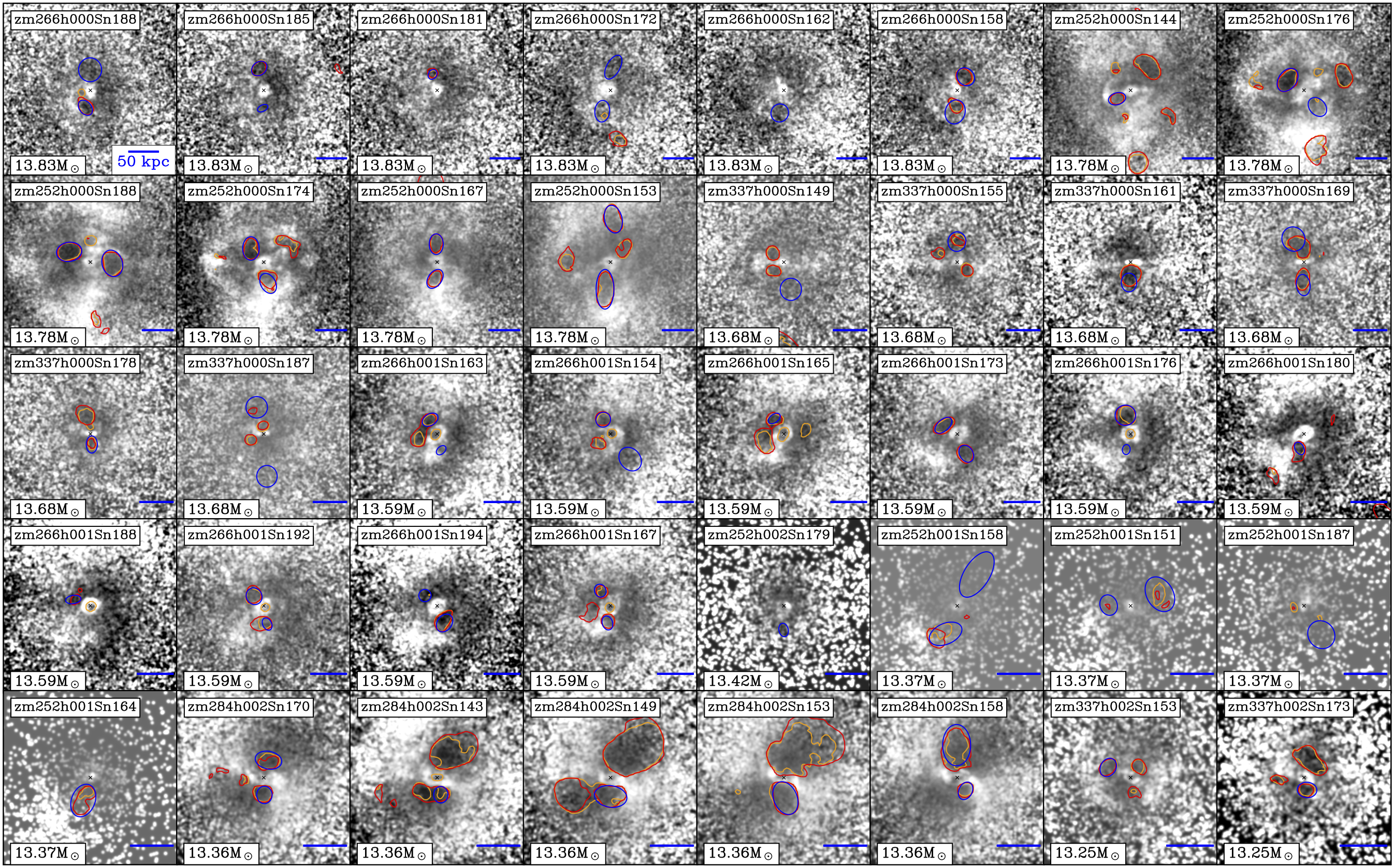}}\hfill
\caption{Same as for Fig. \ref{fig:unsharp mosaic} but for the King-model subtracted X-ray images.}\label{fig:king model mosaic}
 \end{figure*}

In Fig. \ref{fig:unsharp mosaic} and Fig. \ref{fig:king model mosaic} we plot our sample of mock-observed cavities in our catalogue. Shown are the Unsharp-masked and King-model subtracted maps, respectively, over which we overlay \textsc{cadet} confidence contours and our fitted cavity ellipses. Qualitatively, we see that the cavities produced in \hyenas are similar to those seen in real systems \citep[e.g.][]{Hlavacek-LarrondoMcDonald2015}, and are largely well-identified by \textsc{cadet}. Later on in this section we shall demonstrate that they are also quantitatively similar in terms of size and location.

\subsection{Cavity Uncertainties}\label{subsec:Cavity Uncertainties}

\begin{figure*}
	\includegraphics[width=1\textwidth]{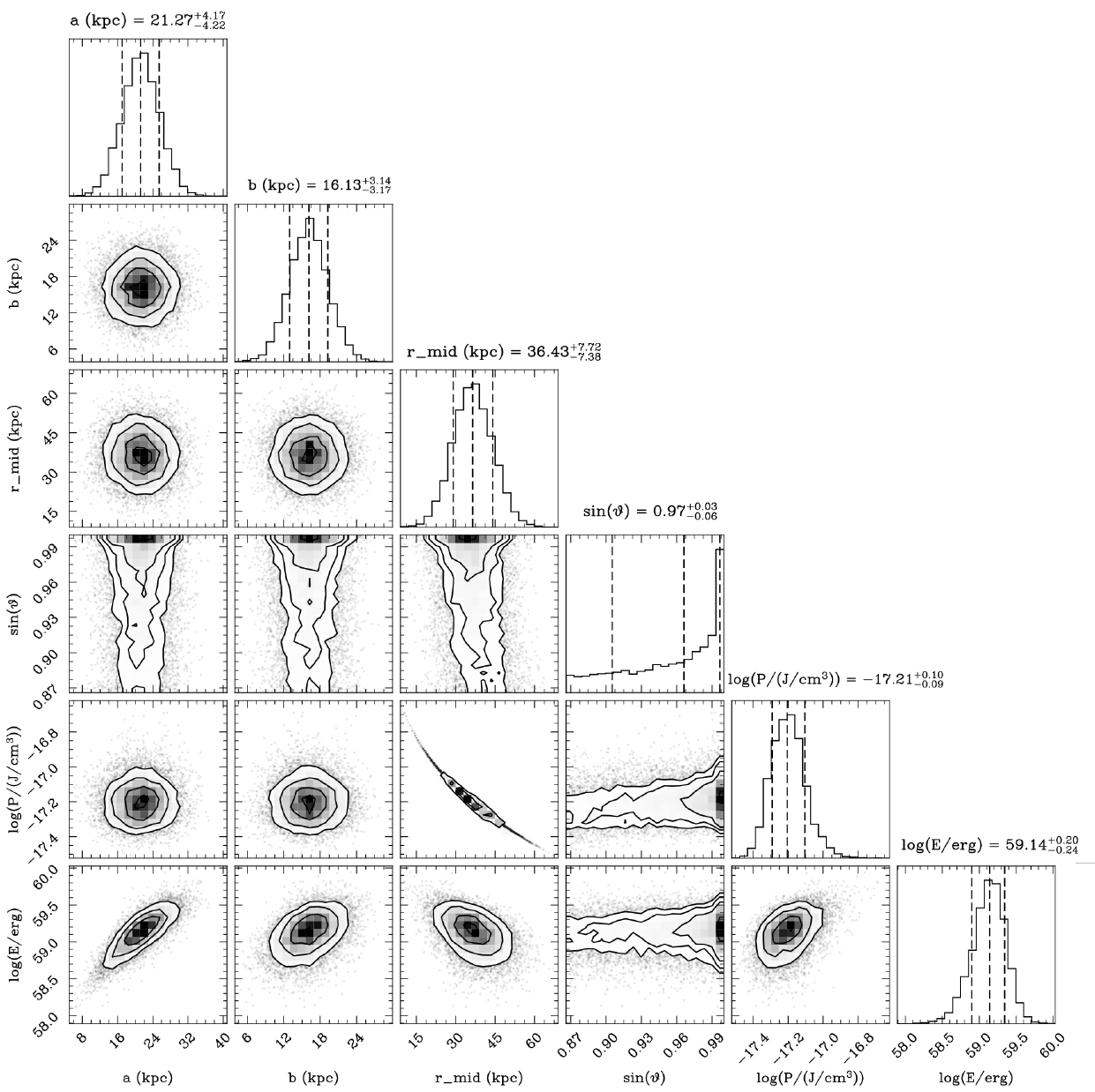}
    \caption{An example (\textbf{Zoom252\_halo0 (snapshot 188)}) of the posterior distribution of the derived enthalpy obtained from our uncertainty estimation method, showing it's dependence on the uncertainty in the bubble measurements and inclination.}
    \label{fig:enthalpy error corner plot}
\end{figure*}

The uncertainties on observed X-ray cavity sizes and, therefore, enthalpies, are generally large. Unfortunately, it is difficult to identify their exact dimensions and, simultaneously, it is difficult to measure the thermodynamic radial profiles of the halo \citep{Hlavacek-LarrondoMcDonald2015}. Therefore, we must quantify these uncertainties accurately.

First, we assign a typical uncertainty of $20\%$ to the minor and major axes of the fitted ellipse, and also to the radial distance of the bubble center from the halo center. These are in line with observational studies (see e.g. \citealt{Hlavacek-LarrondoMcDonald2015}). We assume that the errors on these quantities are Gaussian distributed. 

Second, we consider uncertainties arising from bubble inclination. In general, a nonzero, unknown inclination will move the true bubble center to a larger radius than the \textit{projected} bubble midpoint radius. We assign a uniform distribution to this angle $\theta$, under the assumption that for us to clearly identify a bubble the inclination can be no more than $30 ^{\circ}$ inclined from the plane perpendicular to the line of sight (i.e. the plane of observation). We take $10,000$ enthalpy samples ${H_\text{tot}(\boldsymbol{\Theta}_i)}$ with parameter values drawn from these aforementioned distributions;

\begin{equation}
    H_\text{tot}(\boldsymbol{\Theta}_i) = \frac{5}{2} * P \Big( r_\text{proj,i} / \sin (\theta_i) \Big) * \frac{4}{3} \pi a_i^2 b_i,
\end{equation}

\noindent where $r_\text{proj}$ is the projected bubble midpoint radial distance, and $P$ is the median pressure at the de-projected bubble midpoint radius $r_\text{proj,i} / \sin (\theta_i)$. We assume cylindrical symmetry of the ellipse about the radial unit vector, so $a$ is the tangential ellipse axis and $b$ the radial axis of the ellipse, with the origin at the halo center. Therefore, we have
\begin{equation*}
\begin{aligned}
    r_\text{proj,i} \sim \mathcal{N}\big(r_\text{proj}, (0.2r_\text{proj})^2\big), \\
    a_{i} \sim \mathcal{N}\big(a, (0.2a)^2\big), \\
    b_{i} \sim \mathcal{N}\big(b, (0.2b)^2\big), \mathrm{and} \\
    \theta_i \sim \mathcal{U}\big(\frac{\pi}{3}, \frac{\pi}{2} \big).
\end{aligned}
\end{equation*}

\noindent We assume that the change in the bubble volume due to inclination $\theta_i$ is small compared to the measurement uncertainties, and we also assume that we know the pressure profile $P(r)$ perfectly. In reality, there is a non-negligible uncertainty attached to the 3D de-projected radial profile.

Using this method, we can account for most of the uncertainties in the measurements, and the usual assumption of zero inclination. Most importantly, we account for the uncertainty in the change in the radius at which the pressure is sampled. We show an example corner plot of the enthalpy posterior of one of our cavities in Fig. \ref{fig:enthalpy error corner plot}. We find the resulting posterior distributions for the logarithm of the enthalpy for our cavities are approximately normally-distributed, and so we use the $16^{th}$ and $84^{th}$ percentiles as the $1\sigma$ error for each individual cavity.  If we have two bubbles present, the error on the final enthalpy is the sum in quadrature of the individual enthalpy uncertainties.

\subsection{Cavity Dimensions}

In this Section, we investigate the physical characteristics of the X-ray cavities in our \hyenas sample. In order to examine whether the simulated gas properties are accurate in terms of where the bubbles are injected and how large they grow once seeded, we compare to the observational results of \citet{RaffertyMcNamara2006} who present a sample of $62$ cavities in galaxy clusters, and \citet{ShinWoo2016}, who present a study of 148 detected cavities from a sample of 69 clusters, groups, and elliptical galaxies.

\begin{figure}
	\includegraphics[width=\columnwidth]{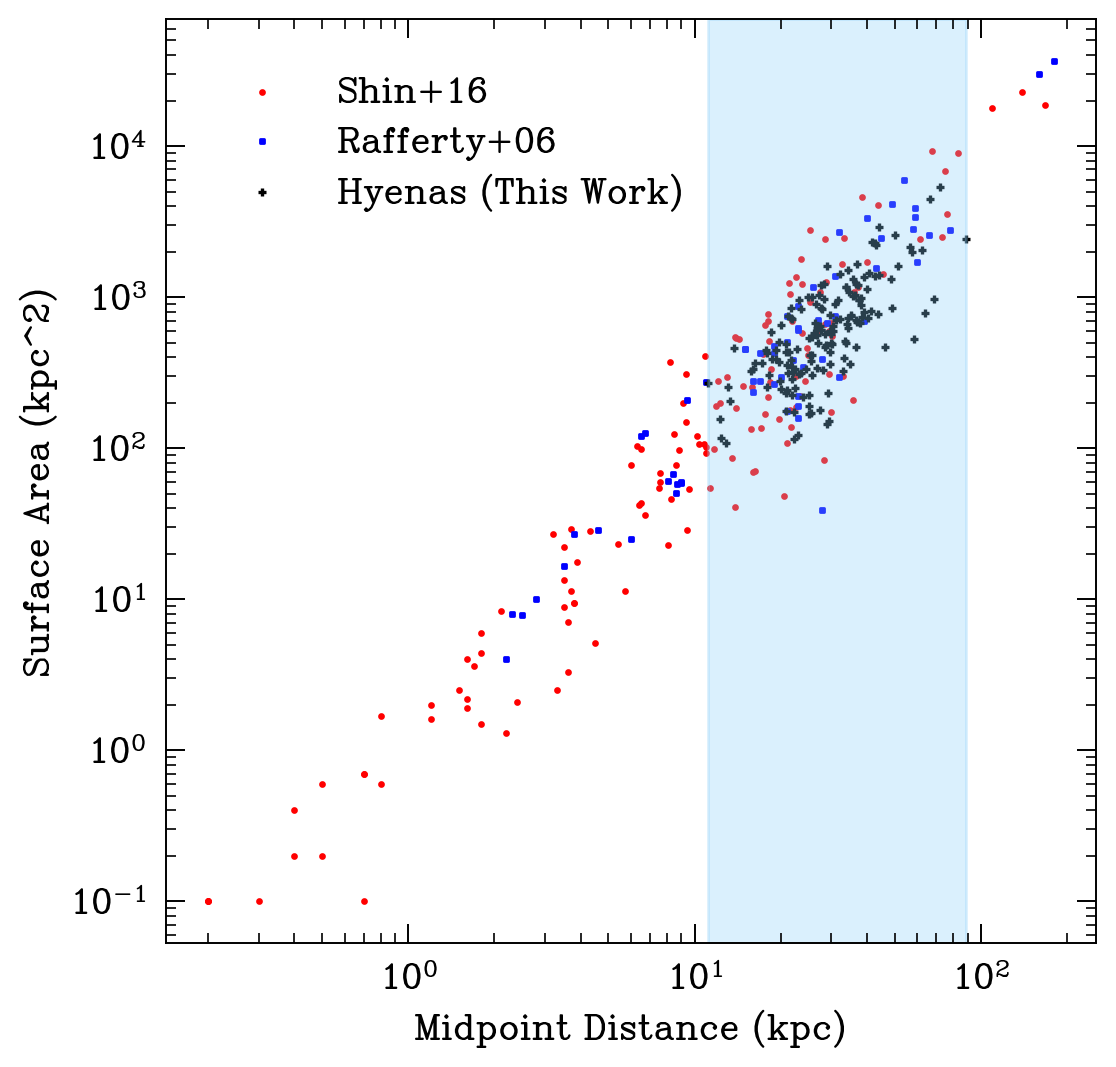}
    \caption{The bubble surface area as seen in the plane of observation compared to the observational data of \citet{RaffertyMcNamara2006} and \citet{ShinWoo2016}. The lower limit of midpoint distance for our sample is set by the decoupling length of the winds and is $\sim 10$kpc. Our bubbles match the relation and spread between area and midpoint distance well. For clarity, no error bars are shown we ascribe an error of $20\%$ to the midpoint distance and$\sqrt{0.2^2+0.2^2} \approx 28 \%$ to the bubble volume. Highlighted in blue is the range of midpoint distances we measure.}
    \label{fig:surface area vs midpoint}
\end{figure}

\begin{figure}
	\includegraphics[width=\columnwidth]{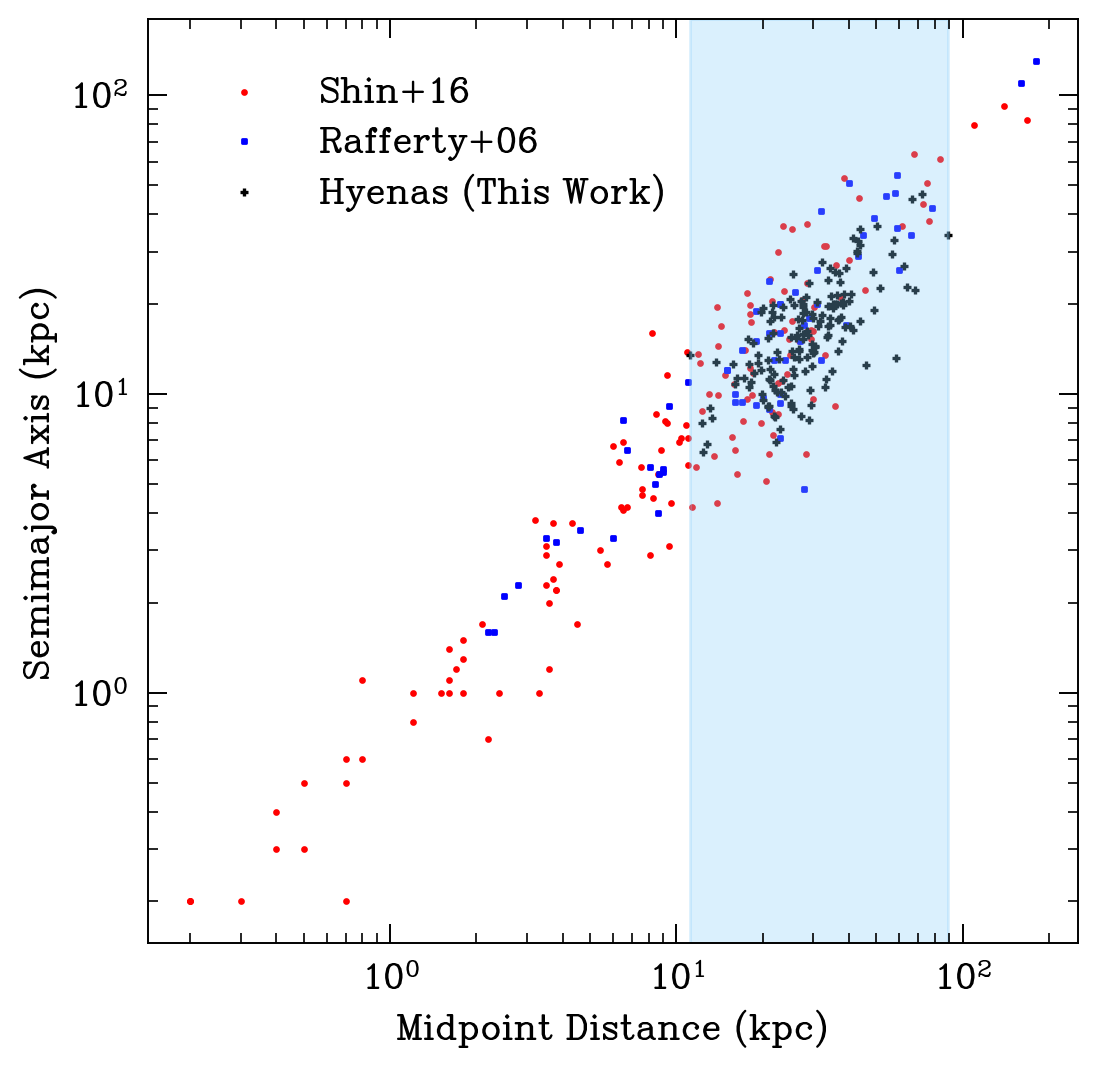}
    \caption{Same as Fig. \ref{fig:surface area vs midpoint} but for the semi-major axis length. Again we match the observed trend of \citet{RaffertyMcNamara2006} and \citet{ShinWoo2016} well.}
    \label{fig:semimajor axis vs midpoint}
\end{figure}

Figs \ref{fig:surface area vs midpoint} and \ref{fig:semimajor axis vs midpoint} show that our cavity sample matches the \citet{RaffertyMcNamara2006} and \citet{ShinWoo2016} data in both the linear trend and the scatter. Hyenas cavities tend to be found at the larger end of the linear trend that is traced by the observational data. We have a lower midpoint-distance limit of approximately $10$kpc - this corresponds to the radius at which fast jets re-couple at these redshifts (the recoupling distance is given by $v_\text{jet} * 10^{-4} t_\text{H(z)}$ where $\text{H(z)}$ is the Hubble time at the redshift at launch - with $t_\text{H(z)} \sim 13$ Gyr this distance is $10-11$ kpc). Even for jets launched with slower velocities which recouple at smaller distances there is a slim chance of spotting a bubble within this $10$kpc radius with the snapshot frequency of $10$ Myr, because the bubbles then have only a short distance to cover before reaching $10$kpc.


It is possible that Hyenas bubbles are slightly too small or injected slightly too far from the centre of the halo, since our points have a tendency to lie slightly below the \citet{RaffertyMcNamara2006} and \citet{ShinWoo2016} data. 

\begin{figure}
	\includegraphics[width=\columnwidth]{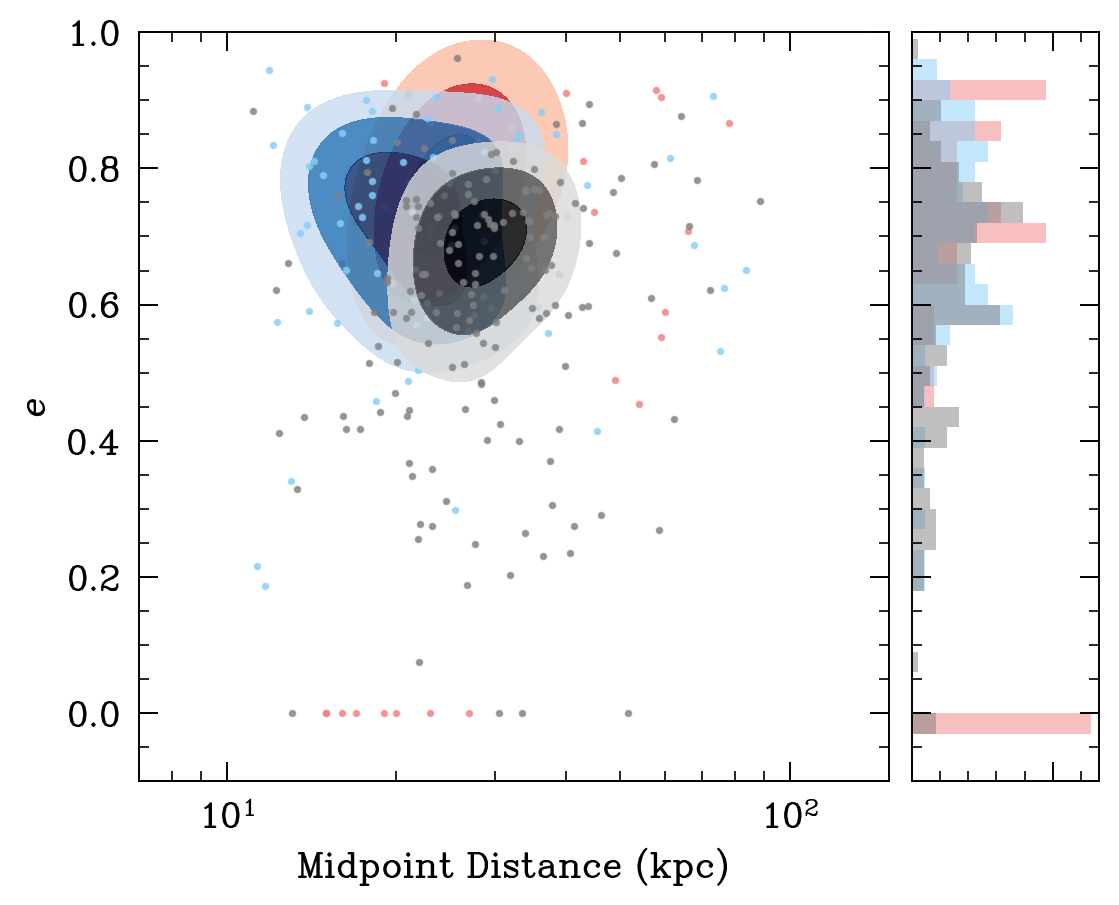}
    \caption{We plot the (projected) eccentricity of the Hyenas bubbles (black) against those of the published data of \citet{RaffertyMcNamara2006} (red) and \citet{ShinWoo2016} (blue), from which we restrict to those corresponding to our range of midpoint-distances (blue band in Figs \ref{fig:surface area vs midpoint}, \ref{fig:semimajor axis vs midpoint}). Contours correspond to $50$th, $70$th, and $90$th percentiles. Also plotted is the normalised histogram. We match observations reasonable well, with an average eccentricity  at $\sim 0.7$, only slightly lower than the observed trends. We tend to have a slight overabundance of cavities at low eccentricities of $0.2-0.4$.}
    \label{fig:eccentricity}
\end{figure}

In Fig. \ref{fig:eccentricity}, we show the eccentricity of Hyenas cavities from our mock observation technique, comparing to data from \citep{RaffertyMcNamara2006} and \citep{ShinWoo2016}, matched to the midpoint distance range of this work. We find promising agreement in the position of the peak of the eccentricity distribution and in the general scatter, though we tend to have a long-tail towards low-eccentricity compared to observations, as we show in the normalised histogram presented alongside. Overall, we demonstrate that \hyenas cavities are found in the right locations, with the correct sizes and shapes when compared to real bubbles.

\simba's feedback model, including the re-coupling timescale of jet outflows, as well as the energy and momentum transfer rates, is tuned to reproduce both the galaxy stellar mass function (GSMF) and the quenched fraction at low $z$. \simba was not calibrated to match the general hot gas properties of halos, let-alone the smaller scales of bubble injection. Therefore, it is encouraging that a realistic X-ray cavity population emerges from a bipolar feedback subscription that is calibrated on galaxy population properties. We suggest that a bipolar feedback model such as \simba's is able, and possibly required, to match observable signatures of \ac{agn} feedback on the \ac{igrm} and \ac{icm} on the scales of tens of kiloparsecs, whilst also producing realistic galaxy populations.

\subsection{Cavity Power Estimation}\label{subsec:Cavity Power Estimation}
In this Section, we investigate the simulated, X-ray determined cavity power compared to the true jet power. Additionally, we compare the fraction of jet power being contributed by the \ac{bhl} and torque-limited accretion modes. For the X-ray measured cavity power, we select a snapshot at which a detected bubble appears clearly in the intensity map and/or in the \textsc{cadet} prediction. We assume that the cavity is in approximate pressure equilibrium with the \ac{igrm}, such that $P_{cav} \approx P_\text{ICM} \equiv P$, and therefore the total enthalpy is \citep{BirzanRafferty2004}:

\begin{equation} \label{eqn: total enthalpy}
H_\text{tot} = U_\text{int} + PV = \frac{1}{\gamma - 1}PV + PV = \frac{5}{2}PV
\end{equation}

\noindent which gives the (minimum) energy that is needed to account for both the internal energy and the energy required for an adiabatic bubble to expand against the \ac{icm} (at a constant pressure), and we assume (as in \simba) that $\gamma=5/3$. Note, this differs slightly from the usual observer assumption that the adiabatic index $\gamma = 4/3$ since the cavity gas in real systems is thought to be relativistic \citep{GrahamFabian2008,Fabian2012}. \footnote{This difference in adiabatic index in \simba will mean that, for given cavity enthalpy, a given bubble in \hyenas will have a volume $8/5=1.6$ times larger \textit{in volume} than a system with the relativistic adiabatic index. This means we have axes lengths which are $1.17$ times larger. This is a relatively small factor (smaller than the measurement error we ascribe to the axis lengths) and would not alter our results in a qualitative way.}

To determine the ambient pressure at the appropriate radius for each halo and cavity event, we take the best-fit radial profiles of the electron density $n_e$ and the plasma temperature $k_BT$ using the \PyXSIM-derived, emissivity-weighted filtered gas, as described in \hyperref[sub:thermodynamic profiles]{\S2}. We assume that the total pressure follows $P = (\mu_e/\mu) *  n_e k_B T$, computed at the bubble radius. $\mu_e = 1.155$ here is the mean molecular weight per electron. The volume of the cavity is measured by associating an ellipse to each bubble based on the size, location, and orientation of the features seen in the flux map and the filtered images, as well as the \textsc{cadet} predictions. We assume symmetry about whichever axis (semi-minor or semi-major) aligns closest to the radial vector pointing out from the halo center. We calculate the enthalpy by computing the pressure at the radius of the bubble centre; unless otherwise stated, this is what we present as being used for the cavity enthalpy/power. For comparison, we also calculate the enthalpy from the pressure at the radius of the outermost point on the bubble perimeter. This gives a lower limit on any enthalpy that could be calculated from a method that instead considers a radially-varying pressure profile within the bubble as opposed to a constant/midpoint value.

We calculate the \ac{agn} energy injected into the halo via kinetic jets directly from the simulation data, with bubble gas being isolated via a cut on the hot gas phase space and its change between the snapshots identified as containing the jets contributing to the energy of the bubble. We identify cavity members in a first sweep as any particle that between any two snapshots ($10$Myr spacing) has satisfied all of the following: the particle has increased its internal energy by at least $40\%$, reduced its Lagrangian density by at least $40\%$, has a post-kick velocity and velocity increase of at least $200$ km/s, and lays within $20$ kpc with respect to the halo center before being heated. A second sweep is performed to also identify particles which may be initially hot (so that their internal energy may not have increased by the above percentage threshold) but which are also entrained in outflows. This sweep finds outflowing particles which have had their internal energy raised by the minimum energy increment found in the particles identified from the first sweep. We visually inspect projections of thermodynamic properties that these criteria result in bubble/jet particles being clearly identified.

\begin{figure*}
\captionsetup[subfigure]{labelformat=empty}
 \subfloat[]{\includegraphics[width=0.48\textwidth]{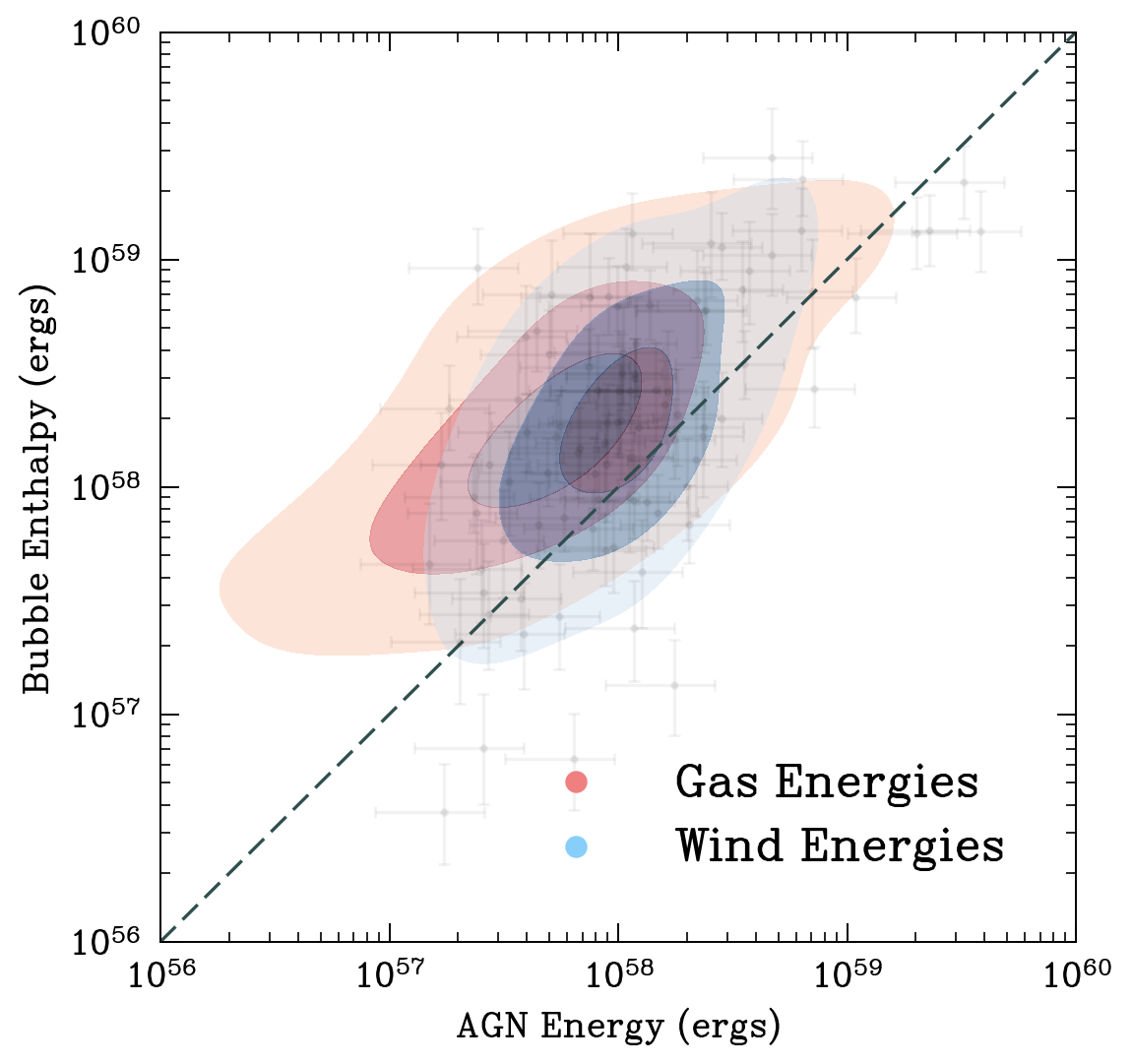}}\hfill
 \subfloat[]{\includegraphics[width=0.48\textwidth]{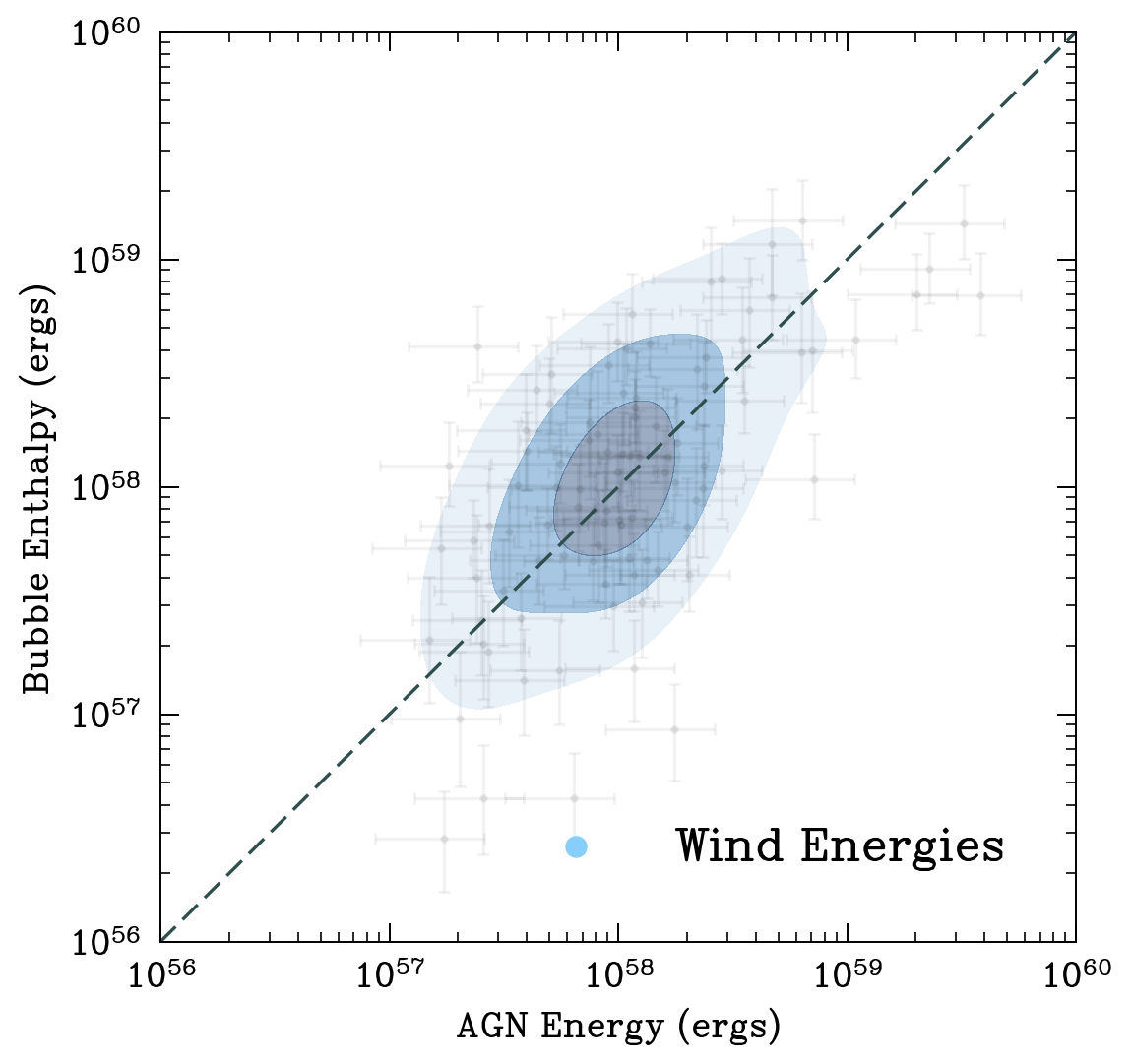}}\\[-2ex]  

 \caption{On the left we plot the mock-observed bubble enthalpies as calculated using the midpoint pressure against the AGN energy, calculated using the two methods described in the text. The scatter shows data for the mock-observed enthalpy versus the \textit{Wind Energies}. On the right we plot the same but for using the enthalpy derived with the pressure at the bubble's outermost tip. Contours correspond to $20$th, $50$th, and $80$th percentiles. For the midpoint energy the bubbles are over-powered, but a 1:1 relation is recovered when using the lower pressure. The small region where the \textit{Gas Energies} appear larger than \textit{Wind Energies} can be explained by additional energy injection by \simbaNG's X-ray feedback, which is active when the jets are but heats particles directly rather than via a wind, for these systems. There may also be a contribution from stellar winds from the central galaxy, which are hard to disentangle from AGN winds. The x-axis of the plot captures both winds and jets but not this additional heating at the very high energies.} \label{fig: True vs observed energies}
\end{figure*}

Furthermore, we calculate the AGN energy via a second method by simply summing the wind kinetic energies (which are saved in \simba at each timestep) over the time range that the jets responsible for inflating the cavities are active. We label these two schemes as "Gas Energies" and "Wind Energies" in Fig. \ref{fig: True vs observed energies}.

The left plot in Fig. \ref{fig: True vs observed energies} shows that the cavity enthalpies (from a \textit{midpoint} pressure) generally \textit{exceed} the apparent energy input by the \ac{agn} over the corresponding jetting time-frame \footnote{The duration of time that jets directly responsible for the inflation of the bubble/s under cosnideration are actively recoupling and driving outflows}. The energy estimated from the change in gas energies around the \ac{bgg} is slightly lower than the value obtained from the winds themselves. We posit that this is caused a combination of cooling losses, and from gentle heating by weak shocks and sound waves, which may not be captured through our tagging scheme. We leave this for future work to quantify.

It is interesting that both of our \ac{agn} energy estimates appear \textit{too low} in power compared to the enthalpy contained in cavities. The fact that the cavity dimensions are consistent with observations (if anything, we produce slightly under-sized bubbles) implies that the cavity volumes are correct. There are two possible explanations. First, there could be a significant decrement in pressure along "channels" in the \ac{igrm} carved by older feedback episodes, that can enhance the apparent energy contained within a freshly inflated bubble by reducing the amount of $PdV$ work it must do in reality compared to inflating within pristine halo gas. Second, the enthalpy prediction for the bubble is too high. To test the latter option, we compute the enthalpy using the lowest pressure that the bubble can "see" --- the pressure at the outermost tip of the bubble. We show this result in the right plot in Fig. \ref{fig: True vs observed energies}. This gives the lowermost limit of the enthalpy within the bubble since, at these radii, the pressure is monotonically decreasing with radius from the halo centre. There is a clear linear trend between the bubble enthalpies and the \ac{agn} output, with very little offset (but a large scatter). This indicates that the more important physical quantity is the enthalpy calculated from the outermost pressure, and not from the midpoint. We argue that the bubble inflation process is dominated by work done along the radial axis into the \ac{igrm} at the bubble tip, where pressure is lowest and where expansion requires the least work. So long as the resistance provided by the "added mass" that the bubble must uplift is comparatively small, this radial direction offers the "path of least resistance" for the bubble to expand along, and it will do $PdV$ work using the lowest pressure the bubble can "see". This preferential direction of expansion also explains why both observed and simulated cavities are so eccentric (Fig. \ref{fig:eccentricity}).

To compute the cavity ages, we use the characteristic timescales based on the instantaneous position and thermodynamic state of the bubble when observed, and via a \textit{Direct} estimate making use of the many snapshots that we actually have access to from the simulation. To transform the mock-observed cavity energy into a power, we calculate the \textit{buoyancy} timescale associated with each cavity \citep{BirzanRafferty2004}

\begin{equation}\label{eqn: buoy_timescale}
    t_\mathrm{buoy} = r_\text{mid}/v_t = r_\text{mid} \sqrt{\frac{SC}{2 g V}},
\end{equation}

\noindent where $v_\mathrm{t}$ is the terminal buoyant rising velocity (obtained simply by equating the ram pressure force and the buoyancy force \citep{ChurazovBruggen2001}, $S$ is the surface area of the bubble, $C = 0.75$ is the drag coefficient \citep{ChurazovBruggen2001}, $g$ is the gravitational acceleration at the bubble midpoint, and $V$ is the bubble volume. We calculate $S$ by assuming a circular cross-section with radius equal to the ellipse axis that aligns best with the tangential unit vector around the group centre. We calculate the gravitational acceleration $g$ by assuming hydrostatic equilibrium and using our best-fit pressure profile. For the error calculation, we calculate this quantity for each sample of our cavity properties which are generated as described in Section \ref{subsec:Cavity Uncertainties}. Note that we treat the thermodynamic profile as exact in that we do not consider any uncertainty around the best-fit fitting parameters (although the midpoint of the bubble at which we sample the profile will be different for each sample) and so our error estimates will be under-estimated compared to actual observational data.

Additionally, we require an estimate of the \textit{sonic timescale} \citep{BirzanRafferty2004}, which is the time-scale for the cavity to travel from the central \ac{agn} to the bubble midpoint radius $r_\mathrm{mid}$.  Therefore,

\begin{equation}\label{eqn: sonic_timescale}
    t_\mathrm{c_s} \equiv r_{\text{mid}}/c_\text{s} = r_{\text{mid}} \sqrt{\frac{\mu m_\mathrm{p}}{\gamma k_B T}},
\end{equation}

\noindent where $c_\mathrm{s}$ is the speed of sound in the \ac{igrm} and we take $\mu = 0.62$.  Compared to the free-fall time or buoyant-rising time, the sonic time-scale does not depend on estimates of the local gravitational acceleration, which can be very uncertain \citep{Hlavacek-LarrondoMcDonald2015}. Furthermore, it is more accurate for high-powered jets, where the bubbles are driven outwards rather than buoyantly inflating \citep{OmmaBinney2004}. Because \simba has a powerful jet model, this timescale may be the more appropriate measure of the cavity lifetime.

 For our \textit{Direct} estimates, we find the time difference between the snapshot where we observe the cavity, and the midpoint of the timescale we identify associated jet activity. For example, if we identify jets contributing to a given bubble over a timescale of $20-30$ Myr before we measure the bubble, we associate a \textit{Direct} cavity timescale of $25$ Myr. We also calculate a \textit{Maximum Direct} timescale, simply the time between the bubble mock observation and the estimated start time of all associated bubble activity (this would be $30$ Myr in the example above). These estimates are based on visual inspection of the density projection movies, with cross-referencing to large spikes in the time-binned wind kinetic energy.

\begin{figure}
	\includegraphics[width=\columnwidth]{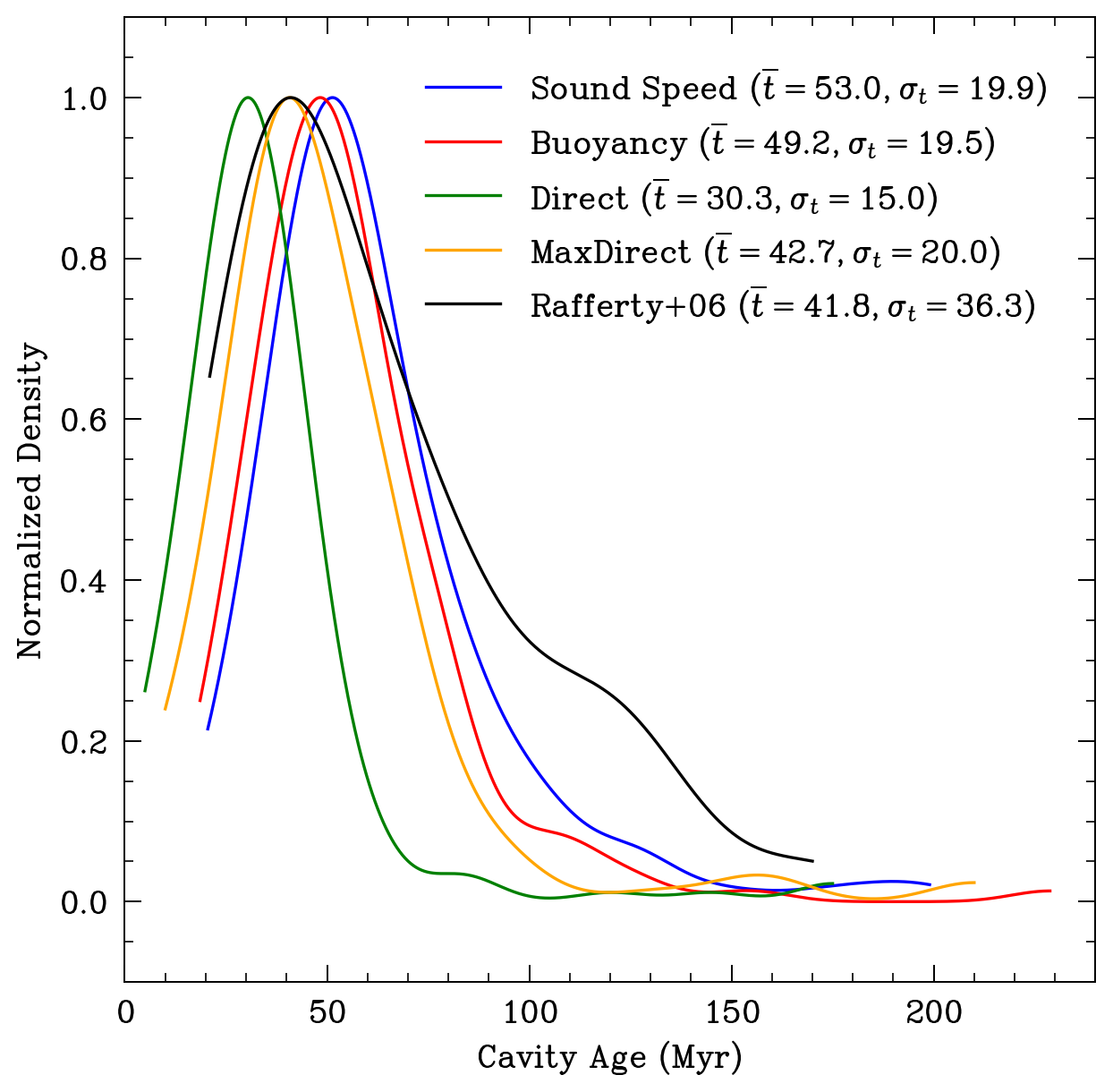}
    \caption{ The Kernel-Density Estimate for the distribution of our estimated cavity ages is shown, with the normalized distribution density on the y-axis plotted against the cavity age. The best-fit Gaussian distribution is calculated for the curves above a normalised density of $0.3$, and its mean and standard deviation stated. We compare to the observational data of \citet{RaffertyMcNamara2006} that lies within our minimum and maximum midpoint distances as plotted in Fig. \ref{fig:surface area vs midpoint}.}
    \label{fig:cavity_timescales_kde}
\end{figure}

 We plot the kernel density estimates (KDEs) for the distributions of our age estimates in Fig. \ref{fig:cavity_timescales_kde} for comparison\footnote{Using the \href{https://scipy.org}{scipy} statistics library.}. Median and $1-\sigma$ errors were estimated from samples drawn from the KDE distribution above a normalized density of $0.3$ and fit to a Gaussian. All distributions are well-fit by a Gaussian above a normalised density of $0.3$. The \textit{direct} estimate is the shortest, with an average timescale of $30.3$ Myr and a standard deviation of $15.0$ Myr. Next are the \textit{Maximum Direct} and buoyancy timescales, followed by the sound-speed/sonic timescale, with best-fit mean values at $42.7, 49.2, 53.0$ Myr respectively. We see there is a relatively large spread in the time taken for cavities to migrate from the central few kpc, even for the most conservative estimate. The fact that the sound speed timescale is significantly larger than the others may be a result of the rather arbitrary definition of the sound-speed, where the bubble is assumed to always travel at a Mach number of unity. In reality, as long as the shock remains weak in order to reproduce the observational dearth of strong cavity shocks and the gentle heating observed in systems, this cavity speed can be larger and the sound-speed timescale can be brought more in line with the buoyancy estimate. Indeed, observations have shown that lobes can possess Mach numbers greater than unity, and of up to $\mathcal{M} \approx 2$ \citep{SimionescuRoediger2009, KraftBirkinshaw2012, SonkambleVagshette2015, VagshetteNaik2017}. Modifying the formula for the sound-speed cavity age by a universal conservative Mach number $\mathcal{M} = 1.1-1.4$ would align $t_{cs}$ with $t_\mathrm{buoy}$ and $t_\textit{MaxDirect}$ resepectively. 
 
 The buoyancy timescale and the maximum direct timescale distributions agree reasonably well in the mean and scatter. The maximum direct timescale corresponds to the bubble launched by the first jet during a period of activity, and so it is this bubble that experience the maximal drag force, which is set by the (undisturbed) hot atmosphere. Subsequent bubbles during the same period of activity, blowing along the same axis of outflow, will experience a lowered drag force and so a shorter cavity lifetime, because of the fact that dense gas has already been evacuated from this channel. Therefore, it makes sense that these two estimates for the cavity lifetime are in good agreement and, furthermore, indicates that the buoyancy timescale is likely close to being an unbiased estimator for the true cavity lifetime, at least for the first bubble launched along an axis. Finally we note that the \hyenas $t_\text{buoy}$ matches the buoyancy timescales as measured by \citet{RaffertyMcNamara2006} very well in terms of the average, although the scatter on our values is smaller by around a third.

\section{Case Studies of Individual Groups}\label{sec:Case Studies}

We begin by considering a few case studies of selected individual galaxy groups that display interesting behaviour.  We particularly highlight features in our X-ray maps that are reminiscent of real systems, to provide qualitative support for the idea that the way in which \hyenas's jets interact with the IGrM is reasonably realistic.

\begin{figure*}
\captionsetup[subfigure]{labelformat=empty}
 \subfloat[]{\includegraphics[width=0.48\textwidth]{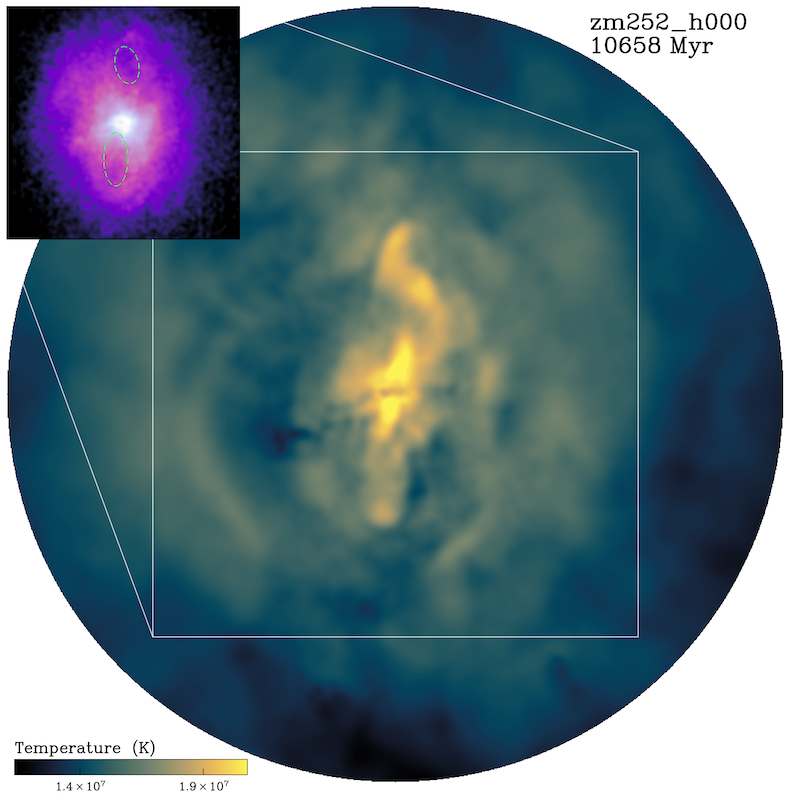}}\hfill
 \subfloat[]{\includegraphics[width=0.48\textwidth]{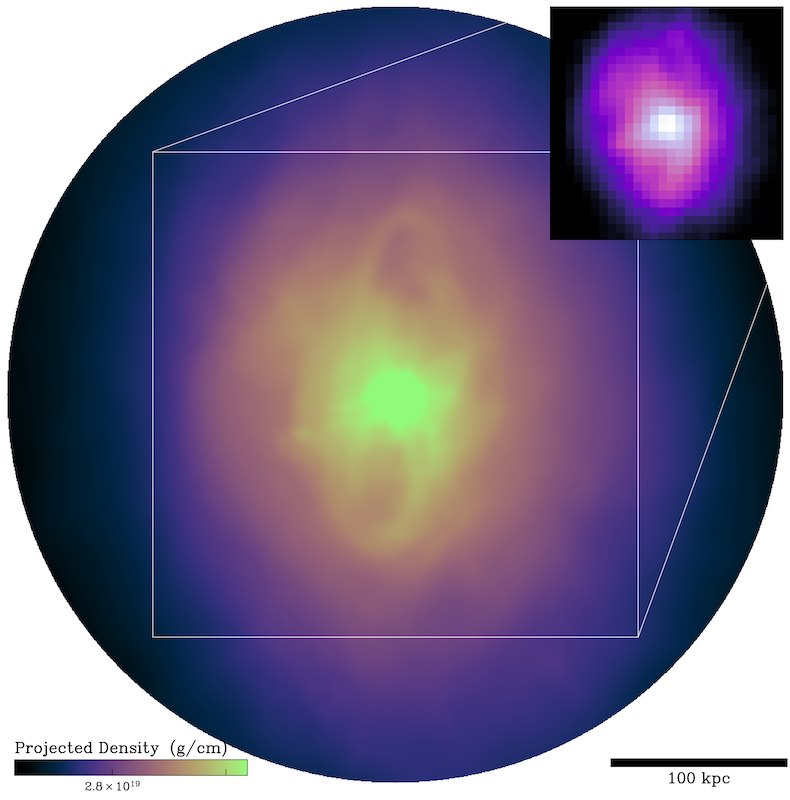}}\\[-2ex]  

 \caption{\textbf{Zoom252\_halo0 ($z_{obs}=0.032$):} Strong bipolar jets are seen in the temperature and density maps. Inset are mock Chandra (left) and \textit{LEM} (right) observations of the central regions, with X-ray cavities identified in green dashed line.} \label{feature:Zoom252_halo0}
\end{figure*}

In Fig. \ref{feature:Zoom252_halo0}, we show temperature and density projection maps of one of our largest halos with an initial $M_{500}$ of $10^{13.77} M_\odot$. This halo has some of the clearest, most well-defined bubbles we see in our sample. A north and south cavity system is clearly visible in both the \textit{Chandra} and \textit{LEM} mock images, which have been lightly smoothed. Large sound waves are emitted during these events, which appear as edges in the pressure projection, and clearly reach at least several tenths of $R_{500}$ ($100$s of kpc) before being significantly damped. The fact that we see these features in our simulations is encouraging, because sound waves are thought to be a significant mechanism for transporting energy (possibly up to $25\%$ of the heating power; \citealt{BambicReynolds2019}) out of the group or cluster core, though this is a topic of current debate and investigation \citep{FabianWalker2017,BambicReynolds2019}. Sound waves have been observed in the cluster environment, notably in the Perseus cluster \citep{FabianSanders2003,SandersFabian2007,GrahamFabian2008}. It is left to future work to quantify the dissipation of energy into the hot gas by compressive acoustic waves in our suite of simulations. 

Strong, bipolar cavities have been observed in several known clusters. For example, in RBS 797 \citep{SchindlerCastillo-Morales2001, DoriaGitti2012}, Abell 3017 \citep{ParekhDurret2017,PandgeSebastian2021}, and also in the Perseus cluster \citep{BoehringerVoges1993, FabianSanders2000, FabianSanders2006}. It is yet to be shown whether such features are realistically reproducible in cosmological simulations which do not employ a bipolar feedback model, although some work has been done on this in the very recent past. For example, \citet{TruongPillepich2024} examine Perseus-like clusters in the \textsc{tng-cluster} suite demonstrate that bubbles can be formed via their isotropic feedback model in which the wind direction at each event is random \citep{WeinbergerSpringel2017}, although it is to be seen whether these \textit{quantitatively} match observed cavities.

\begin{figure*}
\captionsetup[subfigure]{labelformat=empty}
 \subfloat[]{\includegraphics[width=0.48\textwidth]{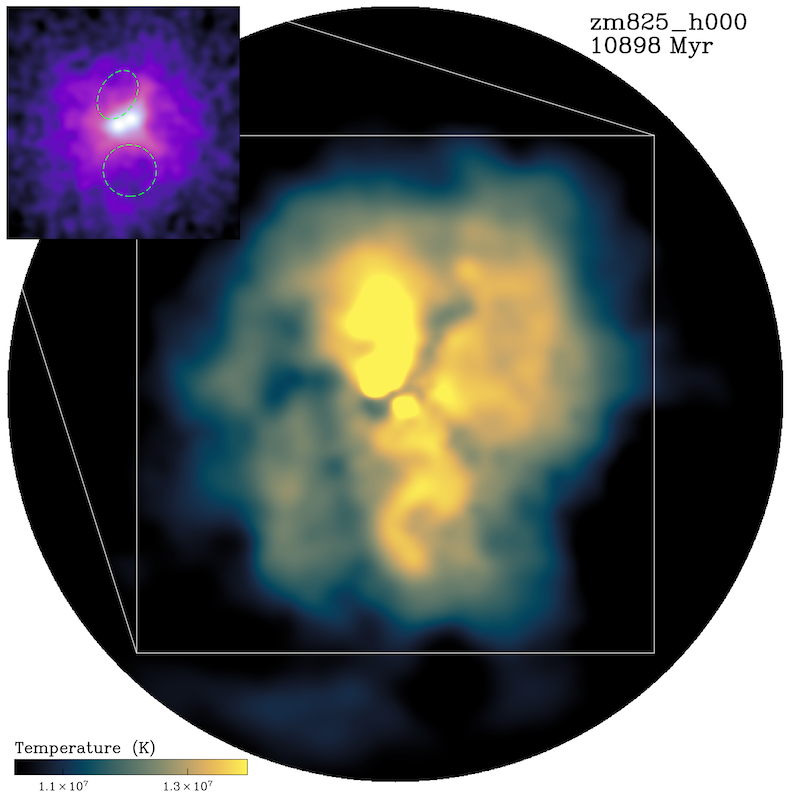}}\hfill
 \subfloat[]{\includegraphics[width=0.48\textwidth]{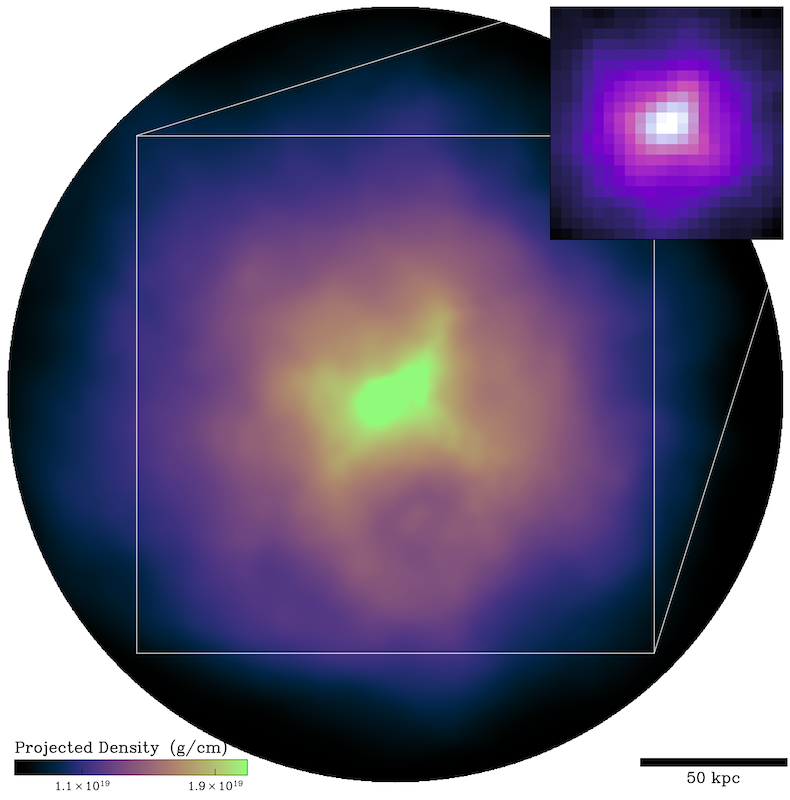}}\\[-2ex]  

 \caption{\textbf{Zoom825\_halo0 ($z_{obs}=0.026$):} A mid-range mass halo in our sample, with an initial mass of $M_{500} = 10^{13.49}M_\odot$. The cavity to the south is clearest, and cooler, yet interestingly these two cavities were inflated by jets launching at about the same time. The south cavity is larger whereas the northern cavity is more compact and contains hotter gas. The southern cavity is especially clear in both \textit{Chandra} and \textit{LEM} flux maps. This group has a complex multi-phase structure, with filaments of cold gas extending into the core, and cold blobs peppering the \ac{igrm}.} \label{feature:Zoom825_halo0}
\end{figure*}

\begin{figure*}
\captionsetup[subfigure]{labelformat=empty}
 \subfloat[]{\includegraphics[width=0.48\textwidth]{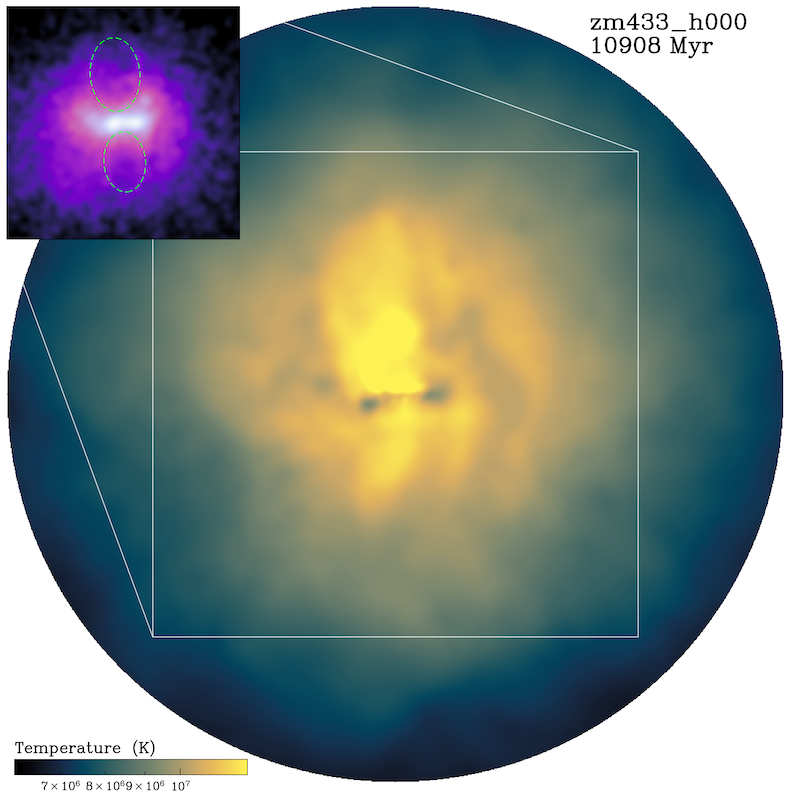}}\hfill
 \subfloat[]{\includegraphics[width=0.48\textwidth]{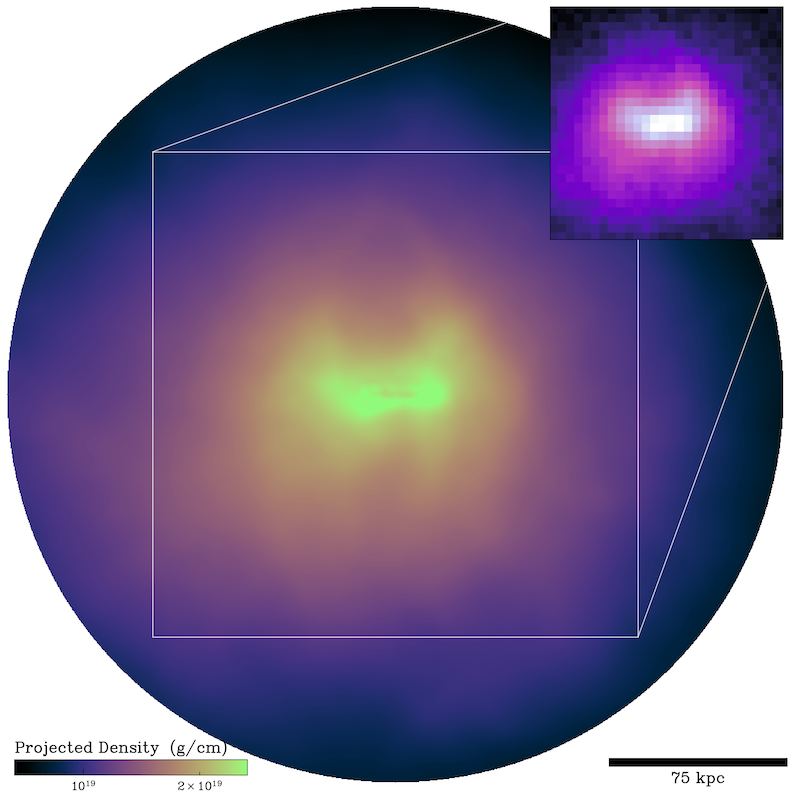}}\\[-2ex]  

 \caption{\textbf{Zoom433\_halo0 ($z_{obs}=0.0237$):} A group with mass $10^{13.39} M_\odot$, which shows the classic x-shaped region corresponding to jet feedback as has been observed in real systems. This corresponds to "wings" of emission which bound the cavities.} \label{feature:Zoom433_halo0}
\end{figure*}

\begin{figure*}
\captionsetup[subfigure]{labelformat=empty}
 \subfloat[]{\includegraphics[width=0.48\textwidth]{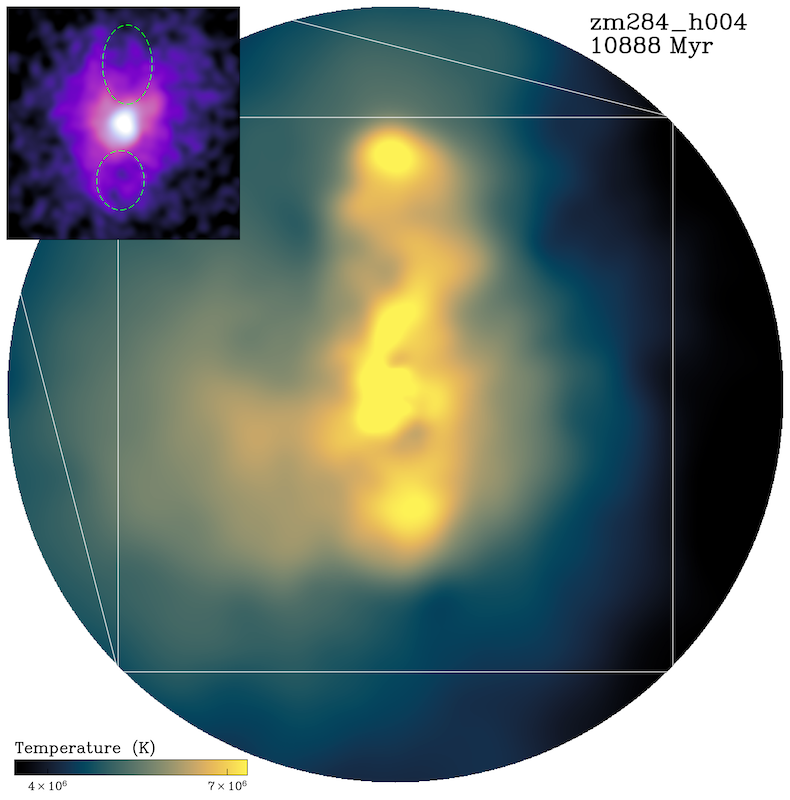}}\hfill
 \subfloat[]{\includegraphics[width=0.48\textwidth]{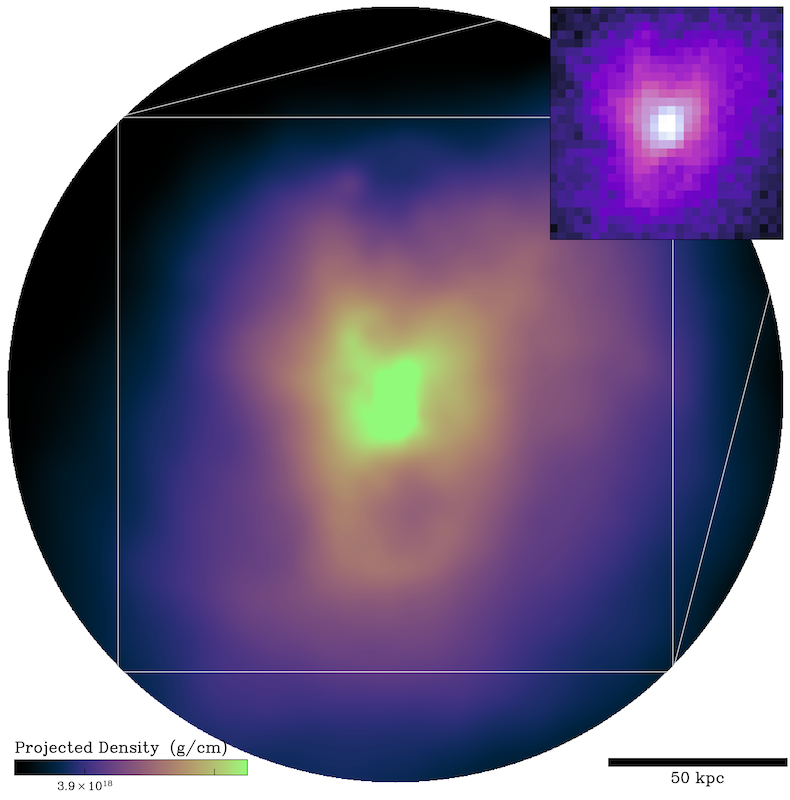}}\\[-2ex]  

 \caption{\textbf{Zoom284\_halo4 ($z_{obs}=0.0179$):} One of the lowest-mass halos we have in our sample at $M_{500} = 10^{13.03} M_\odot$, this group still produces clear bipolar X-ray cavities characterised by a dip in the density and a boosted plasma temperature. Wing-like structures are still observable in the mock X-ray maps.} \label{feature:Zoom284_halo4}
\end{figure*}

In Figs. \ref{feature:Zoom825_halo0}, \ref{feature:Zoom433_halo0}, and \ref{feature:Zoom284_halo4} we show that even halos in the lowest mass bin show clear x-shaped emission structure due to cavities displacing fluid as they grow and rise. This suggests that groups in this respect are similar to clusters, and we should expect to observe similar "wing"-like excess emission around the centres of groups corresponding to the opening of cavities. Such features, on scales of a few tens of kiloparsecs, are ubiquitous in observations of cluster-scale halos, for example in \textit{Abell 133} \citep{FujitaSarazin2002,FujitaSarazin2004,RandallClarke2010}, and in \textit{SDSS 1531} where they may coincide with increased star formation \citep{OmoruyiTremblay2024}.

\begin{figure}
	\includegraphics[width=\columnwidth]{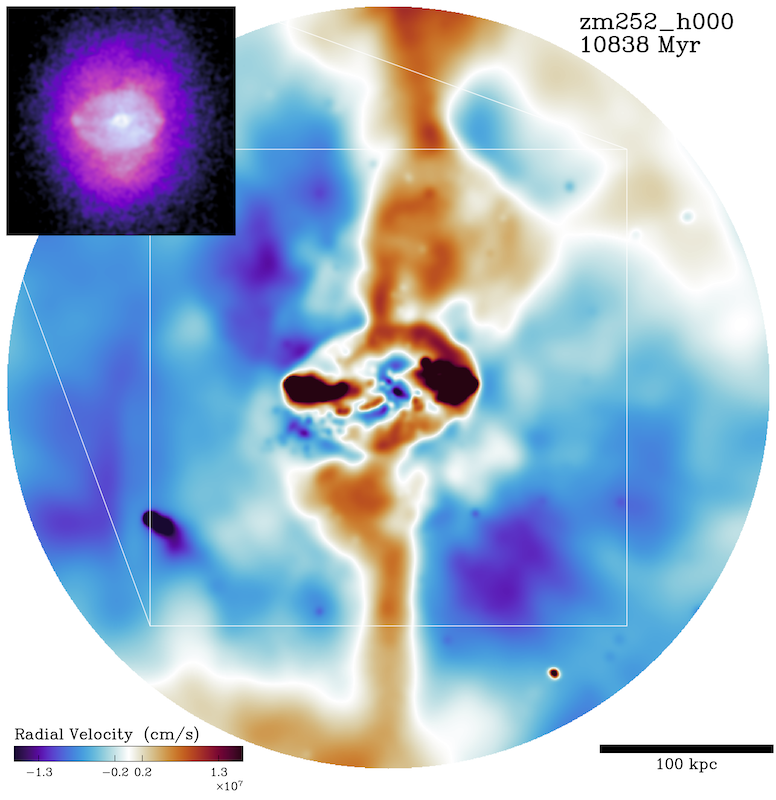}
    \caption{We show how the \simba model can re-produce hot-spots and jet-associated X-ray excess that has been seen in observations, for instance in Cygnus A \citep{DuffyWorrall2018}, as well as on smaller galactic scales \citep{MarshallHarris2001, HardcastleLenc2016}.}
    \label{fig:zm252_h000 hotspot}
\end{figure}

In Figs \ref{feature:Zoom284_halo4} and \ref{fig:zm252_h000 hotspot} we see examples, for two different groups, of "hotspots" in the projected temperature and in the X-ray luminosity respectively occurring at the apexes of each jet lobe. This matches observations of, for example, \textit{Cygnus A} \citep{SteenbruggeBlundell2008,SniosJohnson2020}, and is due to the gas shocking at the head of the jet \citep{SmithNorman1985, SmithDonohoe2021}.

\begin{figure*}
\captionsetup[subfigure]{labelformat=empty}

 \subfloat[]{\includegraphics[width=0.223\textwidth]{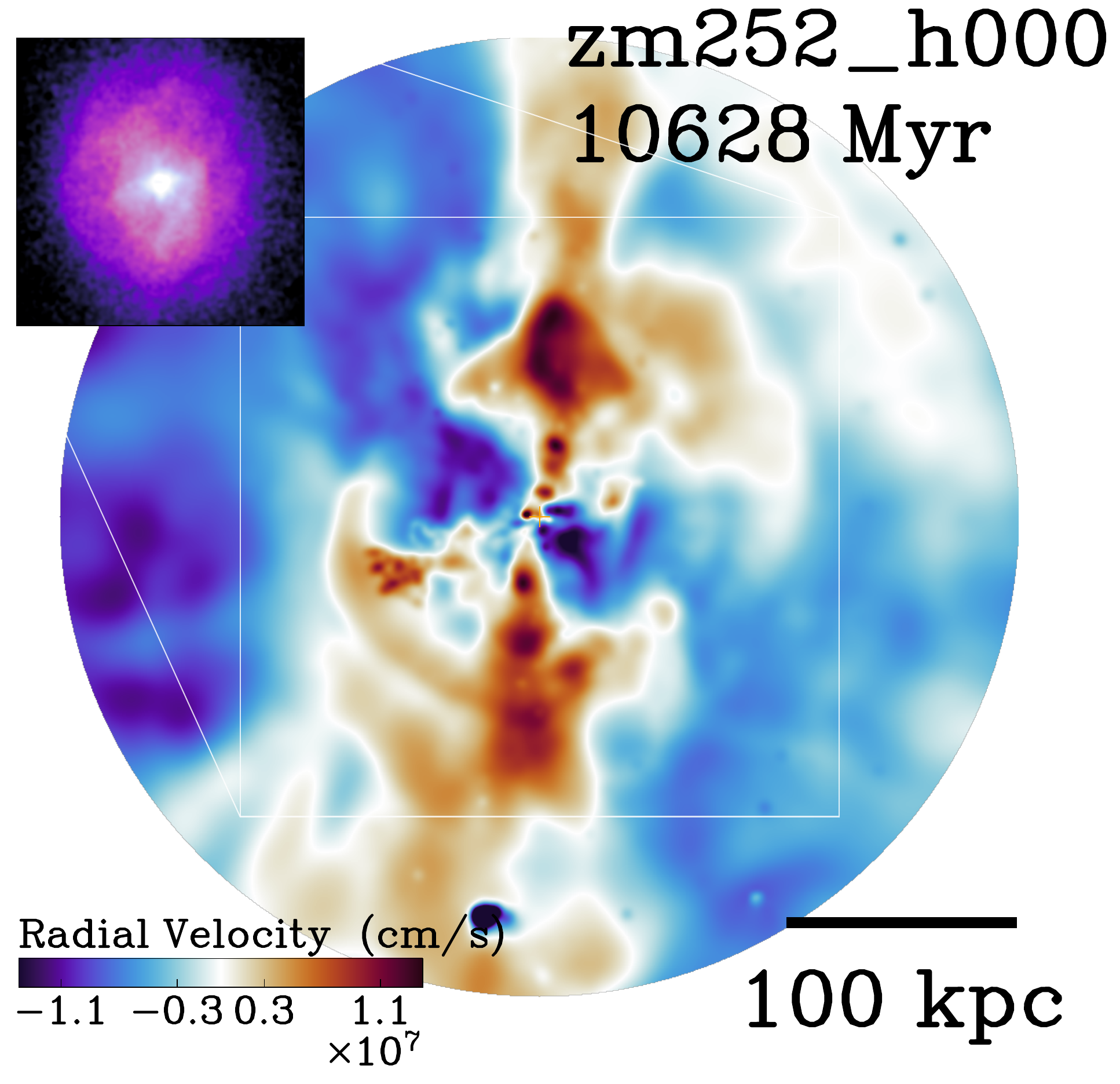}}\hfill
 \subfloat[]{\includegraphics[width=0.223\textwidth]{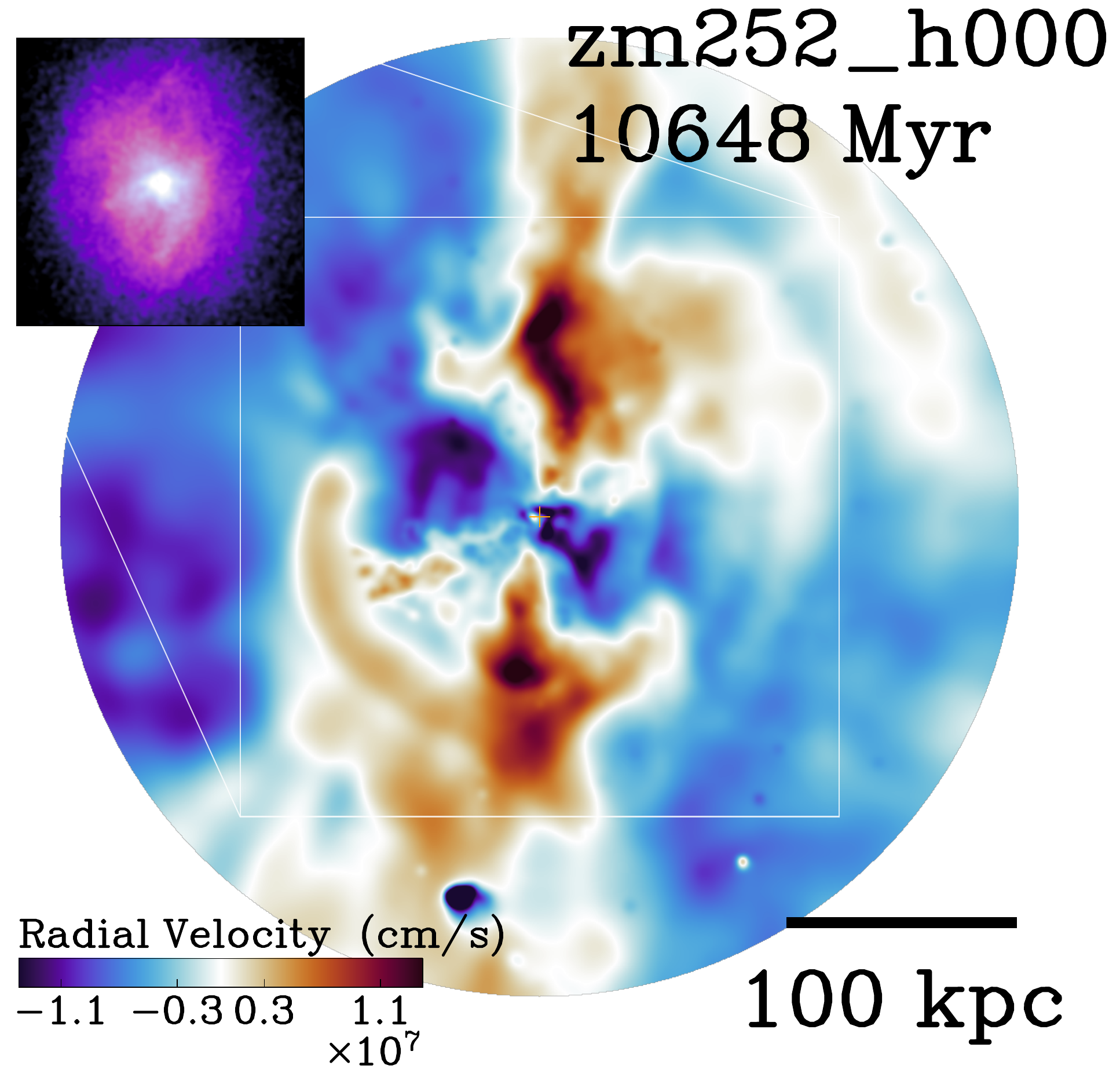}}\hfill
 \subfloat[]{\includegraphics[width=0.223\textwidth]{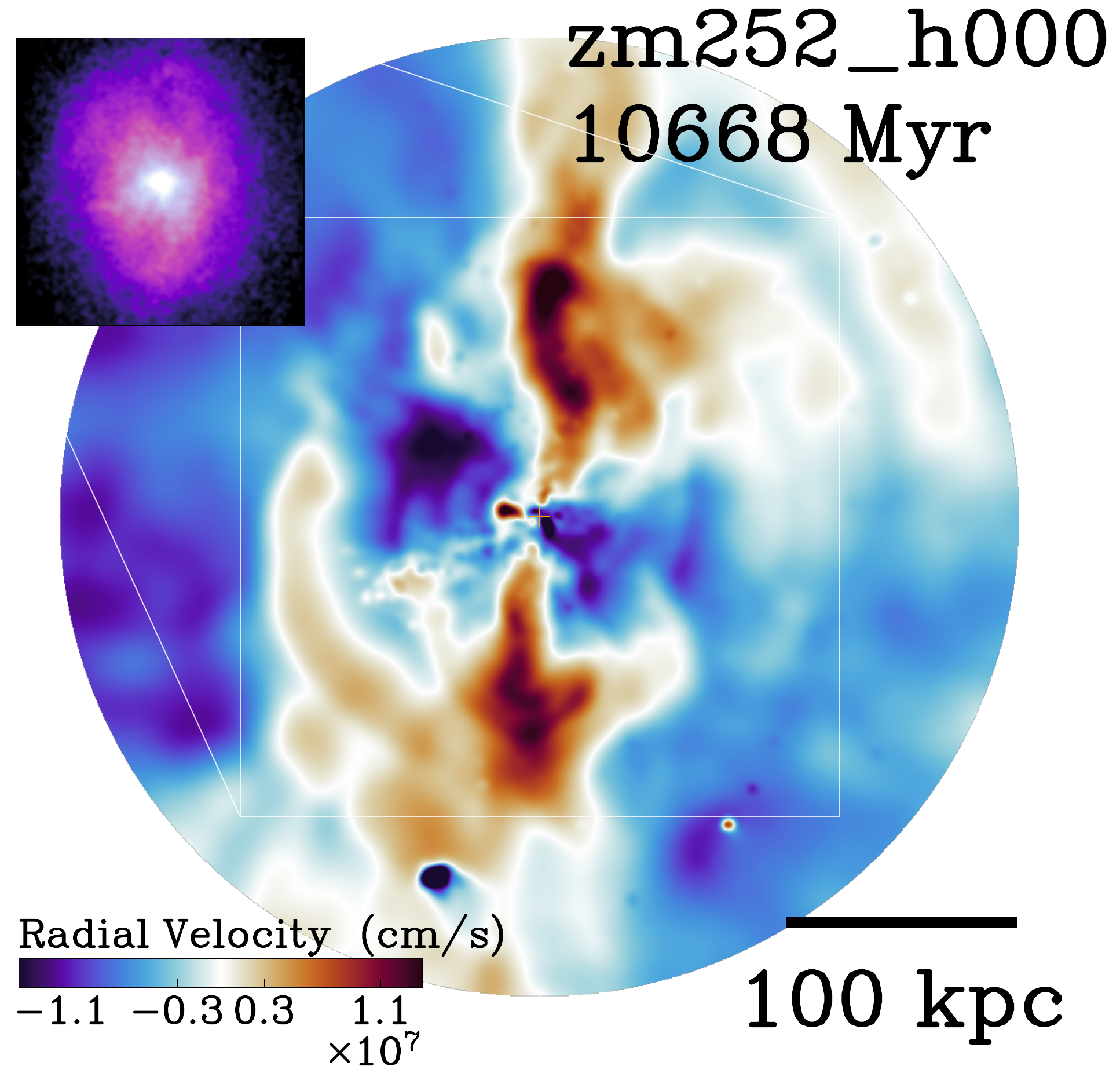}}\hfill
 \subfloat[]{\includegraphics[width=0.223\textwidth]{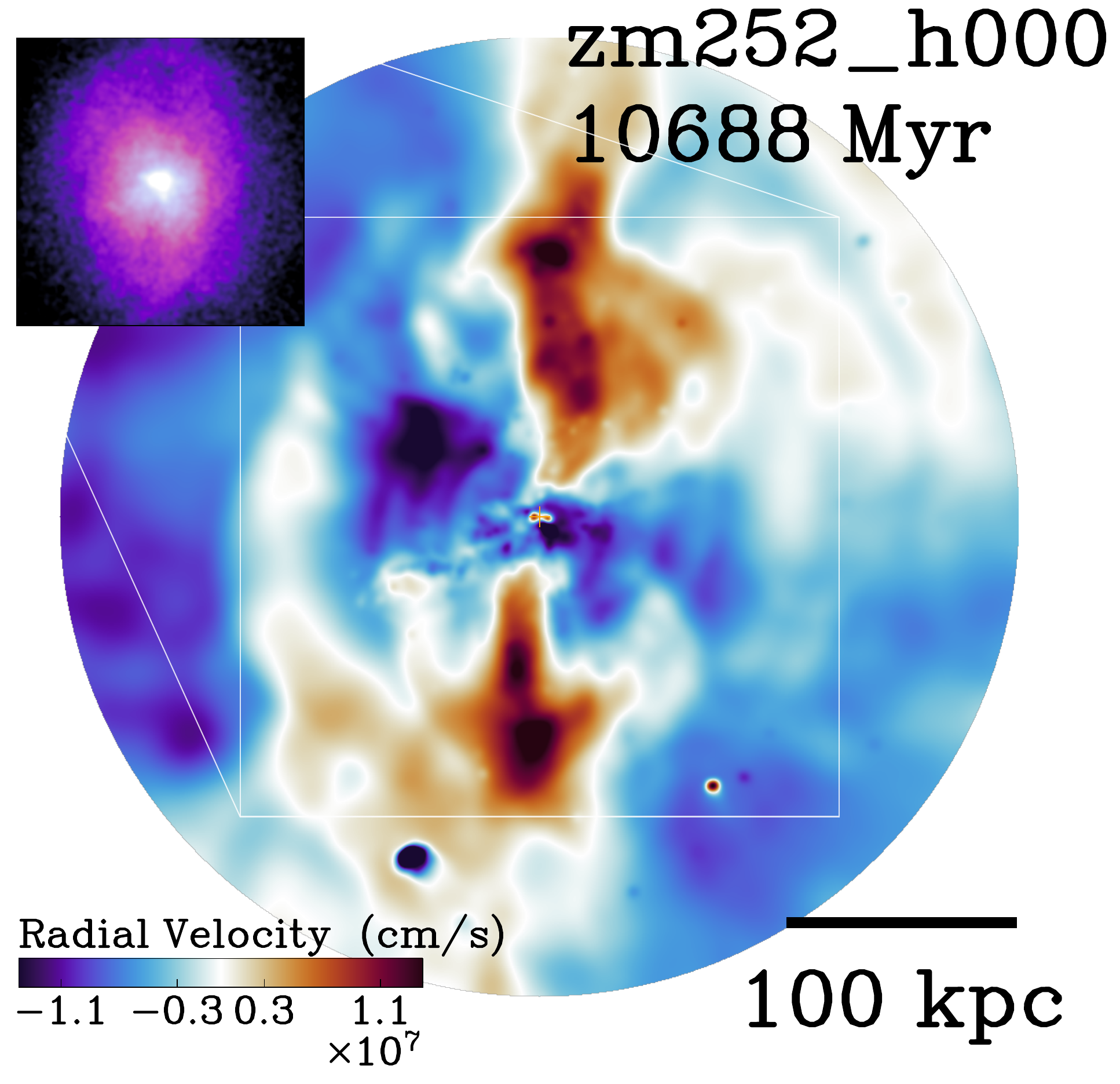}}\\[-6ex]  
 \subfloat[]{\includegraphics[width=0.223\textwidth]{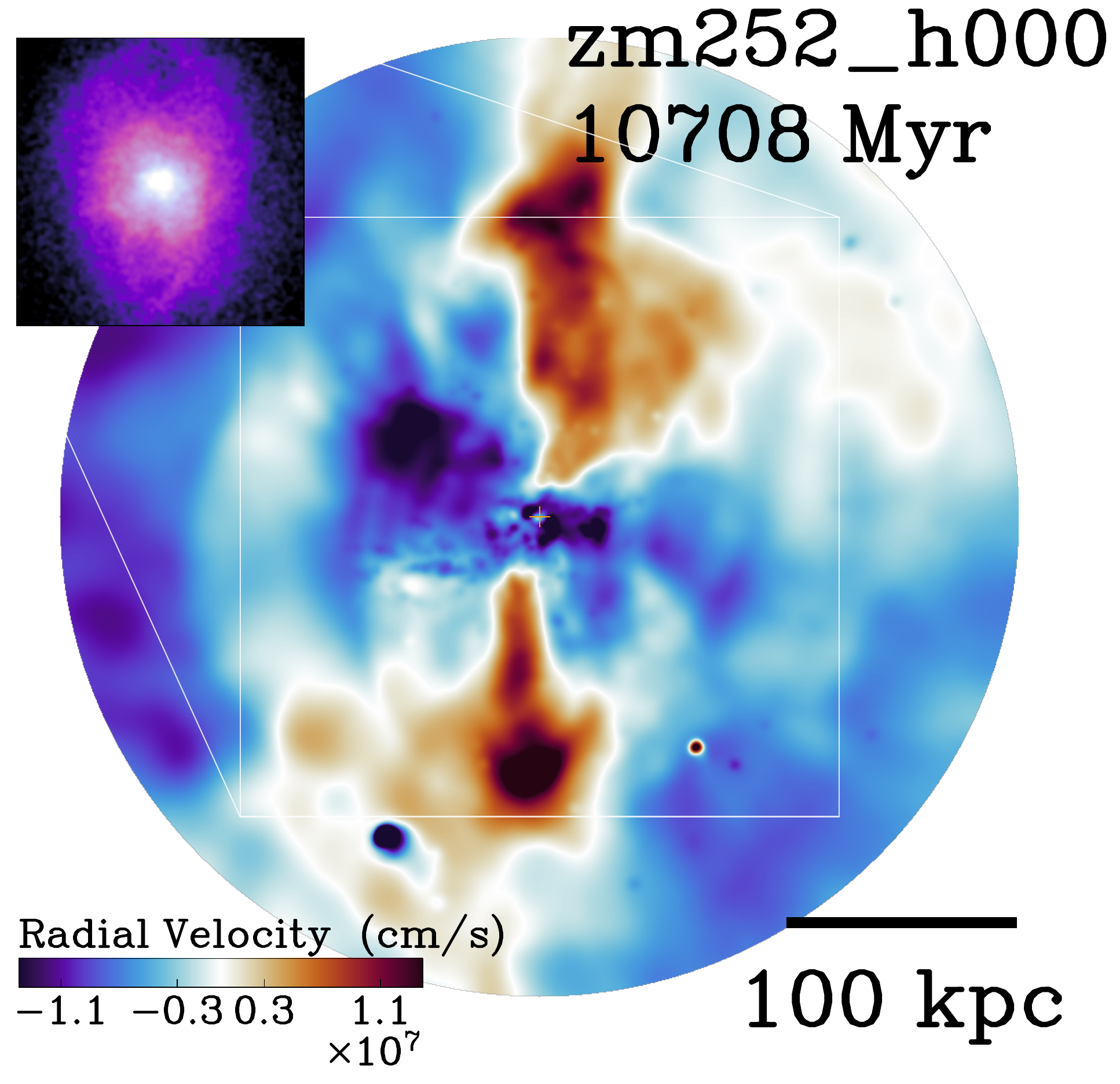}}\hfill
 \subfloat[]{\includegraphics[width=0.223\textwidth]{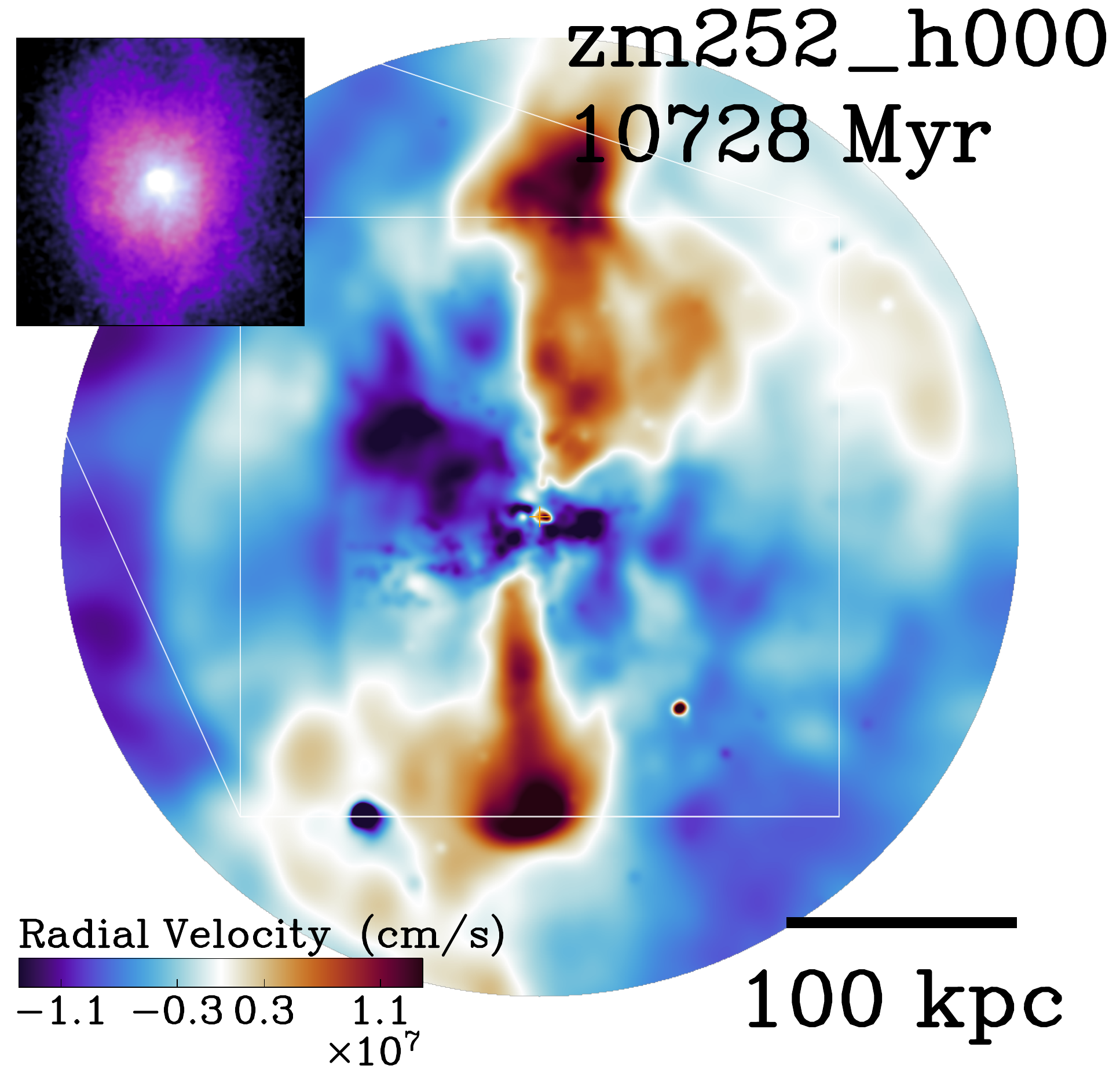}}\hfill
 \subfloat[]{\includegraphics[width=0.223\textwidth]{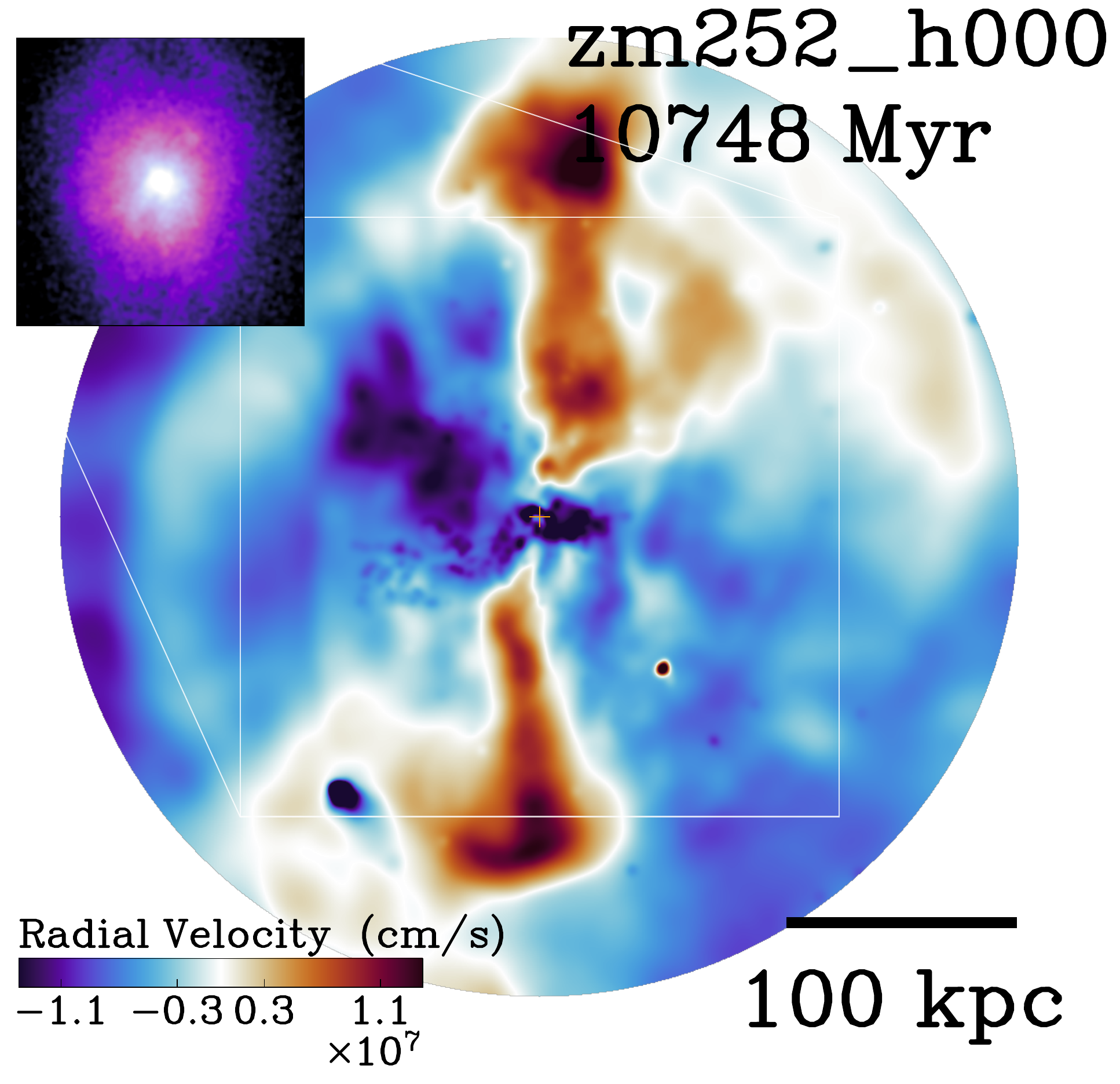}}\hfill
 \subfloat[]{\includegraphics[width=0.223\textwidth]{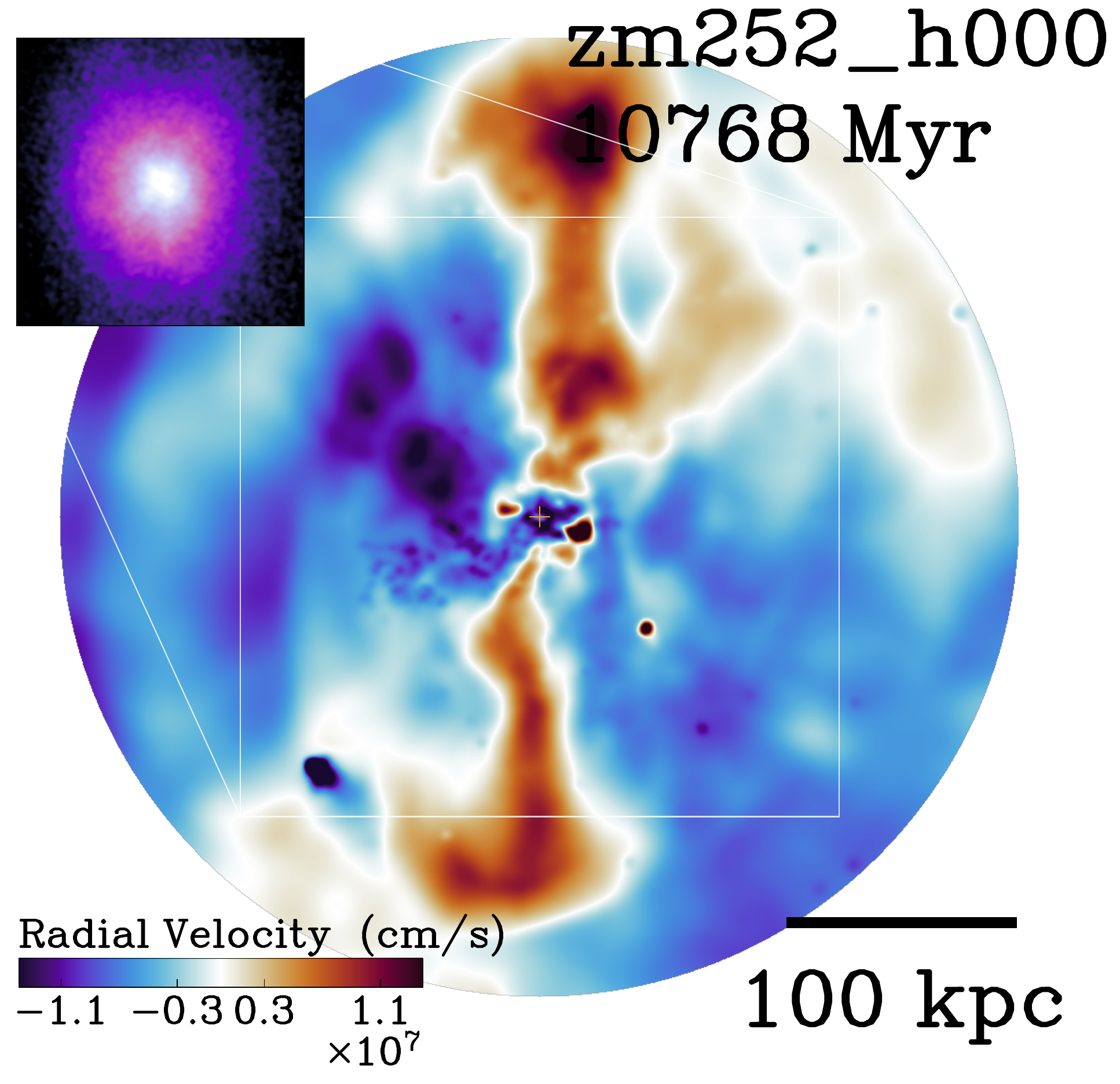}}\\[-6ex]  
 \subfloat[]{\includegraphics[width=0.223\textwidth]{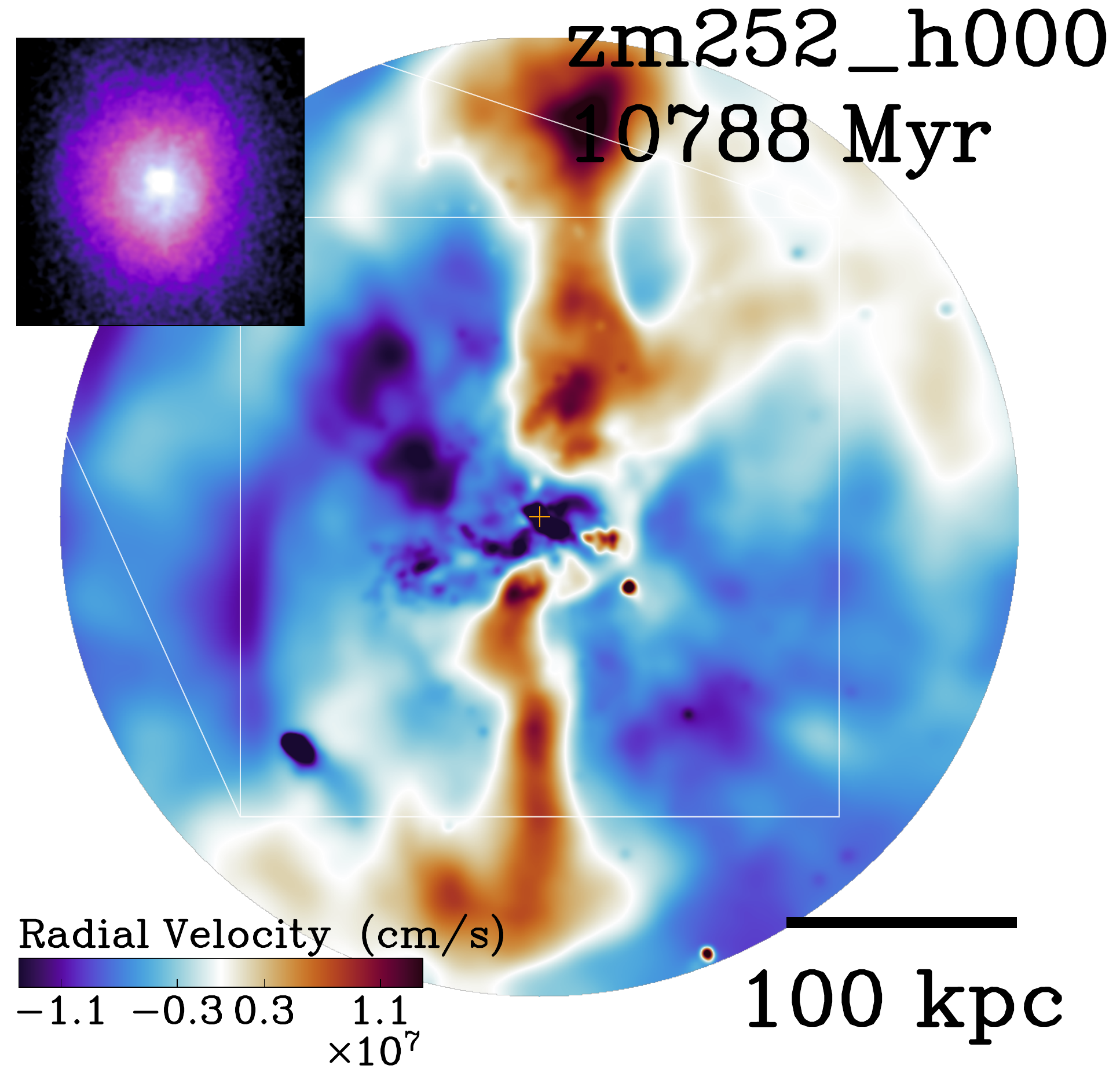}}\hfill
 \subfloat[]{\includegraphics[width=0.223\textwidth]{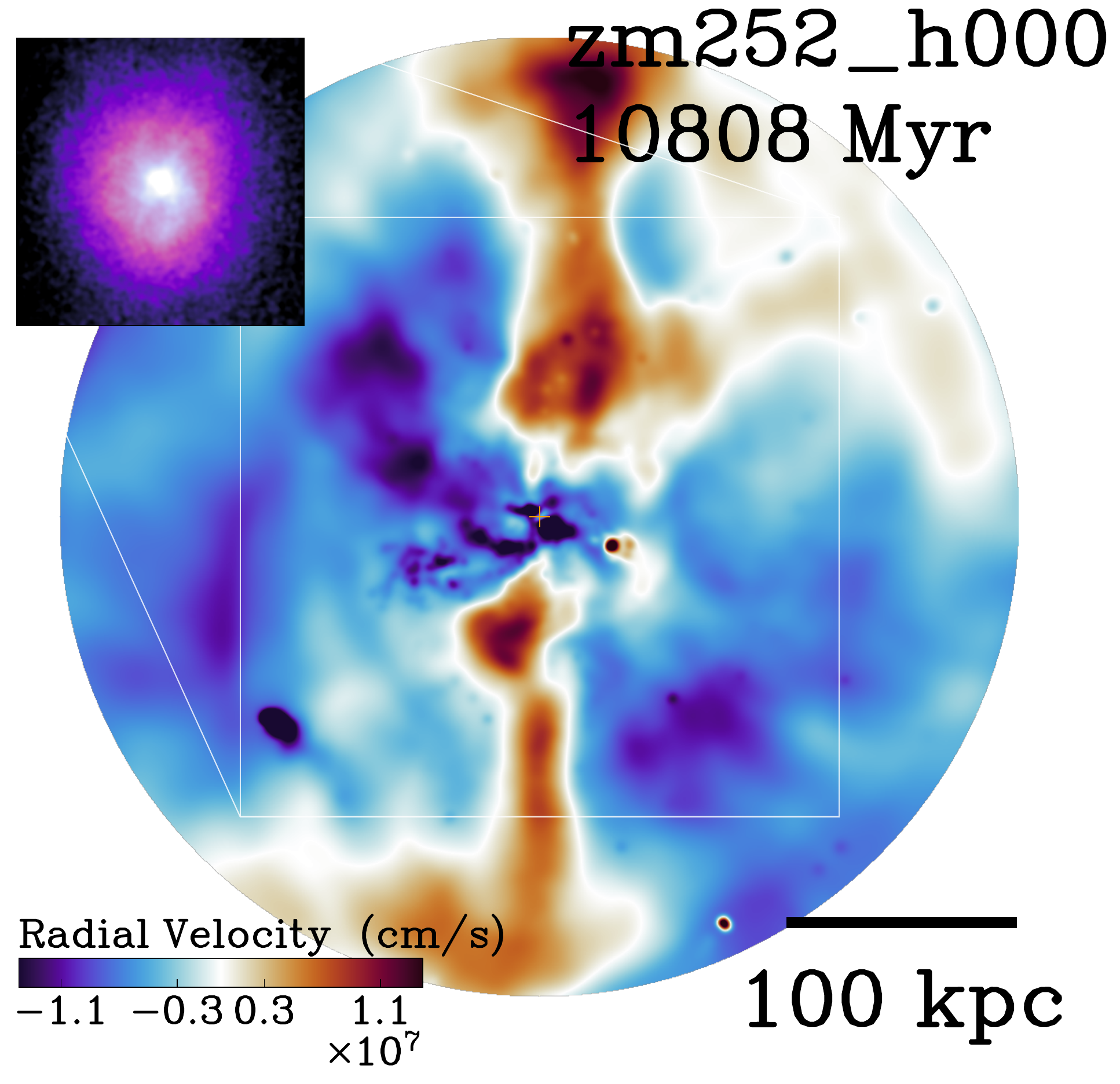}}\hfill
 \subfloat[]{\includegraphics[width=0.223\textwidth]{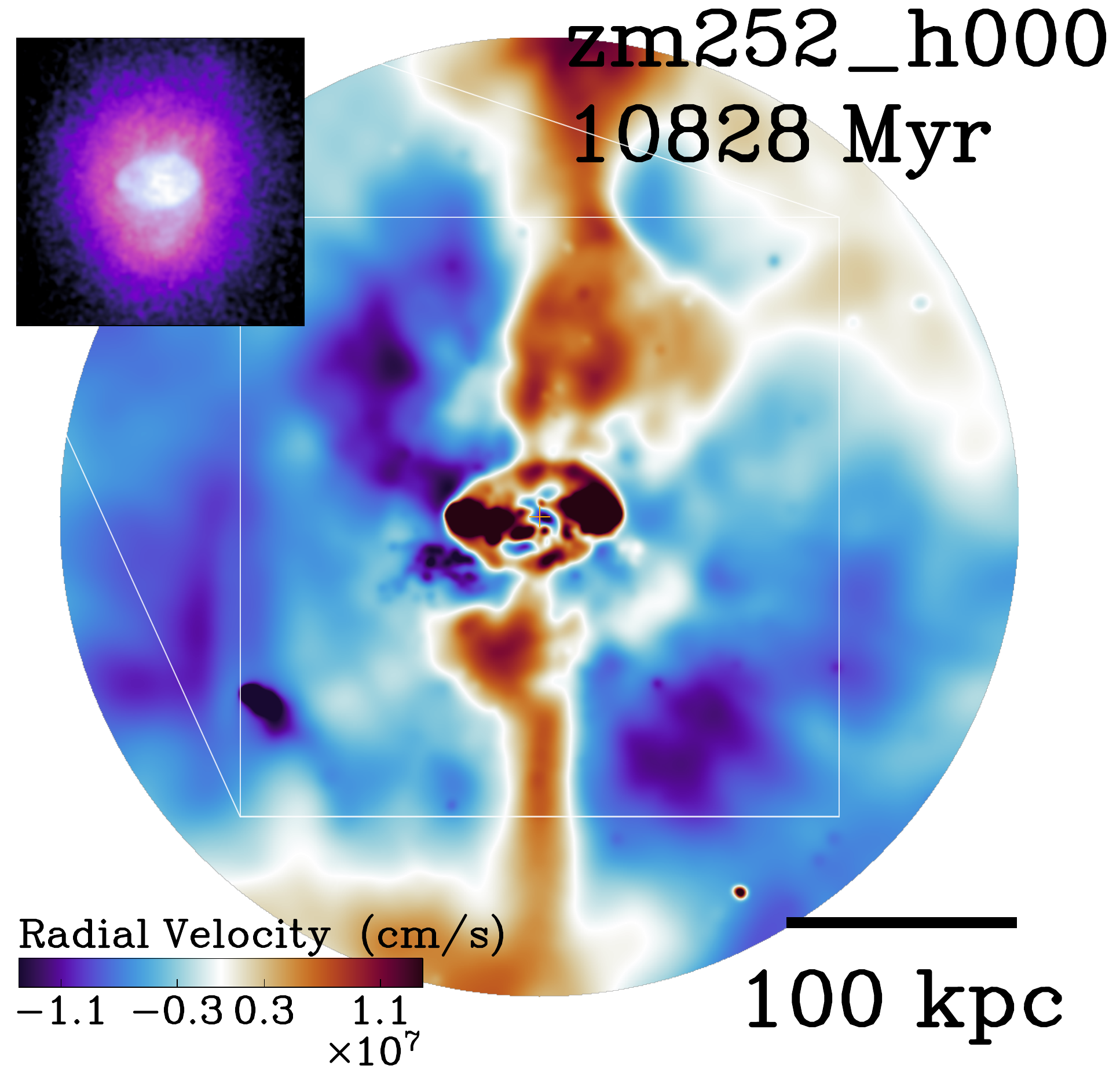}}\hfill
 \subfloat[]{\includegraphics[width=0.223\textwidth]{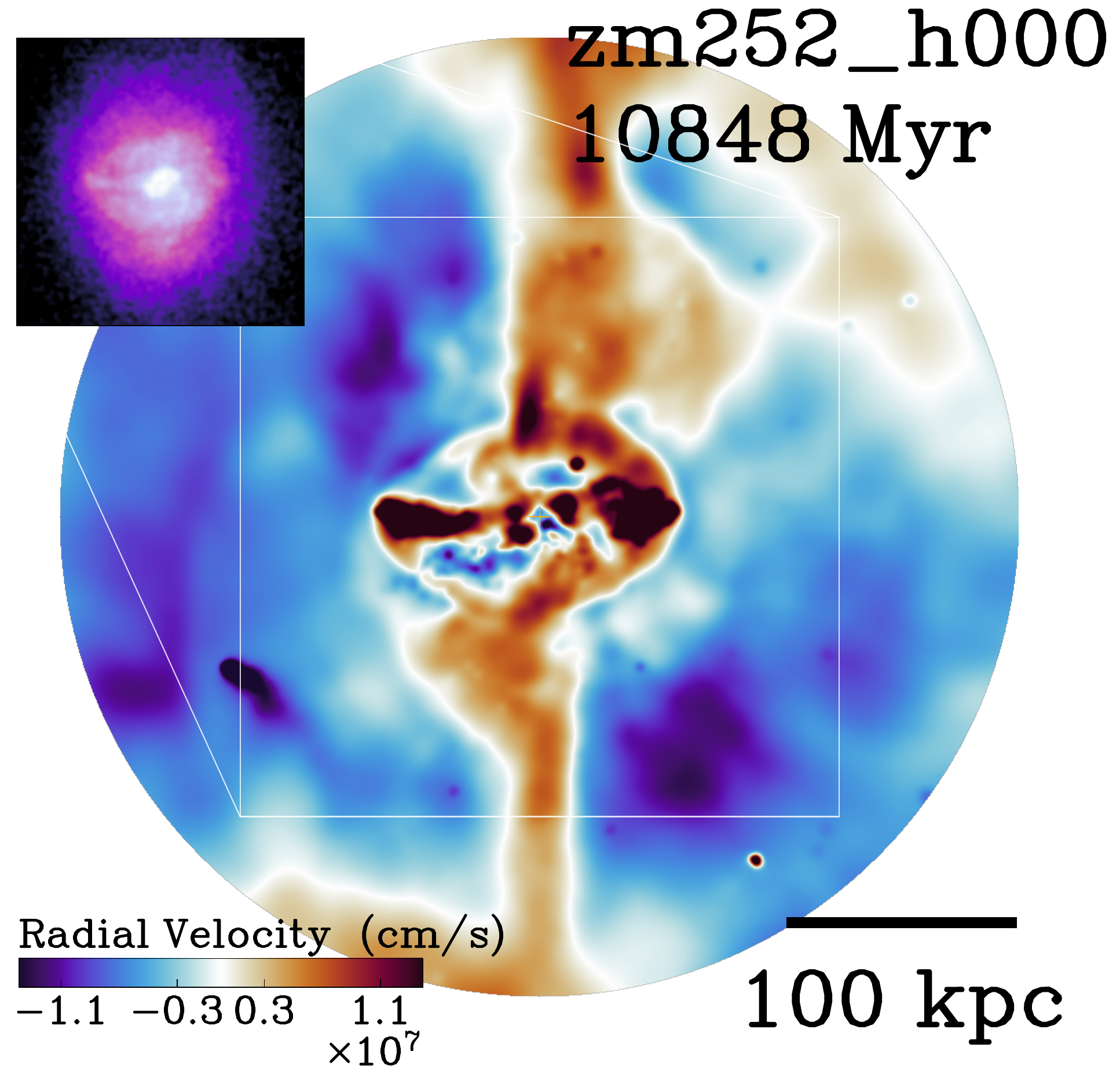}}\\[-6ex]  
 \subfloat[]{\includegraphics[width=0.223\textwidth]{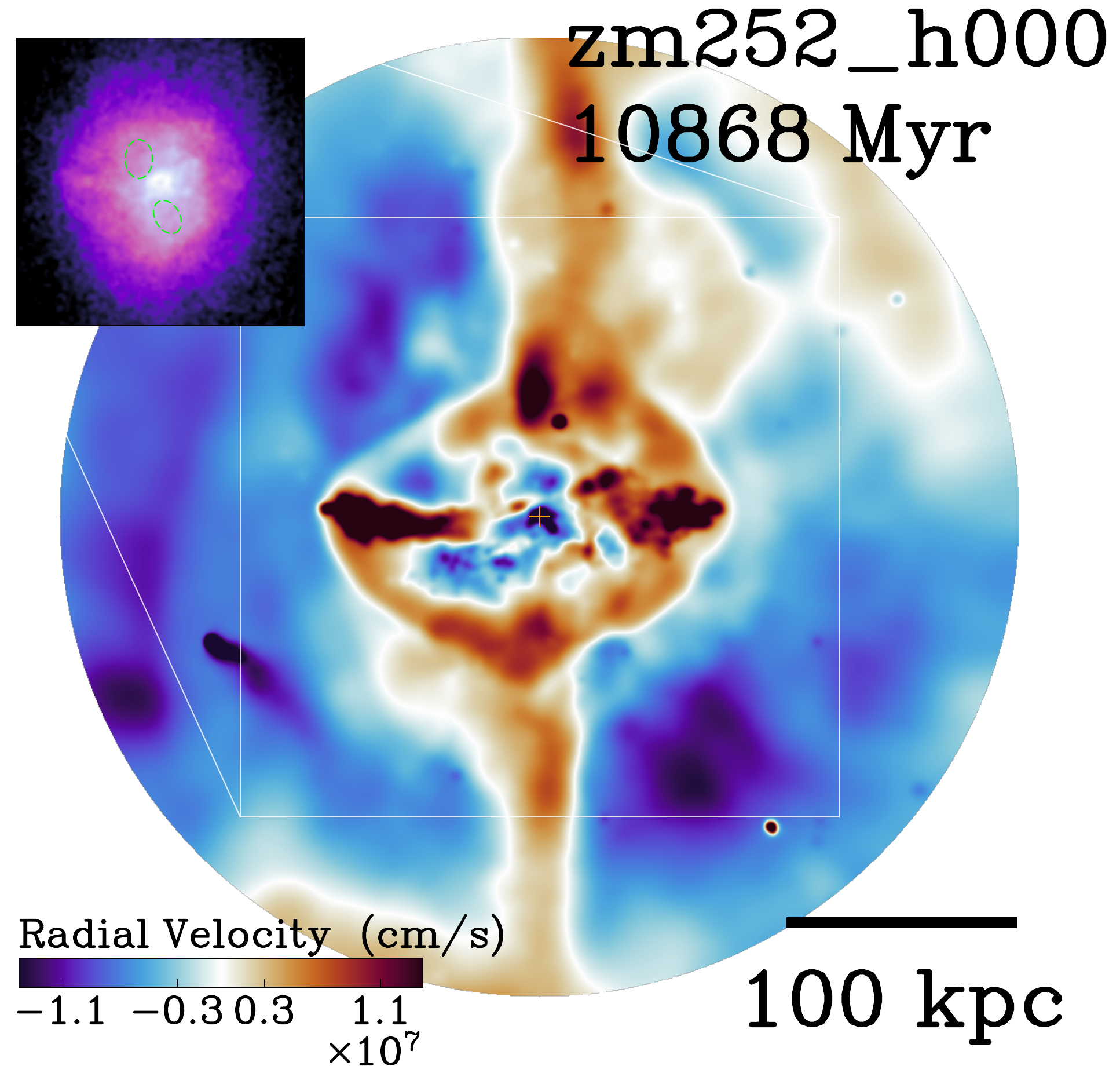}}\hfill
 \subfloat[]{\includegraphics[width=0.223\textwidth]{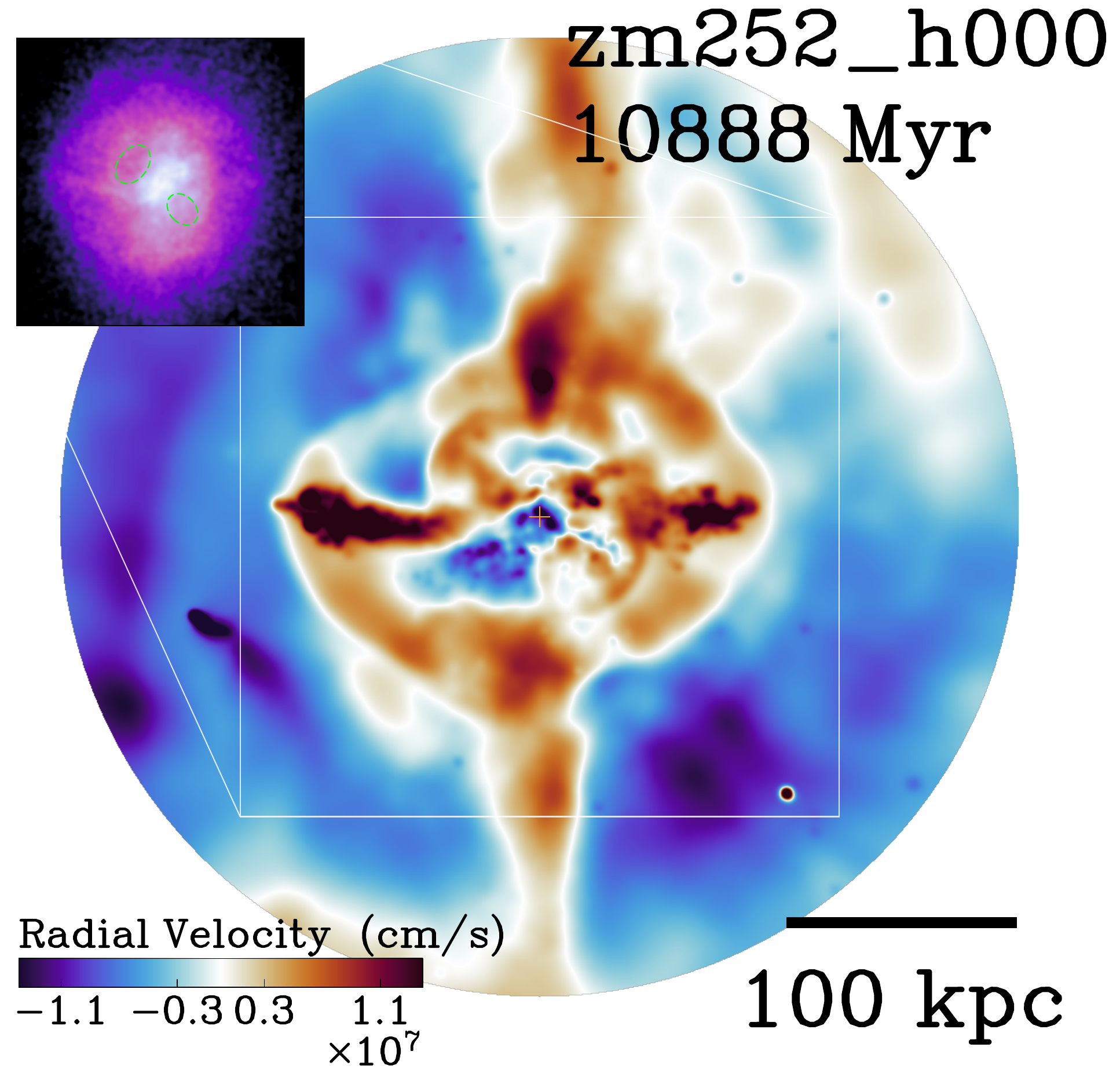}}\hfill
 \subfloat[]{\includegraphics[width=0.223\textwidth]{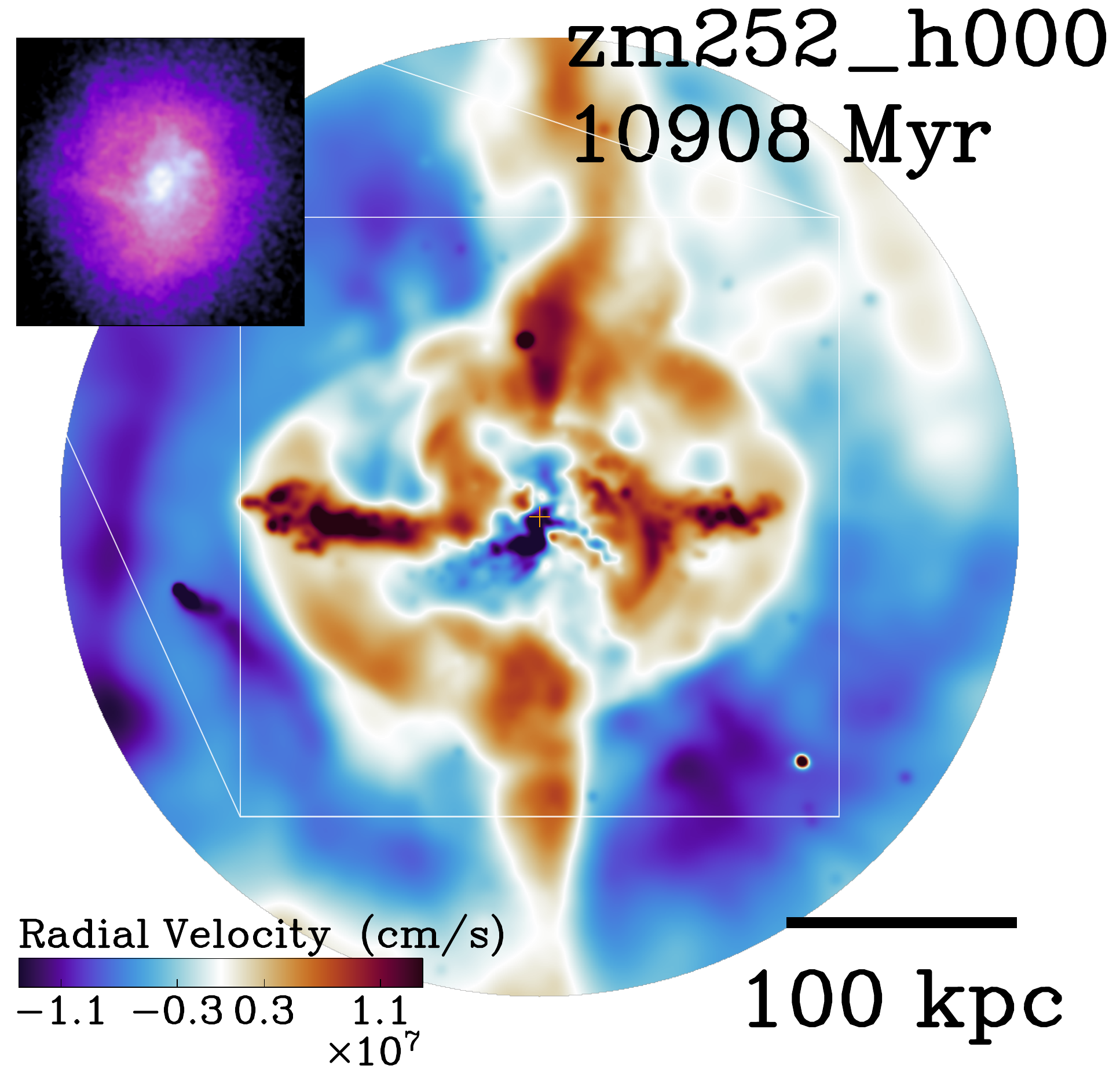}}\hfill
 \subfloat[]{\includegraphics[width=0.223\textwidth]{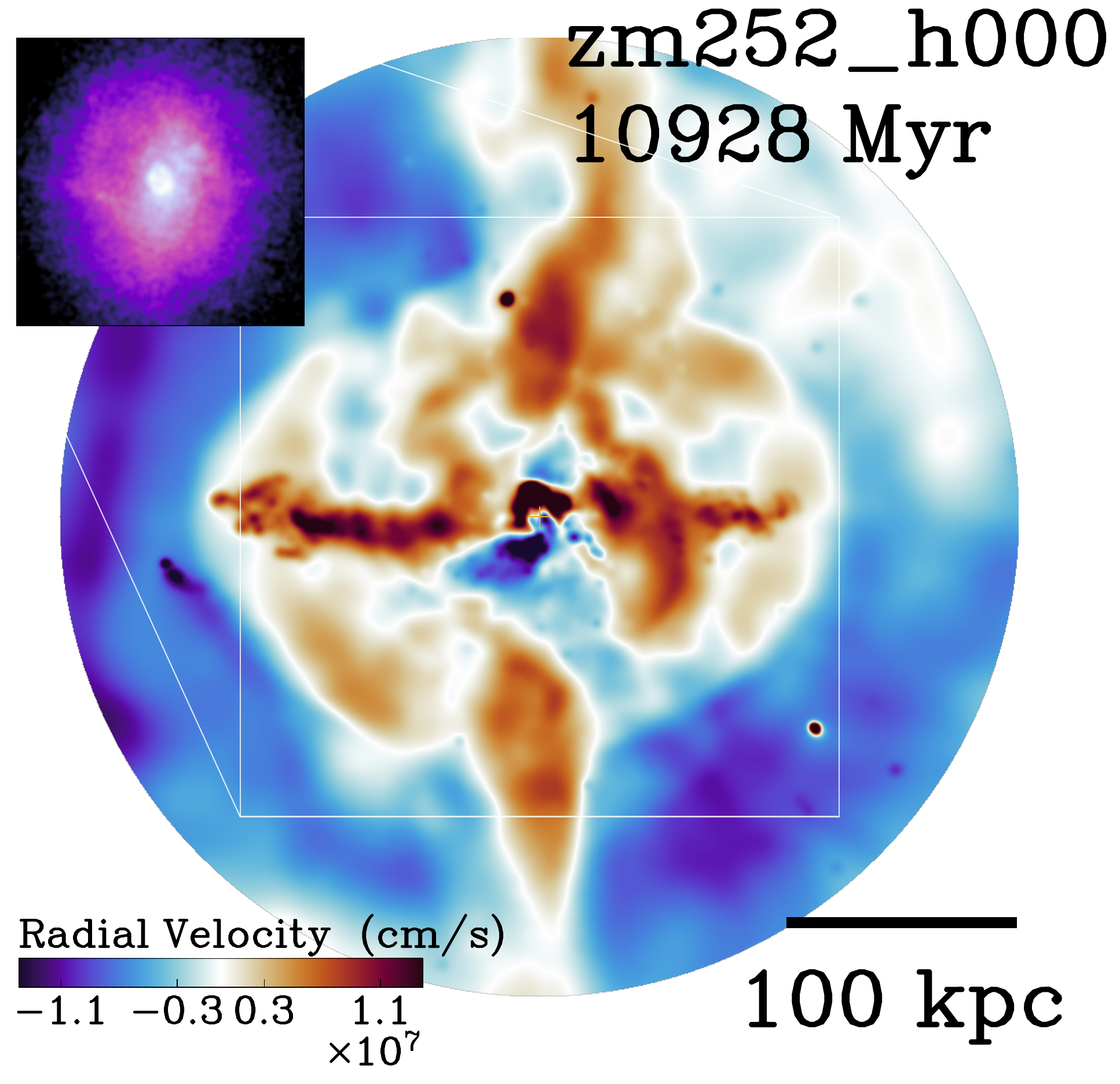}}\\[-6ex]  
 \subfloat[]{\includegraphics[width=0.223\textwidth]{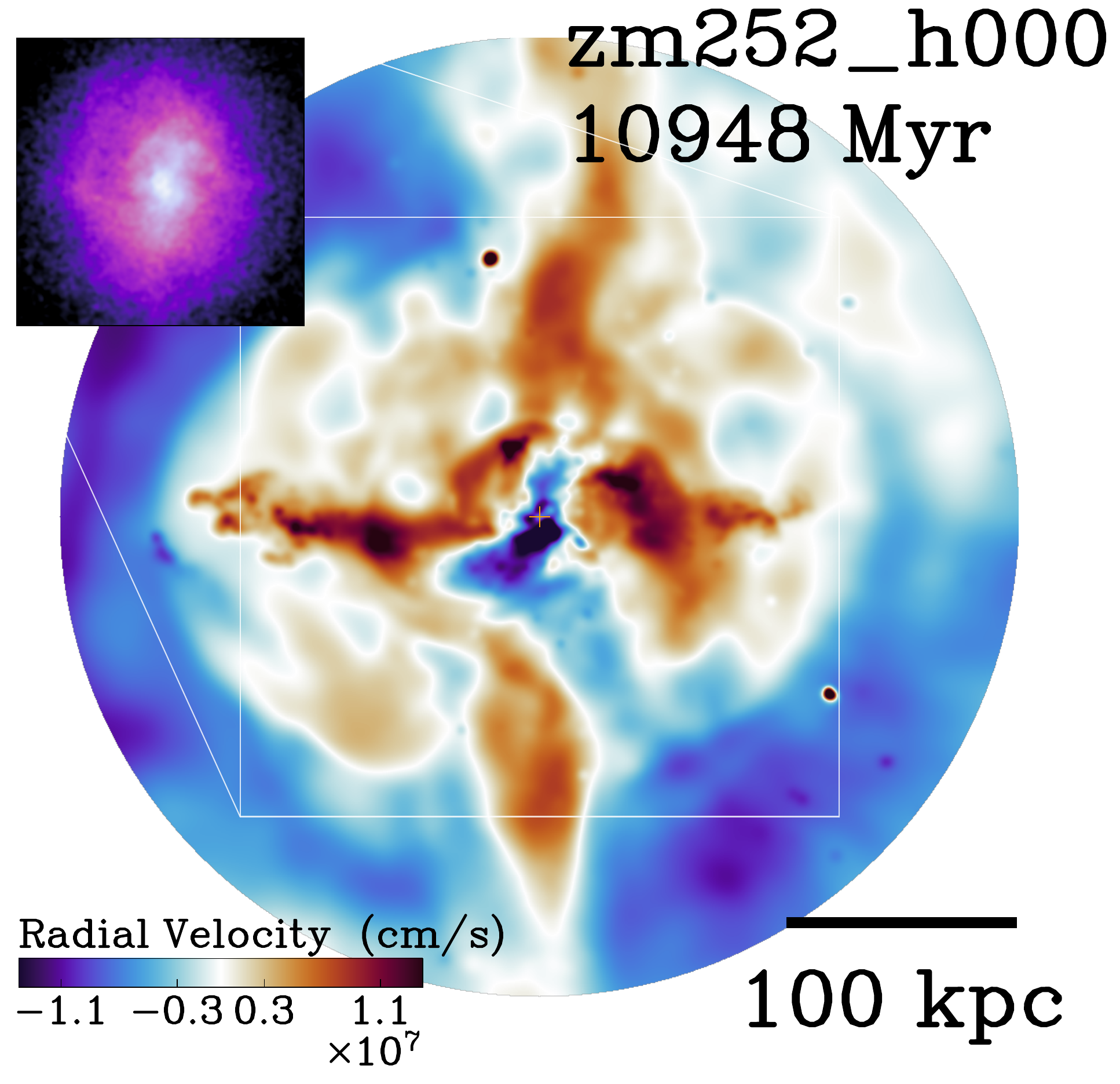}}\hfill
 \subfloat[]{\includegraphics[width=0.223\textwidth]{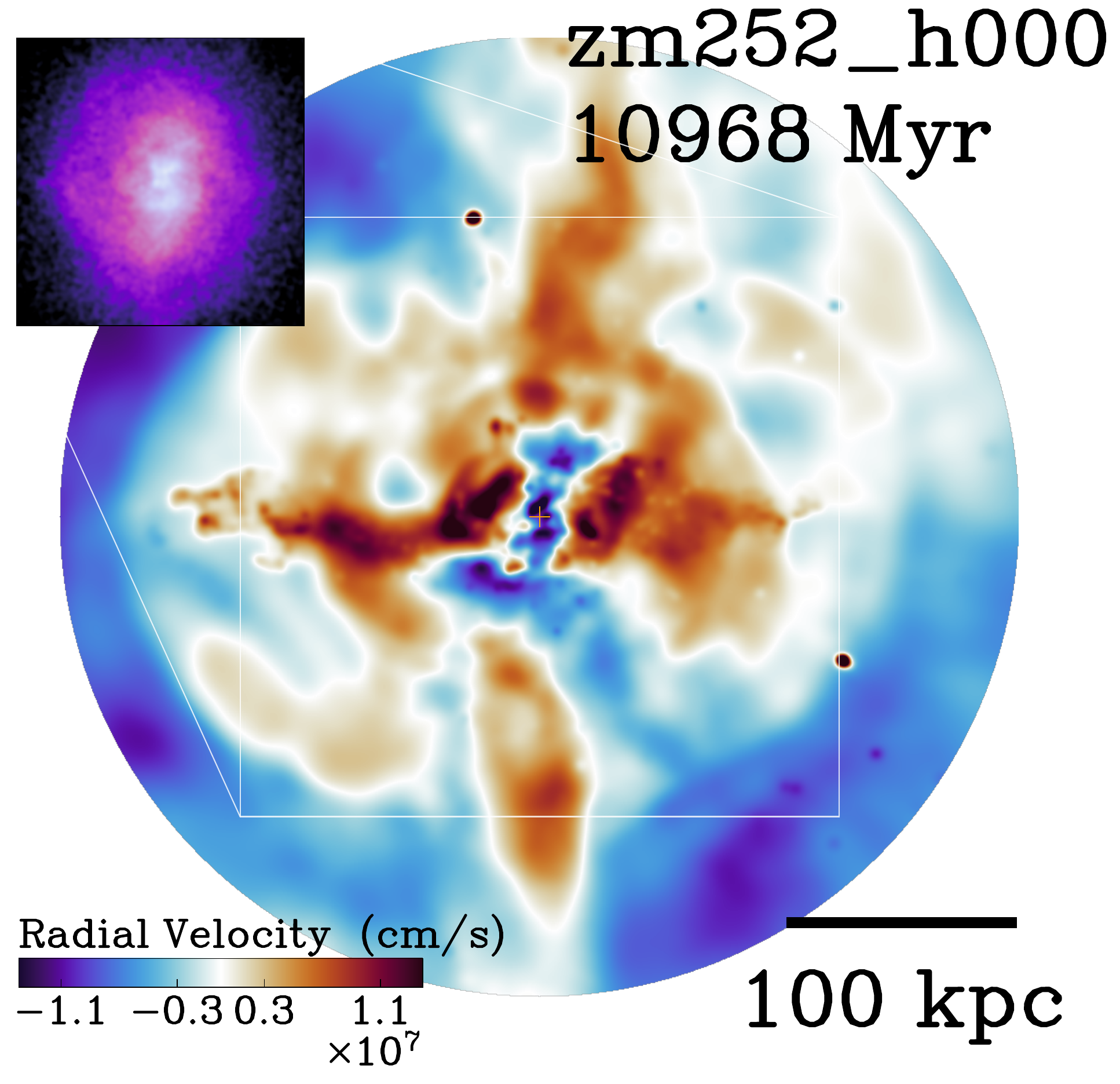}}\hfill
 \subfloat[]{\includegraphics[width=0.223\textwidth]{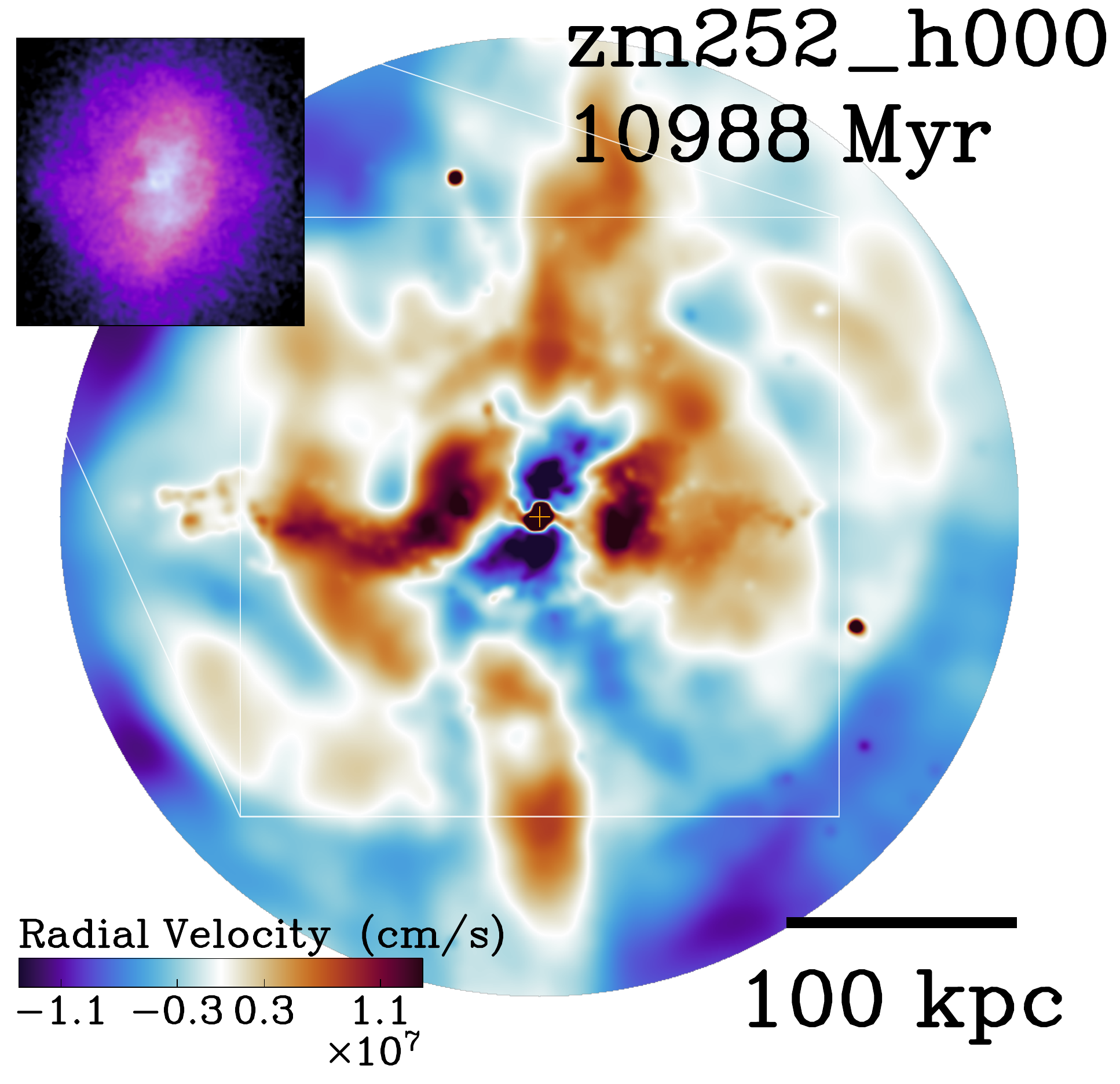}}\hfill
 \subfloat[]{\includegraphics[width=0.223\textwidth]{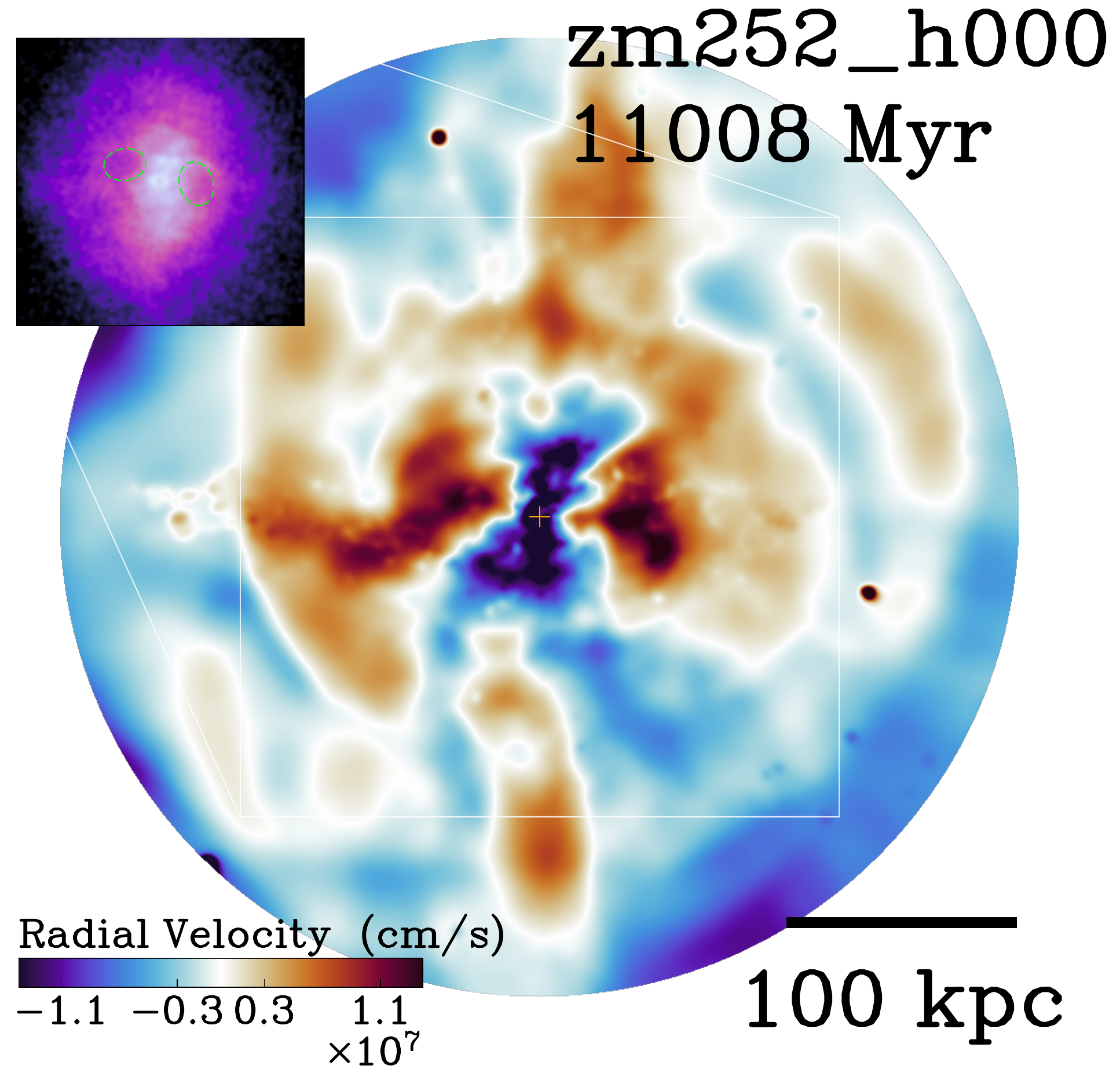}}

 \caption{We show the time-evolution of the radial velocity of all gas in \textbf{Zoom252\_halo0} ($z_{obs}=0.032$) to demonstrate how a strongly-jetting system evolves with time in \textit{Hyenas}. In the earlier snapshots, the jet is North-South and inflates two large bubbles which grow and rise with time. Sound waves in the \ac{igrm} can also be clearly seen emanating for several $10$s of kpc from the central regions, propagating perpendicular to the jet. After $~200$ Myr the jet axis flips and the bubbles inflate along the East-West direction. The jets themselves are sufficiently hot that the X-ray emission is maintained such that in several cases we only observe cavities slightly adjacent to them, and observe the jets themselves in the X-ray (for instance at a simulation time of $10848$ Myr). This matches what is seen occasionally in real systems, for instance in \textit{Cygnus A} \citep{SteenbruggeBlundell2008,SniosJohnson2020}. After the initial burst, the jet cools and cavities become apparent along the jet axes.} \label{fig:zm252_h000_time_series}
\end{figure*}

Finally in Fig. \ref{fig:zm252_h000_time_series}, we show a time-series for the radial velocity evolution of our large halo \textbf{Zoom825\_halo0} spanning $380$ Myr, where the correspondence between outflows and inflated bubbles become visible in the X-ray maps. The bipolar nature of the launching is evident, as seen in both the radial velocity maps (Fig. \ref{fig:zm252_h000_time_series}) and the thermodynamic projections (Fig. \ref{feature:Zoom252_halo0}). Hot "channels" are carved in the \ac{igrm} over several tens of Myr before the jet axis re-orients, at which point the jet starts to work against pristine halo gas, causing an X-ray bright hotspot to appear at the site of the shock. Bubbles tend to live for several tens of Myr and become dimmer as they migrate radially.

\section{Bubbles Suppressing Cooling Flows}

The relationship between the central cooling luminosity and the cavity power is important because it indicates whether \ac{agn} feedback can sufficiently heat the \ac{icm} and prevent large-scale cooling flows. If the cavity power is of order the energy loss rate due to cooling, this is strong evidence towards jet feedback being able to supply sufficient energy (though the exact method of transferring this energy to the hot atmosphere is still poorly understood).

Observations have shown that in a large fraction of observed systems the energy contained within hot cavities is sufficient to offset radiative losses \citep{BirzanRafferty2004,Hlavacek-LarrondoMcDonald2015, OlivaresSu2022}. Heating on sufficient or near-sufficient levels has also been reproduced in idealised simulations of jet-mode feedback within cluster atmospheres, suppressing large-scale cooling flows \citep{SijackiSpringel2006,SijackiSpringel2007,PuchweinSijacki2008,CieloBabul2018}. It has yet to be determined if \ac{agn} jets in \textit{cosmological} simulations can effectively heat the \ac{igrm}, given that their energies are tuned to reproducing larger-scale trends in galaxy quenching and the galaxy stellar mass function, rather than to reproduce hot gas properties in halos. In this Section, we whether this is possible in the \simba model.

\subsection{Calculating the Cooling Luminosity}

\begin{figure}
 \captionsetup[subfigure]{labelformat=empty}
 \subfloat[]{\includegraphics[width=\columnwidth]{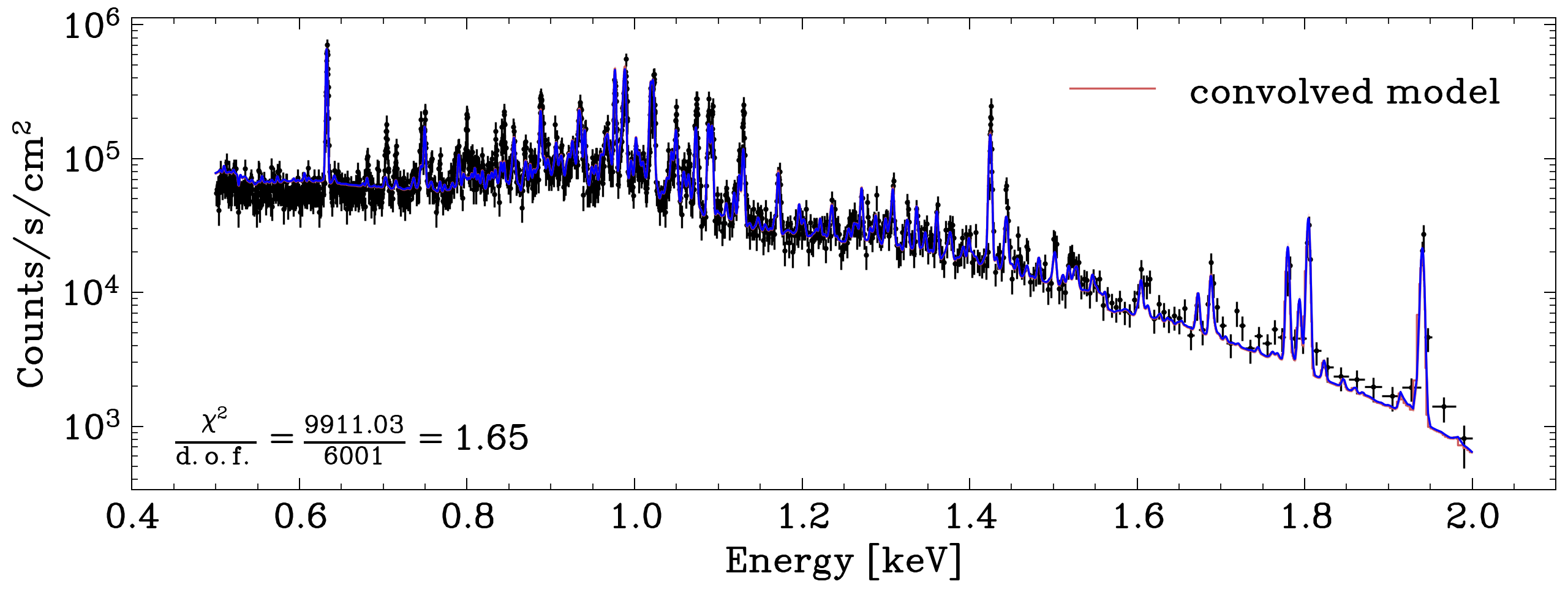}}\hfill
  \caption{The best fit to the $0.5-2$keV (observed-frame) mock spectrum extracted from the central $7.7$ Gyr cooling radius, as observed by \textit{LEM} with $2$ eV spectral resolution for one of the larger halos in our sample \textbf{Zoom266\_halo0 (snapshot 172)}. Posterior samples of the model parameter vector are shown in red and the maximum-likelihood solution is shown in blue.} \label{fig:LEM spec fit}
\end{figure}

\begin{figure}
 \captionsetup[subfigure]{labelformat=empty}
 \subfloat[]{\includegraphics[width=\columnwidth]
{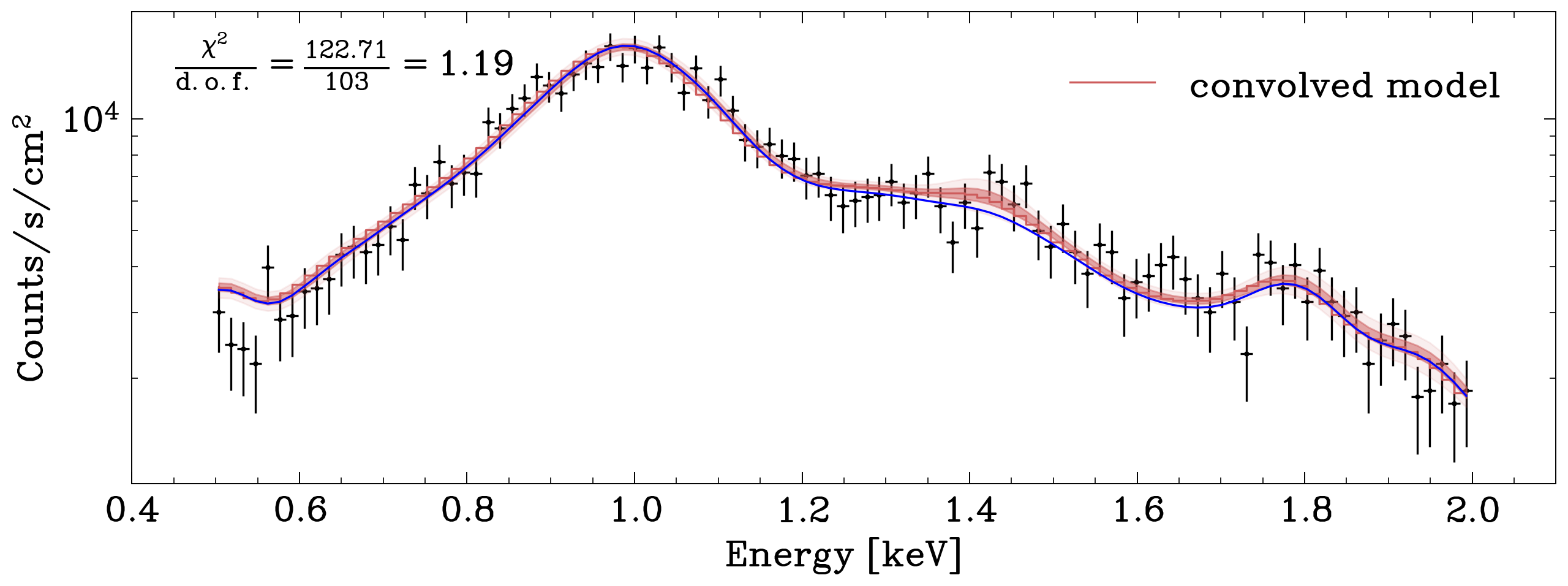}}\\[-2ex]  
 \caption{The same as the above figure, but for the corresponding \textit{Chandra} mock observation.}
 \label{fig:chandra spec fit}
\end{figure}

First, we calculate a central cooling luminosity and plot this against cavity power, to test the energy balance between heating and cooling. The central cooling luminosity is defined as the luminosity within a radius $r_\text{cool}$, where this radius is selected such that the cooling timescale $t_\text{cool}$ within $r_\text{cool}$ is equal to $7.7$ Gyrs \citep{MacconiGrandi2022}. We assume that the halo is approximately in thermodynamic balance, since the luminosity of an actively cooling gas component is typically fairly small compared to the total luminosity ( $\lesssim20\%$ and frequently $<10\%$) \citep{RaffertyMcNamara2006}, meaning that $L_\text{cool}$ is the luminosity which must be matched by mechanical heating. We calculate $r_\text{cool}$ from the emission-weighted radial profiles, through the cooling timescale

\begin{equation}
    t_\text{cool}(r) = \frac{E_\text{th}(<r)}{L_\text{X,bol}(<r)},
\end{equation}

\noindent where $E_\text{th}(<r)$ is the sum of the internal energy (assuming ideal gas with mean molecular weight $\mu = 0.62$) as well as the work done to the gas as it cools isobarically \citep{Hlavacek-LarrondoMcDonald2015}. We calculate $E_\text{th}(<r)$ with the ISM-free filtered field within radius $r$, and $L_{X,bol}(<r)$ is the total \textit{bolometric} ($0.1-100$ keV) luminosity within this radius. We then select $r_\text{cool}$ such that $t_\text{cool}(r_\text{cool}) = 7.7$ Gyrs.

We use two separate methods to calculate the central bolometric cooling luminosity. First, we define $L_\text{cool} \equiv L_{X,bol}(<r_\text{cool})$. Second, to better match the results and spread of real halos, we calculate an \textit{observational} bolometric luminosity with our mock \textit{Chandra} and \textit{LEM} events, $L_\text{cool}^{obs}$. To do this we extract the spectra within $r_\text{cool}$ using \textsc{soxs}, and fit with the Bayesian X-ray parameter estimation code \textsc{bxa} \citep{Buchner2016}, which connects the spectral-fitting software \textsc{(Py)XSPEC} \citep{Arnaud1996} to the nested-sampling code \textsc{Ultranest} \citep{2016S&C....26..383B,2019PASP..131j8005B, 2021JOSS....6.3001B}. We fit the spectra in the observed-frame energy range $0.5-2.0$ keV with a \textit{bvapec * tbabs} absorbed \ac{cie} emission model, using the Cash statistic as is appropriate for spectra which may have low bin-counts. We choose ${kT, n_e, O, Ne, Si, S, Fe}$ and the \textit{velocity} as free parameters for the fit, and freeze the other quantities to their known values for the Milky-Way foreground column density and the redshift, and to $0.3 Z_\odot$ with solar abundances for the remaining elemental parameters. Using \textsc{bxa}, we generate a chain of posterior parameter samples for the free parameters.

For each sample in the chain, we generate a bolometric luminosity for the intrinsic, \textit{unabsorbed bvapec} model for the halo, using the \textsc{XSPEC} commands \textsc{\textbf{setEnergies}} and \textsc{\textbf{calcLumin}}. Therefore, we generate a posterior probability distribution of bolometric luminosities, from which we calculate the median as the maximum-likelihood value, and the $16^{th}$ and $84^{th}$ percentiles as the $1-\sigma$ error.

\begin{figure}
	\includegraphics[width=\columnwidth]{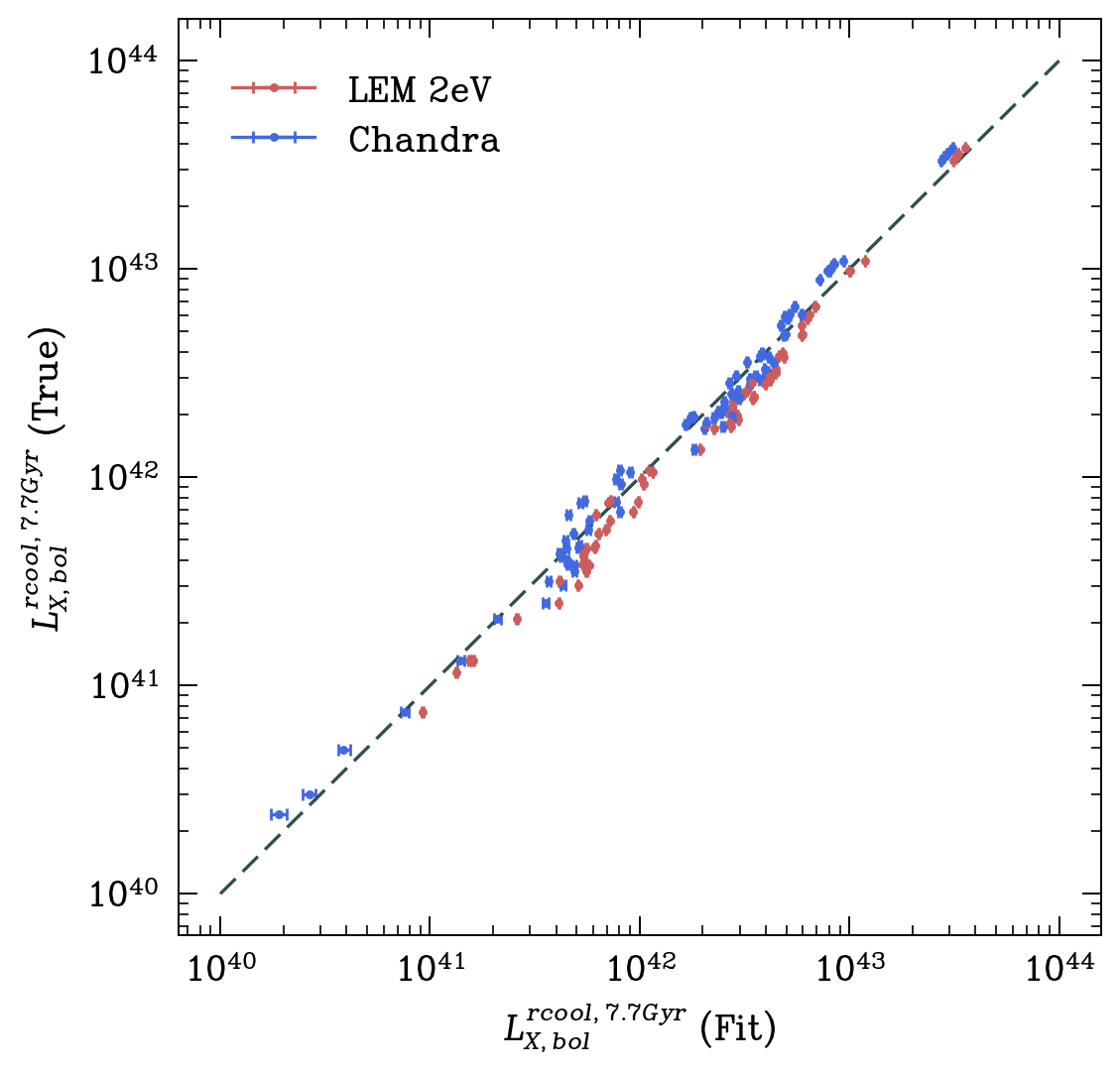}
    \caption{We plot the intrinsic luminosity against the fit-obtained luminosity from an \textit{bvapec*tbabs} model applied to both \textit{Chandra} and \textit{LEM} within $r_\text{cool}$. We plot only fits with $0.5<\chi^2/\text{d.o.f.}<5$. Note that some \textit{LEM} spectra at the low-$L_X$ end were not well fit but we still show the corresponding \textit{Chandra} values. Even for jet-disrupted halos the single-component fit recovers well the true intrinsic source luminosity.}
    \label{fig:lem_vs_chandra_vs_true_Lx}
\end{figure}

It is important as a check to compare the summed intrinsic luminosity, as calculated from the particle properties, with that obtained via a spectral fit to the subsequently projected and observed photons. Since jet feedback has a significant impact on the thermodynamic and emission properties of halos, features like bubbles could compromise the accuracy of a finite (thermodynamic and metal-) component fit. For example, the inflation of cavities can cause asymmetrical heating within the lobes themselves and also compression of the ambient atmosphere at the location of shock fronts, increasing the local density and luminosity. In that case, there would be at least three major components to the gas; hot bubble fluid, compressed gas in and around the cavity shell, and the undisturbed halo atmosphere. In Figs \ref{fig:LEM spec fit} and \ref{fig:chandra spec fit} we show example spectral fits for both \textit{LEM} and \textit{Chandra} respectively and in Fig. \ref{fig:lem_vs_chandra_vs_true_Lx} we plot the ground-truth bolometric luminosity as summed from the particles in our datasets to the luminosities obtained via spectral fits using both of these instruments. We see that even for our sample of jet-disrupted halos the single-component fit recovers the true intrinsic source luminosity. LEM luminosities tend to slightly overpredict, though this may be a coarse-grain effect due to the cutting-out of a circular aperture with larger pixels compared to Chandra. Alternatively, this may be a projection effect due to added light from larger radii than $r_\text{cool}$, and in fact \textit{LEM} is capturing this better than \textit{Chandra}. We proceed in our analysis mainly making use of the intrinsic luminosity quantity with the knowledge that this value will be essentially identical to that gained via a full mock observation and spectral fit. \\

\subsection{Does Cavity Power Offset Cooling?}

\begin{figure}
	\includegraphics[width=\columnwidth]{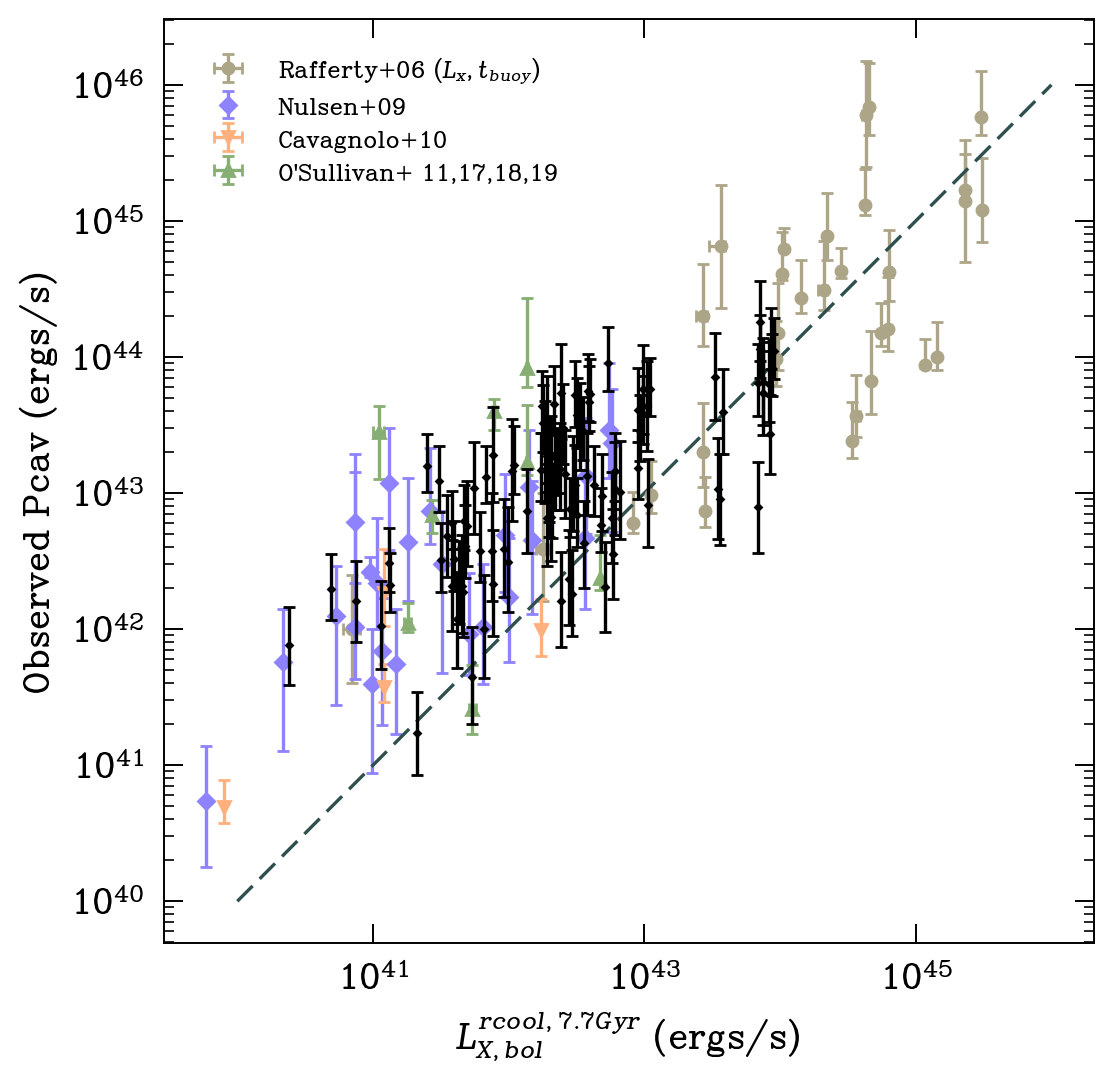}
    \caption{We plot our \hyenas cavity powers (the sum of all cavities seen for each snapshot) in black against the bolometric X-ray luminosity within the cooling radius, and compare to the data of \citet{RaffertyMcNamara2006}, \citet{NulsenJones2009}, \citet{CavagnoloMcNamara2010}, \citet{O'SullivanGiacintucci2011}, \citet{O'SullivanPonman2017}, \citet{O'SullivanKolokythas2018}, and \citet{O'SullivanSchellenberger2019}. Where appropriate we have combined individual cavity powers and their errors for a fair comparison to the \hyenas points. The \hyenas values are using the midpoint pressure. We find very good agreement with real systems, with \hyenas able to sufficiently heat halos at all luminosities, and apparently \textit{overheat} at low luminosities ($\lesssim 10^{43}$ erg/s) in agreement with observations.}
    \label{fig:Pcav_vs_Lx}
\end{figure}

In Fig. \ref{fig:Pcav_vs_Lx}, we show the mock-observed cavity power against the luminosity within the cooling radius, and compare to the observational results of \citet{RaffertyMcNamara2006} and \citet{NulsenJones2009}. To ensure a fair comparison with the \citet{RaffertyMcNamara2006} sample, we take their \textit{total} cavity power for each halo, and plot against their \textit{total} luminosity within the cooling radius. For the data of \citet{CavagnoloMcNamara2010}, \citet{O'SullivanGiacintucci2011}, \citet{O'SullivanPonman2017}, \citet{O'SullivanKolokythas2018}, and \citet{O'SullivanSchellenberger2019}, we also sum individual cavity powers where appropriate, and plot against the cooling luminosity (this data is compiled in \citet{EckertGaspari2021}). 

In all we have a reasonably good match to observational cavity powers. Both the \hyenas and observational data of \citet{NulsenJones2009}, \citet{CavagnoloMcNamara2010}, \citet{O'SullivanGiacintucci2011}, \citet{O'SullivanPonman2017}, \citet{O'SullivanKolokythas2018}, and \citet{O'SullivanSchellenberger2019} appear too shallow below $L_{X,bol} \sim 10^{43}$ erg/s and, therefore, lie above the one-to-one heating/cooling balance. Hence we find that halos in the groups regime are \textit{overheated}. In this scenario, the feedback from the core results in a net heating which over-pressurises the halo, removing cooler gas and reducing the baryon fraction, and quenching star formation \citep[][]{McNamaraBirzan2008}. \citet{EckertGaspari2021} showed that \simba matches the hot gas fraction in the group regime, and it is encouraging that we find further evidence that the feedback model is matching real halos~\citep[see also][]{RobsonDave2020}. 

From Fig. \ref{fig:Pcav_vs_Lx}, we see that at the high-luminosity end the \hyenas cavities \textit{may} be under-luminous, which in this scenario could explain why the halo gas fraction for high-mass groups and clusters in \simba is too high compared to observational constraints; there is insufficient feedback in high-mass, high-luminosity systems. However, we are limited by the small number of halos that we have at these high luminosities and we have several examples at cavity powers near $10^{44}$ erg/s, agreeing with the data and lying on or close to the equality line where heating is more or less exactly balanced by cooling.  Given our \hyenas\ sample we cannot robustly tell whether, had we simulated more massive halos, whether they would lie below the 1:1 line or whether they would veer off and lie on the line as the observations do.

Overall, the \simba feedback model reproduces the $\sim 2-3$ orders of magnitude dynamic range in cavity powers in galaxy groups, roughly from $10^{41}-10^{44}$ erg/s, and matches the observed trend with cooling luminosity. This indicates that the correct coupling of the mechanical and X-ray luminosities is achieved through the \simba model, with heating from the central \ac{agn} via jets providing a substantial fraction, if not the majority, of the required energy to the group gas and preventing large-scale cooling flows. In the low-mass, low-luminosity range, it is likely halos undergo a net heating due to \ac{agn} activity. It is left to future work to conduct a similar analysis in the clusters regime, but the highest luminosity groups we do have in this sample have heating and cooling in balance, much like the observational data.

\subsection{Relationship Between Cavity Power and Accretion Rate}\label{sec:Bondi_vs_Cavity_Power}
To understand how the cavities are powered and how this is related to the dominant mode in which the gas is accreting onto the cental black hole, we consider the relationship between the mass accretion rates (due to both \ac{bhl} and torque-limited accretion) and the cavity power inferred from our mock-observation routine. 

\begin{figure} 
	\includegraphics[width=\columnwidth]{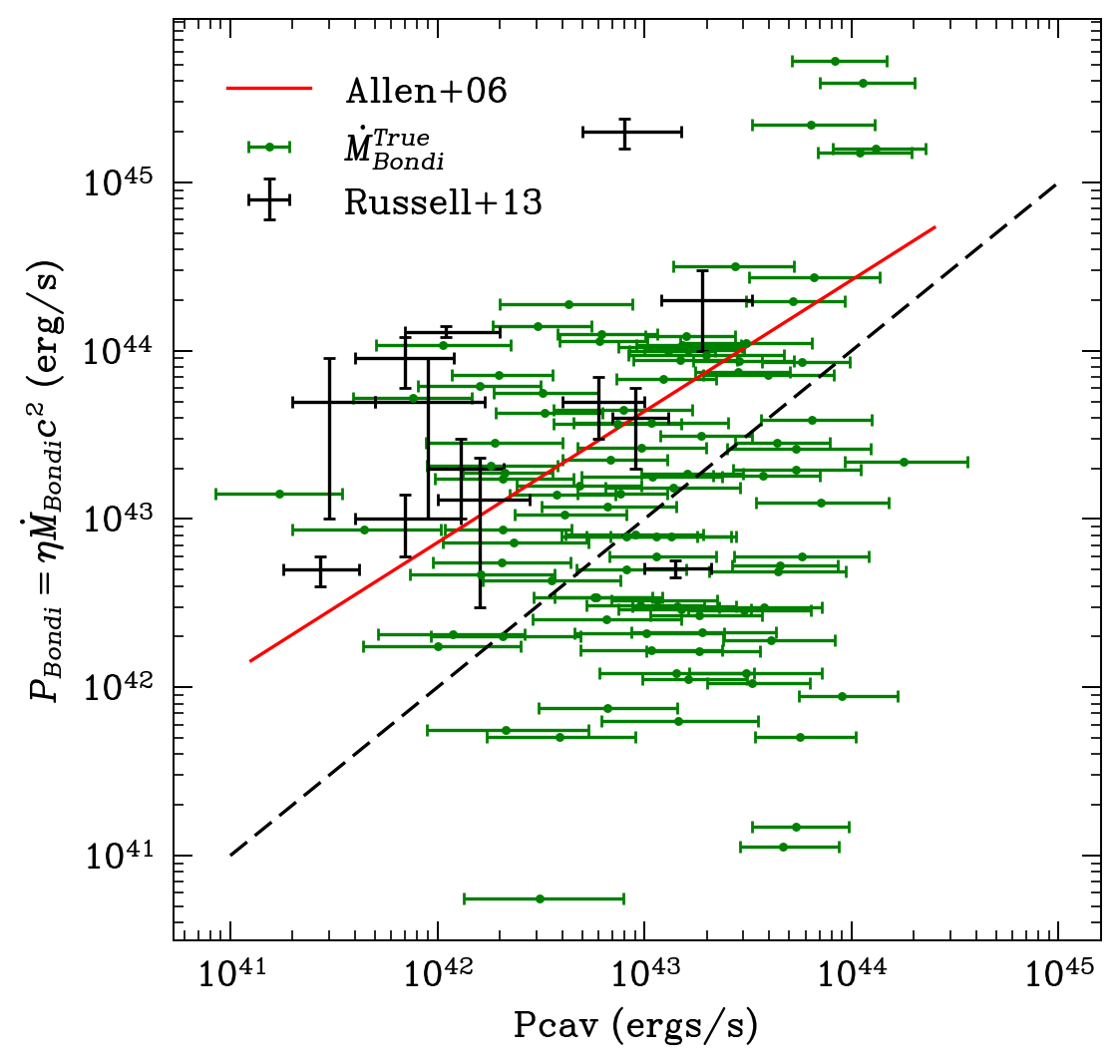}
    \caption{We plot Bondi accretion powers against mock-observed cavity power (using $t_{buoy}$) and compare to the derived observational log-linear relation of \citet{AllenDunn2006} and the datapoints of \citet{RussellMcNamara2013}.}
    \label{fig:Pbondi_Pcav}
\end{figure}

Fig. \ref{fig:Pbondi_Pcav} shows the Bondi accretion power against the cavity power, which is a correlation of interest since it is readily available to observers. There is a large scatter in the \hyenas relationship, with no obvious correlation. This is unlike the strong linear trend found in \citet{AllenDunn2006}, but we do see overlap with the results for individual systems of \citet{RussellMcNamara2013}.

The overlap at the high-power end is encouraging, showing that \hyenas\ can reproduce the observed range of accretion powers.  However, we also predict that there should be a large population of Bondi-dominated systems at low accretion rates hosting powerful cavities.   Such low Bondi powers are not seen in observations, which could owe to observational sensitivity limits.

It may be that those low-accretion cavities are powerful enough to disrupt the hot atmospheres and therefore the supply of hot gas to the central black hole.  However, it could also owe to the large stochasticity in Bondi accretion in the \simba\ model~\citep{ThomasDave2019}, as it is quite sensitive to the number of hot gas fluid elements within the BH kernel which can vary strongly in hot halos due to low resolution.  \citet{ThomasDave2021} mitigated this effect by smoothing the accretion rate over a fixed (short) timescale.  In our case, however, we have a timescale for the jet event, namely the same timescale used for the \textit{Direct} bubble lifetime as introduced in \hyperref[sec:Cavity Analysis]{\S3}. For our purposes this provides a more physically meaningful smoothing scale. We did however check and find that our results are qualitatively the same if we instead used the total accretion power averaged over a narrow timescale at the observation snapshot (i.e. a value obtainable by observers).

\begin{figure*}
\captionsetup[subfigure]{labelformat=empty}
 \subfloat[]{\includegraphics[width=0.48\textwidth]{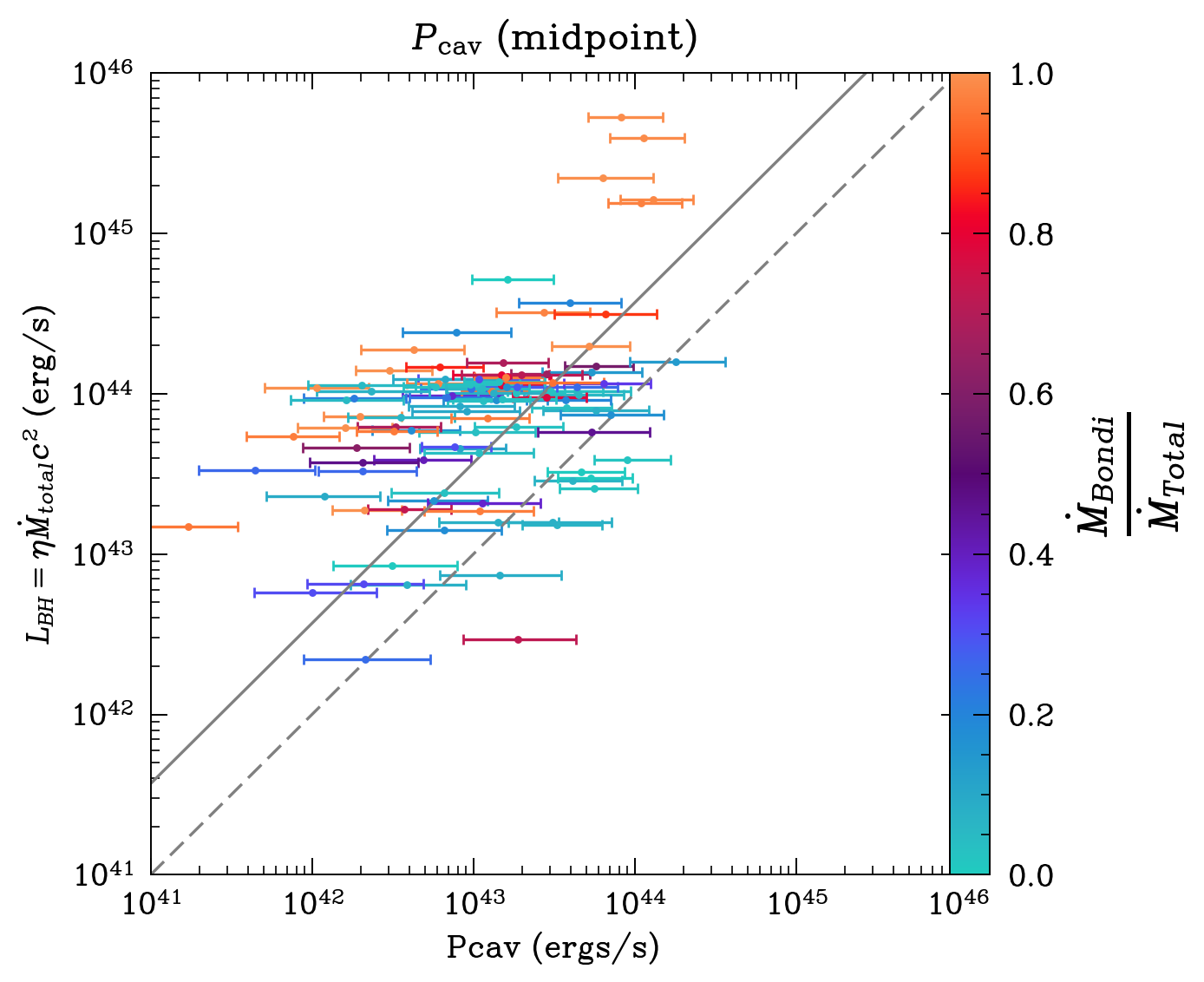}}\hfill
 \subfloat[]{\includegraphics[width=0.48\textwidth]{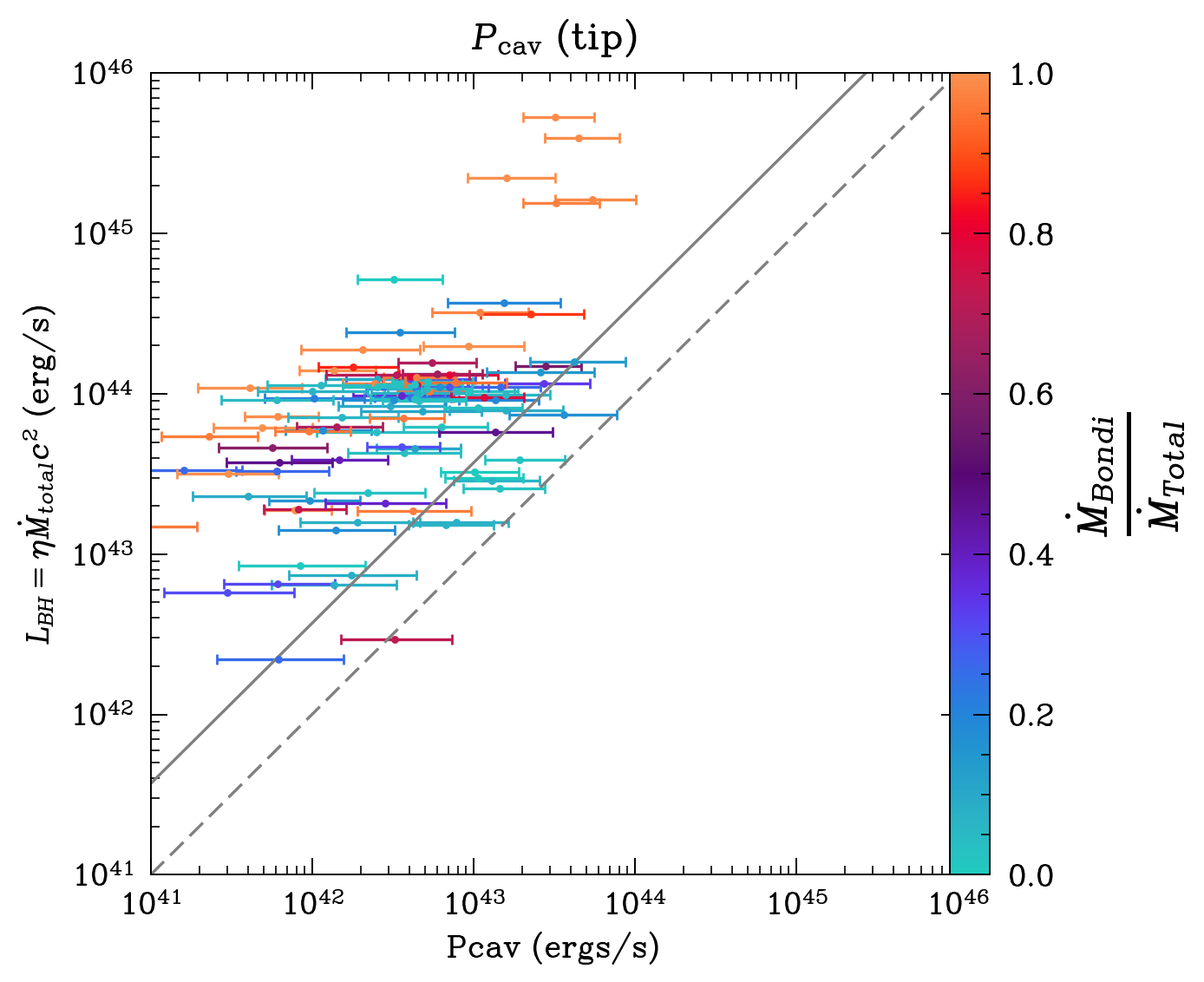}}\\[-2ex]  
 \caption{\textit{Left:} We plot the black hole luminosity associated with the total accretion rate that can go into jets/winds against the mock-observed cavity powers obtained with the midpoint pressure. The dashed line shows equality between the cavity power and the total luminosity ($L_{BH}$), whereas the solid line corresponds to equality between the cavity and the maximum energy put out in winds/jets ($= \epsilon_f L_{BH}$, where $\epsilon_f$ has a maximum value of $0.27$). The accretion power is calculated as the mean over the jetting period, although the results are insensitive to instead using the instantaneous accretion powers at the observation snapshot. Not all cavity powers appear accounted for by the \textit{recent} \ac{agn} accretion. \textit{Right:} As in the left-hand plot but now with the enthalpy calculated from the pressure at the bubble's outermost tip. We see a more encouraging picture with this enthalpy value, with \textit{almost all} cavities able to be accounted for from the recent accretion rate. }\label{fig:P_accretion_total_vs_Pcav}
 \end{figure*}

In the left-hand plot of Fig.~\ref{fig:P_accretion_total_vs_Pcav}, we show the total accretion power (averaged over the aforementioned jetting timescale) as well as the power in jets plotted against the mock-observed cavity power. We indicate the relative contributions of both Bondi and torque-limited accretion rates on the plot.

Almost all cavities lie above the line at which the \ac{agn} can fuel them with recent accretion activity. At the low power range, there are a few anomalies in each plot that appear below the line $P_\text{cav} = P_\text{accretion}$. In the right-hand plot of Fig. \ref{fig:P_accretion_total_vs_Pcav}, we show the same data, except that for the cavity enthalpy we use the enthalpy derived from the pressure at the bubble's outermost tip, instead of at the midpoint. We obtain an almost-perfect partition of points into the upper-half plane, further corroborating our claim made in \hyperref[subsec:Cavity Power Estimation]{\S3.4} --- that this is perhaps the more useful estimate of the bubble enthalpy. Although artificially extending the timescale over which the \ac{agn} winds energy is summed \textit{can} bring the \ac{agn} energy in-line with the cavity enthalpy in \hyperref[subsec:Cavity Power Estimation]{\S3.4} (Fig. \ref{fig: True vs observed energies}), this extension of the time-range considered \textit{does not} bring the accretion power in-line here, indicating that indeed it is the cavity enthalpy that must be altered (through use of the bubble tip pressure) and not the estimate of the timescale over which the jets are actively inflating the cavity. With this estimate (cavity tip pressure), the accretion rate is able to fully explain the rate of energy transfer to the bubble and eventually to the \ac{igrm}. Where the bubble appears underpowered in these plots indicates non-adiabatic losses, for instance from shocks, or from other feedback modes being active as well as the jets.

Looking now at which accretion mode is most active when the bubbles are produced, we see that the very highest-power jet-events ($P_{\dot{M}} \gtrsim 10^{44.5}$ ergs/s) are powered by Bondi accretion, but the remainder are powered by a mix of Bondi and torque-limited accretion, with no apparent relation to cavity power. There are many cavities inflated when the black hole is powered predominantly by torque-limited accretion ($64\%$). This implies that caution must be used in the groups regime when estimating accretion rates; there may be a significant fraction of the accretion provided by the disk, raising the total accretion rate above that of the Bondi rate and adding to the powers of cavities. Cavities which appear significantly under-powered based on their accretion rates are likely to be in a highly active quasar state, with jets contributing less to the total power output at high Eddington fractions. 

Overall, we find that the simulated \hyenas\ groups display jet-driven bubbles that are in good quantitative agreement with observations of bubbles in similar-mass systems.  This is somewhat remarkable given the crude nature of the implementation of jets in the \simba\ model, necessitated by poor resolution.  The bubbles identified in mock X-ray maps using tools analogous to those used by observers show similar cavity powers, and are generally sufficient to counteract cooling in the halo.  In fact, in lower-mass groups they can significantly exceed cooling losses, which explains why \simba's groups are strongly evacuated of baryons~\citep{ApplebyDave2021, SoriniDave2022}.  The power injected by the jet exceeds that seen in the cavity, as expected based on there being some radiative losses and/or energy deposition outside the cavity, and this is only exposed in \hyenas if one measures the pressure at the bubble's leading tip rather than its midpoint.

\section{Conclusions} \label{sec:Discussion}

In this work, we present the first quantitative investigation of X-ray cavities in a large-scale cosmological model suite of simulations, calibrated solely to galaxy properties, that properly straddles simulation and observation. By taking an observer-centric approach to analysing our \hyenas suite of galaxy group zooms, we are able to make predictions and comparisons realistically in the context of existing X-ray catalogues. Using mock X-ray maps generated from the \PyXSIM and \SOXS software packages, we fully mimic the analyses carried out by observers on real data. We use model-subtracted residual maps, unsharp-masked images, and smoothed X-ray flux maps to identify and measure cavity properties. In addition, we use the CADET machine-learning cavity-identifying package and find that it is broadly very useful for identifying potential cavities for follow-up. We create spectra within $r_\text{cool}$ for each system using \SOXS, and measure bolometric luminosities through a Bayesian fitting approach.  Our overarching conclusion is that we demonstrate, for the first time within a fully cosmological setting, how jets from black hole activity can inflate bubbles that deposit sufficient energy in their environments to offset radiative cooling, thereby suppressing cooling flows to yield quenched galaxies. This in a sense "completes the loop" between galaxies, their black holes, and their circum-galactic media in order to allow holistic studies of the process of galaxy transformation.

Based on our analysis in this work, we draw the following conclusions:
\begin{itemize}
    \item \hyenas successfully produces many X-ray cavities. These coincide with depressions in the density maps and with high outgoing velocities in the radial velocity projection maps. Comparing to the sample of \citet{RaffertyMcNamara2006} and \citet{ShinWoo2016}, we find that the dimensions of the \hyenas cavities in terms of semi-major axis, surface area, and eccentricity in relation to the bubble midpoint distance match very closely both the trend and scatter of observed cavities. Furthermore our cavity timescales also agree well with observations, with an average buoyancy timescale of $t_\text{buoy} = 49.2 \pm 19.5$ Myr.
    \item We present King model-subtracted residual maps and unsharp-masked maps in \hyperref[sec:Cavity Analysis]{\S3} and also case-studies of a small number of selected halos in \hyperref[sec:Case Studies]{\S4}. We demonstrate that the \simba model is in the very least qualitatively accurate in reproducing the observed cavity populations of real systems. Through comparison with the pressure maps and radial velocity maps obtained directly from the particle data, we find that cavities are associated with pressure-fronts, and large radial velocities relative to the ambient medium. We also observe large sound waves emanating from the central \ac{agn}, similar to those seen in the Perseus cluster. We leave it to future work to quantify these various features, and most importantly the relative importance of each mechanism in transferring energy from the \ac{agn} to the hot atmosphere.
    
    \item Our bubble enthalpies (for a given snapshot) tend to appear over-powered compared to the energy injected by recent \ac{agn} feedback. This is remedied remarkably well if we instead simply calculate an enthalpy using the pressure at the bubble's \textit{outermost tip} for the $PdV$ work estimate, hinting that this may give a closer estimate of the "effective" pressure of the bubble over its expansion. This prediction that the bubble should do work preferentially along the 'axis of least resistance' is supported by our observation of high bubble eccentricities (of which $65\%$ have the semi-major axis aligned radially) in agreement with observations. Our simulations show excellent agreement with the observed data when we use the bubble \textit{midpoint} pressure in terms of the $P_{cav} = L_{X}$, indicating that this result (that the bubble tip pressure is actually more accurate) should generalise to estimating the energy of the \ac{agn} in real systems.

    \item We compute the mock-observed cavity powers and compare to the bolometric X-ray luminosity within the $7.7$ Gyr cooling radius. We find good agreement at the higher luminosity end with the observational data of \citet{RaffertyMcNamara2006} and excellent agreement at the lower-$L_X$ end with that of \citet{NulsenJones2009}, \citet{CavagnoloMcNamara2010}, \citet{O'SullivanGiacintucci2011}, \citet{O'SullivanPonman2017}, \citet{O'SullivanKolokythas2018}, and \citet{O'SullivanSchellenberger2019}. This relation lies on the line $P_{cav} = L_{X}$, except at lower luminosities ($L_{X,bol} \lesssim 10^{43}$ erg/s) where both the data and our results lie to higher cavity powers.  Therefore, we find that the \simba feedback model is sufficiently powerful in terms of energy output to balance cooling in the \ac{igrm} and prevent a large-scale cooling-flow, and potentially overheats the halo gas at low halo masses. This may be the reason why \simba tends to have a lower hot gas fraction in the groups regime compared to other simulations, which is in line with observational data \citep{RobsonDave2020, EckertGaspari2021}.
    \item  Plotting the Bondi power against the cavity power we observe a very large scatter around the one-to-one relationship. This is unlike the strong correlation found in \citet{AllenDunn2006} but similar to the results in \citet{RussellMcNamara2013}. However, we predict a large population of powerful cavities in systems with low Bondi accretion rates, hinting at a bias in observations, and suggesting alternate modes of accretion onto the \ac{smbh} such as described in \citet{Angles-AlcazarDave2017, PrasadSharma2017}.  In \hyenas\ we find that Bondi accretion dominates in the very most massive systems, but outside of those there are roughly comparable numbers of torque-dominated and Bondi-dominated jet-driving black holes.

\end{itemize}

We demonstrate the feasibility of a large survey of X-ray halos in a cosmological suite of simulations, and how they should be analysed in order to properly compare with catalogues of real systems. We emphasise the importance of this kind of analysis, and the need for feedback models to be tested on small (group or cluster) scales and not just compared against summary galaxy population properties. We envision the simulated cavity population can act as a useful calibration barometer for large-scale simulations in the future, especially as simulation resolution increases, and as the real cavity population is better understood with upcoming  X-ray missions. The main barrier to this is the time required to manually identify and measure cavities. However, this can at least partly be automated with machine-learning pipelines, such as \textsc{cadet}.

\section*{Software}
This project made use of the following external software packages for analysis;
\textsc{pyXSIM 4.3.0} \href{https://github.com/jzuhone/pyxsim}{ \color{orange} \color{orange} \faGithub} \citep{ZuHoneHallman2016}, \textsc{SOXS 4.6.0} \href{https://github.com/jzuhone/soxs}{ \color{orange} \faGithub} \citep{ZuHoneVikhlinin2023}, \textsc{yT Project 4.2.2} \href{https://github.com/yt-project/yt}{ \color{orange} \faGithub} \citep{TurkSmith2011}, \textsc{BXA 4.1.1} \href{https://github.com/JohannesBuchner/BXA/tree/master}{ \color{orange} \faGithub} \citep{Buchner2016} and \textsc{UltraNest 3.6.2} \href{https://github.com/JohannesBuchner/UltraNest}{ \color{orange} \faGithub} \citep{2021JOSS....6.3001B}, \textsc{Scipy 1.11.2} \href{https://github.com/scipy/scipy}{ \color{orange} \faGithub} \citep{2020SciPy-NMeth}, \textsc{Numpy 1.24.3} \href{https://github.com/numpy/numpy}{ \color{orange} \faGithub} \citep{harris2020array}, \textsc{mpi4py 3.1.4} \href{https://github.com/mpi4py/mpi4py}{ \color{orange} \faGithub} \citep{DalcinPaz2011}, \textsc{CAESAR 0.2b0} \href{https://github.com/dnarayanan/caesar}{ \color{orange} \faGithub} \citep{Thompson2015}, \textsc{PyGadgetReader 2.6} \href{https://github.com/jveitchmichaelis/pygadgetreader}{ \color{orange} \faGithub} \citep{Thompson2014}, \textsc{AstroPy 5.3.3} \href{https://github.com/astropy/astropy}{ \color{orange} \faGithub} \citep{AstropyCollaborationRobitaille2013}, \textsc{Corner 2.2.2} \href{https://github.com/dfm/corner.py}{ \color{orange} \faGithub} \citep{Foreman-Mackey2016}, \textsc{(Py)CADET 0.1.4} \href{https://github.com/tomasplsek/CADET}{ \color{orange} \faGithub} \citep{PlsekWerner2024}, \textsc{Matplotlib 3.7.3} \href{https://github.com/matplotlib/matplotlib}{ \color{orange} \faGithub} \citep{Hunter2007}, \textsc{smplotlib 0.0.9} \href{https://github.com/AstroJacobLi/smplotlib}{ \color{orange} \faGithub} \citep{Li2023}, \textsc{cmasher 1.6.3} \href{https://github.com/1313e/CMasher}{ \color{orange} \faGithub} \citep{vanderVelden2021}, \textsc{pandas 2.1.0} \href{https://github.com/pandas-dev/pandas}{ \color{orange} \faGithub} \citep{ThepandasdevelopmentTeam2023}, \textsc{h5py 3.9.0} \href{https://github.com/h5py/h5py}{ \color{orange} \faGithub} \citep{ColletteKluyver2022}, \textsc{XSPEC 12.13.1} \href{https://heasarc.gsfc.nasa.gov/xanadu/xspec/}{ \color{orange} \faChain} \citep{ArnaudDorman1999}, and \textsc{SAOImageDS9 8.4.1} \href{https://sites.google.com/cfa.harvard.edu/saoimageds9}{ \color{orange} \faChain} \citep{SmithsonianAstrophysicalObservatory2000}. We are grateful to their respective authors for making them available.

\section*{Acknowledgements}
We thank the anonymous referee for an insightful and constructive referee report.
FJJ would like to acknowledge the support of the Science and Technology Facilities Council. He would also like to thank Ewan O'Sullivan and Nhut Truong for useful conversations, Britton Smith for advice regarding \yT\, and John ZuHone for advice regarding the \PyXSIM and \SOXS software packages, as well as for making these packages publicly available. 

AB acknowledges support from the Natural Sciences and Engineering Research Council of Canada (NSERC) through its Discovery Grant program. He also acknowledges support from the Infosys Foundation via an endowed Infosys Visiting Chair Professorship at the Indian Institute of Science, and from the Leverhulme Trust via a Visiting Professorship at the Institute for Astronomy, University of Edinburgh.
RD is supported by the STFC RC grant ST/Y001117/1.
WC is supported by the STFC AGP Grant ST/V000594/1, the Atracci\'{o}n de Talento Contract no. 2020-T1/TIC-19882 was granted by the Comunidad de Madrid in Spain, and the science research grants were from the China Manned Space Project. He also thanks the Ministerio de Ciencia e Innovación (Spain) for financial support under Project grant PID2021-122603NB-C21 and HORIZON EUROPE Marie Sklodowska-Curie Actions for supporting the LACEGAL-III project with grant number 101086388.  DR is supported by the Simons Foundation.

These original \hyenas\ simulations were performed using the DiRAC@Durham facility managed by the Institute for Computational Cosmology on behalf of the STFC DiRAC HPC Facility (www.dirac.ac.uk) with the DiRAC Project: ACSP252, titled Simba Zoom Simulations of Galaxy Groups. The equipment was funded by BEIS capital funding via STFC capital grants ST/P002293/1, ST/R002371/1 and ST/S002502/1, Durham University, and STFC operations grant ST/R000832/1. DiRAC is part of the National e-Infrastructure. Computations and analysis specific to this work were made possible with support from SciNet and the Niagara supercomputing cluster, operated by the Digital Research Alliance of Canada (\href{https://alliancecan.ca/en}{alliancecan.ca}).

For the purpose of open access, the author has applied a Creative Commons Attribution (CC BY) licence to any Author Accepted Manuscript version arising from this submission.

\section*{Data Availability}
The relevant data will be shared upon reasonable request to the corresponding author.



\bibliographystyle{mnras}
\bibliography{example} 




\appendix


\bsp	
\label{lastpage}
\end{document}